\documentclass[preprint,prd,aps,showpacs,preprintnumbers,amsmath,amssymb]{revtex4-1}
\usepackage{graphics,epsfig,subfigure}
\usepackage{diagbox}
\usepackage[usenames]{color}
\usepackage{color}
\usepackage{graphicx}
\usepackage{amsfonts}
\usepackage{txfonts}
\usepackage{indentfirst}
\usepackage{booktabs}
\usepackage{hyperref}
\hypersetup{hypertex=true,backref=true,colorlinks=true,linkcolor=blue,anchorcolor=blue,citecolor=blue}
\usepackage{float}
\usepackage{multirow}
\usepackage[section]{placeins}
\usepackage{array}
\usepackage{amsmath}

\usepackage{CJK}   

\begin{document}
\begin{CJK}{UTF8}{gbsn}
\title{\Large \bf Ellis--Bronnikov wormhole in Quasi-topological Gravity}
\author{Gen Li}
\author{Yong-Qiang Wang\footnote{yqwang@lzu.edu.cn, corresponding author}}
\affiliation{ $^{1}$Lanzhou Center for Theoretical Physics, Key Laboratory of Theoretical Physics of Gansu Province,
	School of Physical Science and Technology, Lanzhou University, Lanzhou 730000, China\\
	$^{2}$Institute of Theoretical Physics $\&$ Research Center of Gravitation, Lanzhou University, Lanzhou 730000, China}

\begin{abstract}
We construct higher-dimensional traversable wormholes in quasi-topological gravity (QTG) supported by a phantom scalar field.
Using a static, spherically symmetric ansatz, we numerically analyze how quasi-topological gravity corrections affect the geometry and physical properties of the wormhole solutions.
The resulting wormhole solutions are symmetric about the throat.
Negative mass can arise for certain choices of parameters.
For certain parameter ranges, the scalar charge $\mathcal{D}$ of the phantom field rapidly decreases with increasing the higher-curvature coupling parameter $\alpha$ and approaches zero.
Moreover, by changing $\alpha$, the overall level of the Kretschmann scalar is also lowered.
Finally, for sufficiently large $\alpha$, $-g_{tt}$ becomes close to zero near the throat, exhibiting a ``horizon''-like structure.
\end{abstract}

\maketitle

\section{Introduction}\label{Sec1}

Wormholes are hypothetical spacetime structures predicted by general relativity. 
Owing to their potential to connect distant regions of spacetime---or even different universes---they have long attracted significant interest in theoretical physics. 
The concept of a wormhole can be traced back to 1916, when L.~Flamm, in analyzing the Schwarzschild solution of Einstein's field equations, first pointed out the possibility of a ``tunnel''-like geometry in the corresponding spacetime \cite{Flamm:2015ogy}. 
In 1935, Einstein and Rosen proposed the celebrated Einstein--Rosen bridge to address singularities appearing in general relativity and classical electrodynamics \cite{Einstein:1935tc}. 
Through an appropriate coordinate transformation of the Schwarzschild metric, they obtained a geometric ``bridge'' connecting two asymptotically flat regions. 
Subsequent studies, however, showed that the throat of such a bridge is non-traversable: any attempt to traverse it would lead to a rapid collapse \cite{Kruskal:1959vx,Fuller:1962zza}. 

In 1957, Wheeler coined the term ``wormhole'' for this type of spacetime structure \cite{Misner:1957mt}, thereby establishing wormholes as a formal topic in theoretical physics. 
Later, in 1973, Ellis and Bronnikov constructed traversable wormhole solutions \cite{Ellis:1973yv,Ellis:1979bh,Bronnikov:1973fh,Kodama:1978dw}. 
An important feature of these solutions is that keeping the wormhole throat open typically requires a ``phantom'' scalar field with a negative kinetic term. 
In 1988, Morris and Thorne provided a systematic analysis \cite{Morris:1988cz}, emphasizing that traversable wormholes generally demand so-called exotic matter. 
Such matter components violate the energy conditions , posing a serious challenge to classical general relativity.

To address this issue, subsequent efforts have proceeded along several directions. 
On the one hand, one may attempt to construct wormhole models that reduce the need for exotic matter, or replace it by other matter sources. 
Examples include thin-shell wormholes, in which exotic matter is confined to an arbitrarily thin neighborhood of the throat in order to minimize the amount of exotic matter required \cite{Poisson:1995sv}, as well as wormhole solutions found in alternative frameworks such as Einstein--Dirac--Maxwell (EDM) theory, which may avoid the need for exotic matter \cite{Blazquez-Salcedo:2020czn}. 
On the other hand, a particularly important direction is to explore wormholes in modified gravity, hoping that appropriate modifications of Einstein gravity itself can avoid or at least alleviate energy-condition violations \cite{Lobo:2009ip,Banerjee:2021mqk}. 
 A variety of modified-gravity theories have been developed, including scalar--tensor theories, $f(R)$ gravity, Horndeski-type models, and related extensions \cite{Moffat:2005si,Famaey:2011kh,deRham:2014zqa,Hinterbichler:2011tt,DeFelice:2010aj,Cai:2015emx,Maartens:2010ar,Horndeski:2024sjk}. 
While these theories have notable advantages in explaining phenomena such as cosmic acceleration and galactic rotation curves, it remains crucial to identify which frameworks can genuinely reduce (or eliminate) the need for exotic matter in wormhole spacetimes.

Among the many candidates, quasi-topological gravity (QTG) exhibits several distinctive advantages \cite{Bueno:2019ltp,Bueno:2019ycr,Bueno:2022res,Bueno:2025qjk}. 
QTG modifies the gravitational dynamics by adding suitably constructed higher-curvature invariants to the action. 
On static, spherically symmetric backgrounds, QTG yields second-order field equations, and the field equations for spherically symmetric black holes are algebraic, as for Lovelock theory. However, whereas Lovelock gravity in five dimensions has nontrivial higher-curvature contributions only up to the Gauss--Bonnet term, QTG admits higher-curvature corrections of arbitrarily high order already in $D>4$. 
Explicit quasi-topological gravities have been constructed at cubic order \cite{Myers:2010ru,Oliva:2010eb}, quartic order \cite{Dehghani:2011vu}, and quintic order \cite{Cisterna:2017umf}, as well as at arbitrary order \cite{Bueno:2019ycr}. 
Within this framework, analytic solutions of regular black holes in $D \ge 5$ have been obtained \cite{Bueno:2024dgm}, and it has been shown that finite-order corrections reduce the degree of divergence of a black hole singularity; under certain summation conditions for all-order corrections, spacetime singularities may even be completely removed, leading to globally smooth black-hole geometries with a finite Kretschmann invariant.

This naturally raises the question of whether such higher-curvature corrections can not only smooth out black-hole singularities, but also allow for traversable wormhole solutions. 
Motivated by this, we investigate wormhole spacetimes in quasi-topological gravity. 
As a starting point, we consider the static, asymptotically flat Ellis wormhole, whose geometry is simple and numerically tractable \cite{Dzhunushaliev:2013jja}. 
The supporting matter sector is also relatively simple, consisting of a phantom scalar field with a negative kinetic term, which makes the Ellis wormhole a standard model for studies of traversable wormholes
 \cite{Lobo:2005us,Sushkov:2005kj,Lobo:2005yv,Bronnikov:2012ch}. 
This model has been further generalized to include rotation, higher dimensions, and couplings to additional matter fields \cite{Kleihaus:2014dla,Dzhunushaliev:2014bya,Hoffmann:2017jfs,Yue:2023ela,Ding:2023syj,Hao:2023igi,Su:2023zhh,Nozawa:2020gzz}. 
Building on these developments, we investigate the Ellis wormhole within the framework of quasi-topological gravity and perform a systematic study. 
In particular, we focus on how the higher-curvature coupling parameter $\alpha$, the truncation order N, and the spacetime dimension $D$ affect the wormhole geometry, its physical characteristics.

The paper is organized as follows. 
In Sec.~II, we briefly review the essentials of quasi-topological gravity, present the action, derive the field equations relevant for our ansatz, and introduce the definitions of the physical quantities. 
In Sec.~III, we discuss the boundary conditions required for the numerical construction. 
In Sec.~IV, we analyze the numerical results and classify the wormhole solutions for different parameter choices, and we also present embedding diagrams to visualize the wormhole geometry. Finally, Sec.~V summarizes our conclusions and outlines directions for future work.

\section{THE MODEL}\label{sec2}

\subsection{Action}

We consider an action in $D$ spacetime dimensions consisting of a quasi-topological gravity sector and a matter sector described by a phantom scalar field. It takes the form
\begin{equation}\label{action}
I=\int \mathrm{d}^D x\,\sqrt{|g|}\left[\frac{D-2}{16\pi G}\,\mathcal{L}_g+\mathcal{L}_m\right],
\end{equation}
where $g\equiv\det(g_{ab})$. Since the matter field is minimally coupled to the geometry, we may consider the gravitational Lagrangian density $\mathcal{L}_g$ and the matter Lagrangian density $\mathcal{L}_m$ separately. The gravitational sector is taken to be
\begin{equation}\label{Lg}
\mathcal{L}_g
=R+\sum_{n=2}^{n_{\mathrm{max}}}\alpha_n\,\mathcal{Z}_n,
\end{equation}
where $\mathcal{Z}_n$ denotes the $n$th-order quasi-topological curvature density and $\alpha_n$ are the corresponding coupling constants. The matter sector is given by a phantom scalar field $\Phi$ with Lagrangian density
\begin{equation}\label{lagrangian_matter}
\mathcal{L}_m=\frac{1}{2}\,g^{\mu\nu}\,\partial_\mu\Phi\,\partial_\nu\Phi .
\end{equation}

To facilitate the derivation and analysis of the differential equations arising in quasi-topological gravity, we consider the following static, spherically symmetric (SSS) ansatz
\begin{equation}\label{metric}
\mathrm{d}s^2
=-N(l)^2 f(l)\,\mathrm{d}t^2+\frac{\mathrm{d}l^2}{f(l)}+l^2\,\mathrm{d}\Omega_{D-2}^2,
\end{equation}
where $N(l)$ and $f(l)$ are two functions to be determined. We also assume that the ansatz for the phantom scalar field depends only on the radial coordinate,
\begin{equation}\label{anzte_matter}
\Phi=\phi(l).
\end{equation}

Following the reduction method of Ref.~\cite{Bueno:2024dgm}, we reduce the gravitational action
\begin{equation}
I_g=\frac{(D-2)}{16\pi G}\int \mathrm{d}^D x\,\sqrt{|g|}\,\mathcal{L}_g
\end{equation}
by integrating over the angular coordinates on the SSS ansatz \eqref{metric}. The resulting effective action depends only on $N(l)$ and $f(l)$ and can be written as
\begin{equation}\label{reduced_action}
I_{\mathrm{g}}[N(l),f(l)]
=\frac{\Omega_{D-2}(D-2)}{16\pi G}
\int \mathrm{d}t\,\mathrm{d}l\;
N(l)\,\big[\,l^{D-1}h(\psi)\,\big]',
\end{equation}
where $\Omega_{D-2}=2\pi^{(D-1)/2}/\Gamma\!\left(\frac{D-1}{2}\right)$ is the area of the unit $(D-2)$-sphere and a prime ${}'$ denoting a derivative with respect to $l$. The function $h(\psi)$ accounts for the higher-curvature corrections and can generally be written as
\begin{equation}\label{hpsi}
h(\psi)=\psi+\sum_{n=2}^{N}\alpha_n\,\psi^n,
\end{equation}
with
\begin{equation}\label{psi_def}
\psi(l)=\frac{1-f(l)}{l^2}.
\end{equation}

In Ref.~\cite{Bueno:2024dgm}, in order to eliminate curvature singularities and obtain regular black-hole solutions, a sufficient set of conditions ensuring regularity was imposed on the coefficients $\{\alpha_n\}$:
\begin{equation}\label{eq:regularity_conditions}
\alpha_n \ge 0 \quad (\forall n), \qquad
\lim_{n\to\infty}\big(\alpha_n\big)^{1/n}=C>0 .
\end{equation}
In the present work, however, we focus on wormhole geometries, for which such restrictions are not necessary.
We therefore enlarge the parameter space and investigate how allowing negative coefficients, $\alpha_n<0$, affects the wormhole geometry and the associated energy-condition violations.
Similar explorations of negative-coupling (or negative-parameter) regions have also appeared in other modified-gravity settings \cite{Pavluchenko:2024lcl,Mehdizadeh:2019qvc,Mehdizadeh:2015jra,Baibhav:2016fot,Ilijic:2020vzu}.

Similarly, the matter action $I_m=\int \mathrm{d}^D x\,\mathcal{L}_m$ can also be reduced on the metric \eqref{metric} by integrating over the angular coordinates, yielding the action
\begin{equation}\label{eq:reduced_action_m}
I_m[\phi(l),N(l),f(l)]
=\frac{\Omega_{D-2}}{2}\int \mathrm{d}t\,\mathrm{d}l\;
N(l)\,f(l)\,l^{D-2}\,\big(\phi'(l)\big)^2.
\end{equation}

\subsection{Ordinary differential equations (ODEs)}

Varying the total action
$I[N,f,\phi]=I_{\mathrm{g}}[N,f]+I_m[\phi,N,f]$
with respect to $N(l)$, $f(l)$, and $\phi(l)$ yields three Euler--Lagrange equations. 
The first two can be expressed in terms of the gravitational equations $E_{\mu\nu}$ and the stress-energy tensor $T_{\mu\nu}$ as
\begin{equation}\label{eq:variation_components}
\begin{aligned}
\frac{16\pi G}{\Omega_{D-2} l^{D-2}} \frac{\delta I}{\delta N}
&= \frac{2 E_{tt}}{f N^2} - \frac{2 T_{tt}}{f N^2}, \\
\frac{16\pi G}{\Omega_{D-2} l^{D-2}} \frac{\delta I}{\delta f}
&= \frac{E_{tt}}{N f^2} + N E_{ll}
-\left( \frac{T_{tt}}{N f^2} + N T_{ll} \right).
\end{aligned}
\end{equation}
Variation with respect to the phantom field yields the covariant Klein--Gordon equation,
\begin{equation}\label{eq:variation_scalar}
\nabla^2 \phi = 0.
\end{equation}

Substituting the SSS metric \eqref{metric} and the radial ansatz $\phi=\phi(l)$ into Eqs.~\eqref{eq:variation_components} and \eqref{eq:variation_scalar}, we obtain the following system of ordinary differential equations:
\begin{equation}\label{eq:1}
\frac{(D-2)}{l^{D-2}\,}\frac{\mathrm{d}}{\mathrm{d} l}\!\left[l^{D-1} h(\psi)\right]
=-8 \pi G\, f(l)\, \,\bigl(\phi'(l)\bigr)^{2},
\end{equation}
\begin{equation}\label{eq:2}
\frac{(D-2)}{l\,}\frac{\mathrm{d}h(\psi)}{\mathrm{d}\psi}\,\frac{\mathrm{d} N}{\mathrm{d} l}
=-8 \pi G\, N(l)\, \bigl(\phi'(l)\bigr)^{2},
\end{equation}
\begin{equation}\label{eq:3}
\frac{\mathrm{d}}{\mathrm{d} l}\!\left[N(l)\, f(l)\, l^{D-2}\, \phi'(l)\right]=0.
\end{equation}
Equation \eqref{eq:3} can be integrated immediately, yielding an integration constant. 
We define this constant $\mathcal{D}$:
\begin{equation}\label{eq:D_charge_def}
N(l)\,f(l)\,l^{D-2}\,\phi'(l)=\sqrt{\mathcal{D}}.
\end{equation}
where $\mathcal{D}\ge 0$ is defined by convention. Combining Eq.~\eqref{eq:D_charge_def} with the field equations \eqref{eq:1} and \eqref{eq:2}, one may express $\mathcal{D}$ purely in terms of geometric quantities:

\begin{equation}\label{eq:D_squared}
\mathcal{D}=\frac{(D-2)\,f(l)^2\,N(l)\,l^{2D-4}}{4\pi G}
\left[
\frac{N(l)}{2f(l)\,l^{D-2}}
\frac{\mathrm{d}}{\mathrm{d}l}\!\left[l^{D-1}h(\psi)\right]
-\frac{\mathrm{d}h(\psi)}{\mathrm{d}\psi}
\frac{\mathrm{d}N(l)}{\mathrm{d}l}
\right].
\end{equation}
The $\mathcal{D}$ provides a useful measure of numerical accuracy and quantifies the degree to which the wormhole spacetime relies on the phantom scalar field.

Having obtained the corresponding ODE system, we note that in the coordinate system \eqref{metric} the metric component $g_{ll}$ exhibits a pathological behavior. We attribute this apparent pathological behavior to the choice of coordinates. Therefore, as in Ref.~\cite{Kanti:2011jz}, we therefore adopt an alternative coordinate system in which the metric is free of such pathologies. The line element then takes the form
\begin{equation}\label{eq:wormhole_metric}
\mathrm{d}s^2
=-e^{(D-3)A(r)}\,\mathrm{d}t^2
+\frac{p(r)}{e^{A(r)}}\Big[\mathrm{d}r^2+(r^2+r_0^2)\,\mathrm{d}\Omega_{D-2}^2\Big].
\end{equation}
Here $A(r)$ and $p(r)$ are functions to be determined, $r_0$ denotes the throat parameter, and $r\in(-\infty,+\infty)$.

The ADM mass M can be read from the asymptotic sub-leading behaviour of the metric functions:
\begin{equation}\label{eq:gtt_asym}
g_{tt} =-e^{(D-3)A(r)}=-1+\frac{16\,\pi\,G\,M}{(D-2)\,\Omega_{D-2}\,r^{D-3}\,}+\cdots .
\end{equation}
In practice, the mass M can also be computed via the Komar integral \cite{Wald:1984rg}:
\begin{equation}\label{eq:komar_mass}
M = \frac{(D-2)}{(D-3)8\pi G} \int_{\Sigma} R_{\mu\nu} n^{\mu} \xi^{\nu} \,\mathrm{d}V,
\end{equation}
where $\Sigma$ is an asymptotically spacelike hypersurface, $n^{\mu}$ is its unit normal (satisfying $n_{\mu} n^{\mu}=-1$), $\mathrm{d}V$ is the induced volume element on $\Sigma$, and $\xi^{\mu}$ is the asymptotically timelike Killing vector.

\section{Boundary Conditions}\label{sec3}

To numerically solve the ODE system given by Eqs.~\eqref{eq:1}--\eqref{eq:3} derived in Sec.~II, we rewrite the equations in the coordinate system defined by the metric~\eqref{eq:wormhole_metric}. Using these three ODEs, we eliminate the derivative of phantom field and obtain two coupled differential equations for the metric functions $A(r)$ and $p(r)$ only. To solve these equations, four boundary conditions are required. We impose these boundary conditions,
\begin{equation}\label{eq:bc_asym}
A(\pm\infty)=0, \qquad p(\pm\infty)=1 .
\end{equation}
These conditions ensure that both asymptotic regions are flat and provide the necessary boundary constraints for the numerical integration.

\section{NUMERICAL RESULTS}\label{sec4}

Since the radial coordinate $r$ ranges over an infinite domain, we compactify it via the transformation
\begin{equation}\label{transform}
x=\frac{2}{\pi}\arctan\!\left(\frac{r}{r_1}\right),
\end{equation}
which maps $r\in(-\infty,+\infty)$ to $x\in(-1,1)$. This allows us to perform the numerical integration on a finite interval. 

In this work, all physical quantities are presented in dimensionless form. 
We introduce the following rescalings:
\begin{equation}\label{eq:dimensionless_defs}
\bar{\psi}=\sqrt{4\pi G}\,\psi,
\qquad
r=\tilde r\, r_1,
\qquad
\alpha= \frac{\alpha}{r_1^{2}}.
\end{equation}
For convenience, we fix $r_1=1$ and $4\pi G=1$, which does not affect the generality of our results. In our implementation, we use a uniform grid of $10{,}000$ points and control the relative error to be below $10^{-4}$.

In the numerical analysis, we treat the coupling parameter $\alpha$ as the primary control parameter and fix the throat radius to $r_0=1$ throughout. 
With this setup, we investigate the impact of quasi-topological gravity on wormhole spacetimes from two complementary perspectives.
First, at fixed dimension $D=5$, we compare wormhole solutions obtained from two coefficient configurations: $\alpha_n=\alpha^{\,n-1}$ and $\alpha_n=\frac{1-(-1)^n}{2}\alpha^{\,n-1}$. 
Second, fixing the coefficient configurations: to $\alpha_n=\alpha^{\,n-1}$, we investigate how the solutions depend on the spacetime dimension (e.g., $D=5,6$).
Moreover, owing to the reflection symmetry of the solutions about the throat, we present only one side of the wormhole.

\subsection{Comparison of coefficient configurations: $\alpha_n=\alpha^{\,n-1}$ and \ $\alpha_n=\frac{1-(-1)^n}{2}\alpha^{\,n-1}$}

\begin{figure}[H]
\centering
\setlength{\tabcolsep}{6pt}
\renewcommand{\arraystretch}{1.1}
\begin{tabular}{cc}
\includegraphics[width=0.45\textwidth]{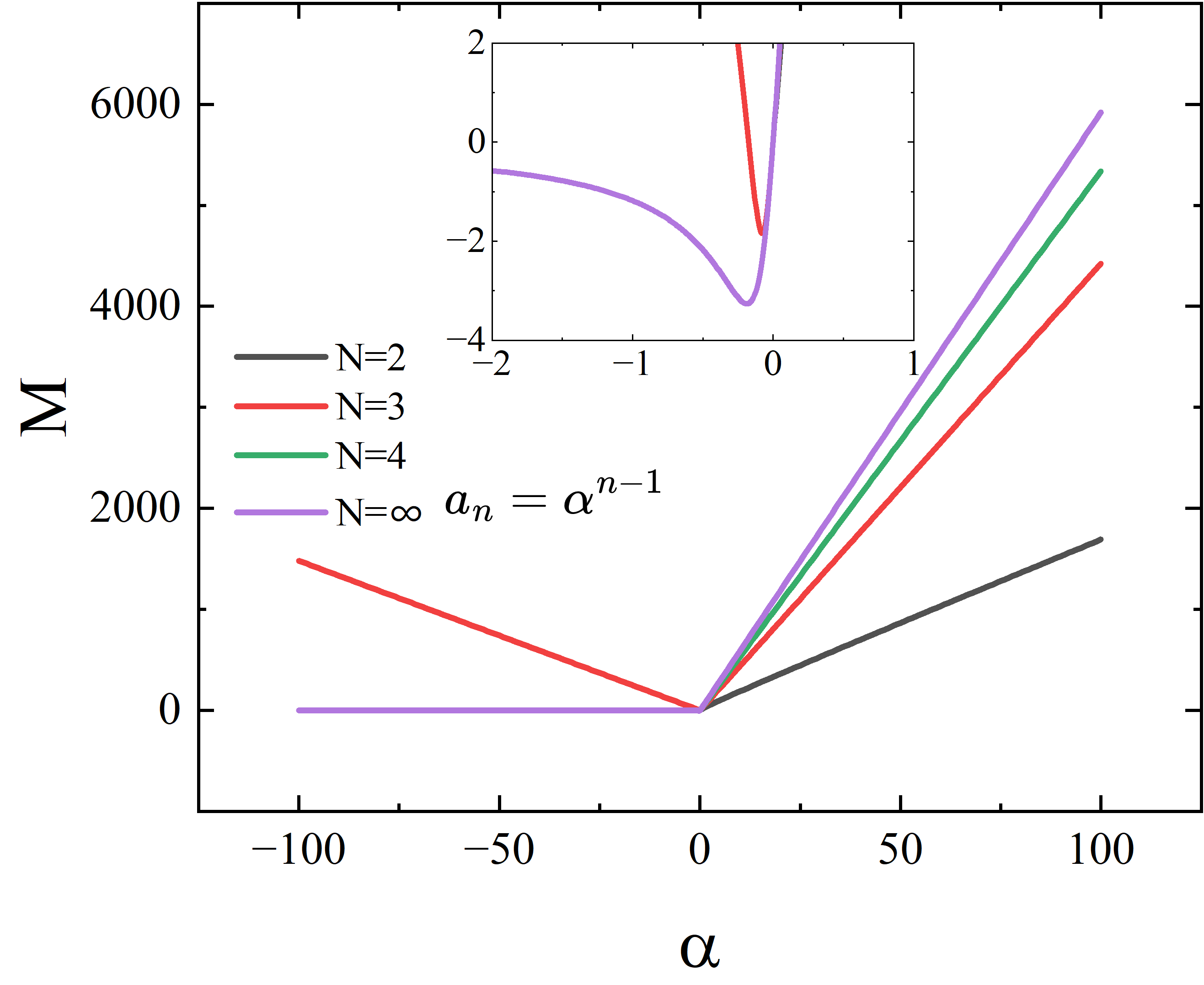} &
\includegraphics[width=0.45\textwidth]{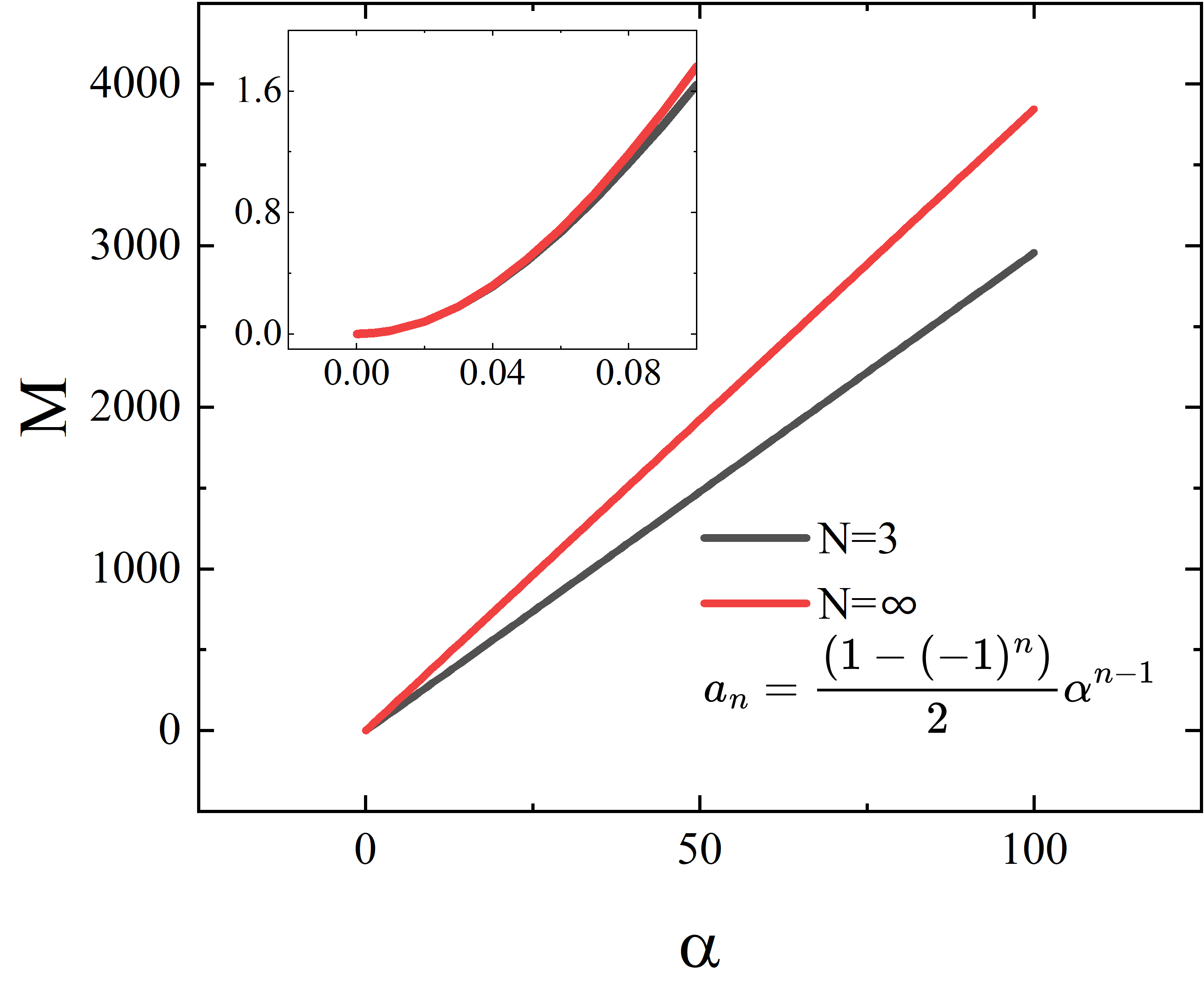} \\
\textit{(a)} & \textit{(b)} \\[0.7ex]
\includegraphics[width=0.45\textwidth]{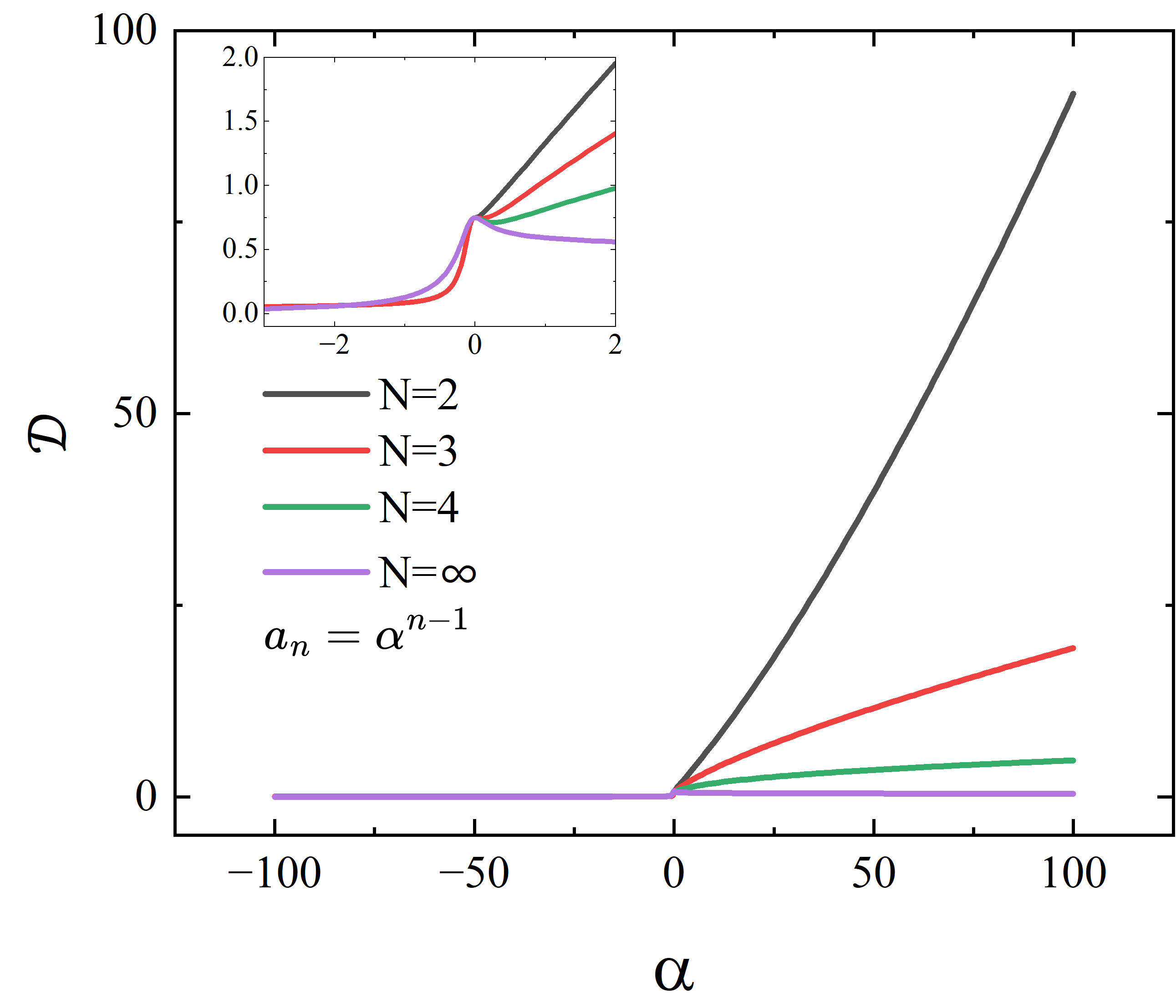} &
\includegraphics[width=0.45\textwidth]{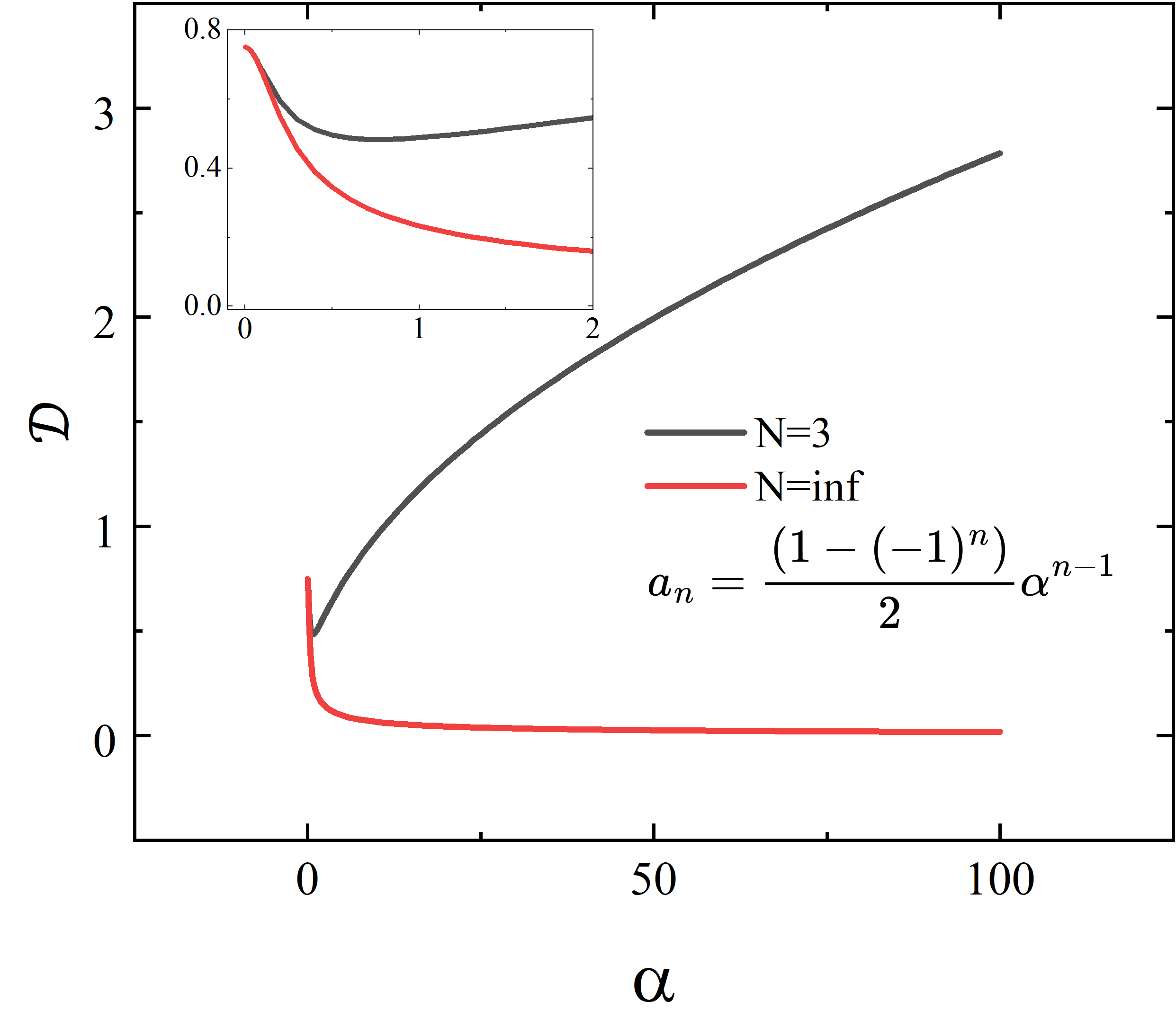} \\
\textit{(c)} & \textit{(d)}
\end{tabular}
\caption{For $D=5$, the total mass $M$ (top row) and the scalar charge $\mathcal{D}$ (bottom row) as functions of the coupling parameter $\alpha$. Different curves correspond to different truncation orders $N$ and different coefficient configurations $\{\alpha_n\}$.}
\label{fig:MorD_vs_a}
\end{figure}

As shown in Fig.~\ref{fig:MorD_vs_a}, we analyze how the total mass $M$ (top row) and the scalar charge $\mathcal{D}$ (bottom row) depend on the coupling parameter $\alpha$. The left panels of Fig.~\ref{fig:MorD_vs_a} correspond to the configuration including corrections at all orders, whereas the right panels of Fig.~\ref{fig:MorD_vs_a} show the case in which only odd-order corrections are retained (i.e., $\mathcal{Z}_3,\mathcal{Z}_5,\ldots$).

We first discuss the behavior of the mass $M$. 
In Fig.~\ref{fig:MorD_vs_a}(a), for $\alpha>0$, $M$ starts near zero and increases approximately monotonically with $\alpha$, reaching values of order $10^3$ at large coupling parameter. 
Moreover, in the large-$\alpha$ regime we observe that increasing the truncation order $N$ tends to decrease $M$ at fixed $\alpha$. 
For $\alpha<0$, the cases $N=3$ and $N=\infty$ can typically be extended to much smaller values, with the admissible range depending on the throat $r_0$. 
Unless otherwise stated, we restrict to $\alpha\in[-100,100]$, which already captures the main $\alpha$-dependent trends considered here.
In particular, for $N=3$ we finds $M<0$ near $\alpha\simeq 0$, and as $\alpha$ decreases the mass first decreases and then turns around and increases. 
For $N=\infty$, the mass remains negative throughout the $\alpha<0$ branch; as $\alpha$ decreases it first becomes more negative and then increases again, eventually approaching zero asymptotically at large negative coupling. Wormhole solutions with negative mass have also been reported in Refs.~\cite{Chew:2019lsa,Hao:2023kvf,Su:2024gxp}.

Fig.~\ref{fig:MorD_vs_a}(b) shows the results for the odd-order-only configuration. 
Since this configuration is symmetric under $\alpha\to -\alpha$, we display only the $\alpha\ge 0$ branch. 
The mass still increases with $\alpha$, but the growth is milder than in Fig.~\ref{fig:MorD_vs_a}(a). 
In addition, the $N=\infty$ curve lies below the $N=3$ curve throughout the displayed interval.

As introduced above, the integration constant $\mathcal{D}$ characterizes the scalar charge carried by the phantom field and also serves as a useful consistency check of the numerical solutions, since for fixed parameters it should be independent of the radial position. 
We therefore use $\mathcal{D}$ to quantify the phantom field and to assess the dependence on $\alpha$. In Fig.~\ref{fig:MorD_vs_a}(c), for $\alpha>0$ the finite-order truncations (e.g., $N=3,4$) typically exhibit a non-monotonic behavior: $\mathcal{D}$ first decreases and then increases, whereas the all-order case $N=\infty$ decreases monotonically and gradually approaches zero. 
Overall, $\mathcal{D}$ tends to be smaller for larger $N$. 
For $\alpha<0$, $\mathcal{D}$ rapidly approaches zero as $\alpha$ decreases. 
This indicates that, compared to the $\alpha>0$ case, less exotic matter is required to keep the wormhole throat open when $\alpha<0$.

Finally, Fig.~\ref{fig:MorD_vs_a}(d) displays behavior that differs appreciably from Fig.~\ref{fig:MorD_vs_a}(c). 
For $N=3$, $\mathcal{D}$ still shows a decrease followed by an increase, but the overall magnitude is significantly smaller. 
For $N=\infty$, when $\alpha>0$ we also observe that $\mathcal{D}$ decreases more rapidly with increasing $\alpha$.

\begin{figure}[H]
\centering
\setlength{\tabcolsep}{6pt}
\renewcommand{\arraystretch}{1.1}
\begin{tabular}{cc}
\includegraphics[width=0.45\textwidth]{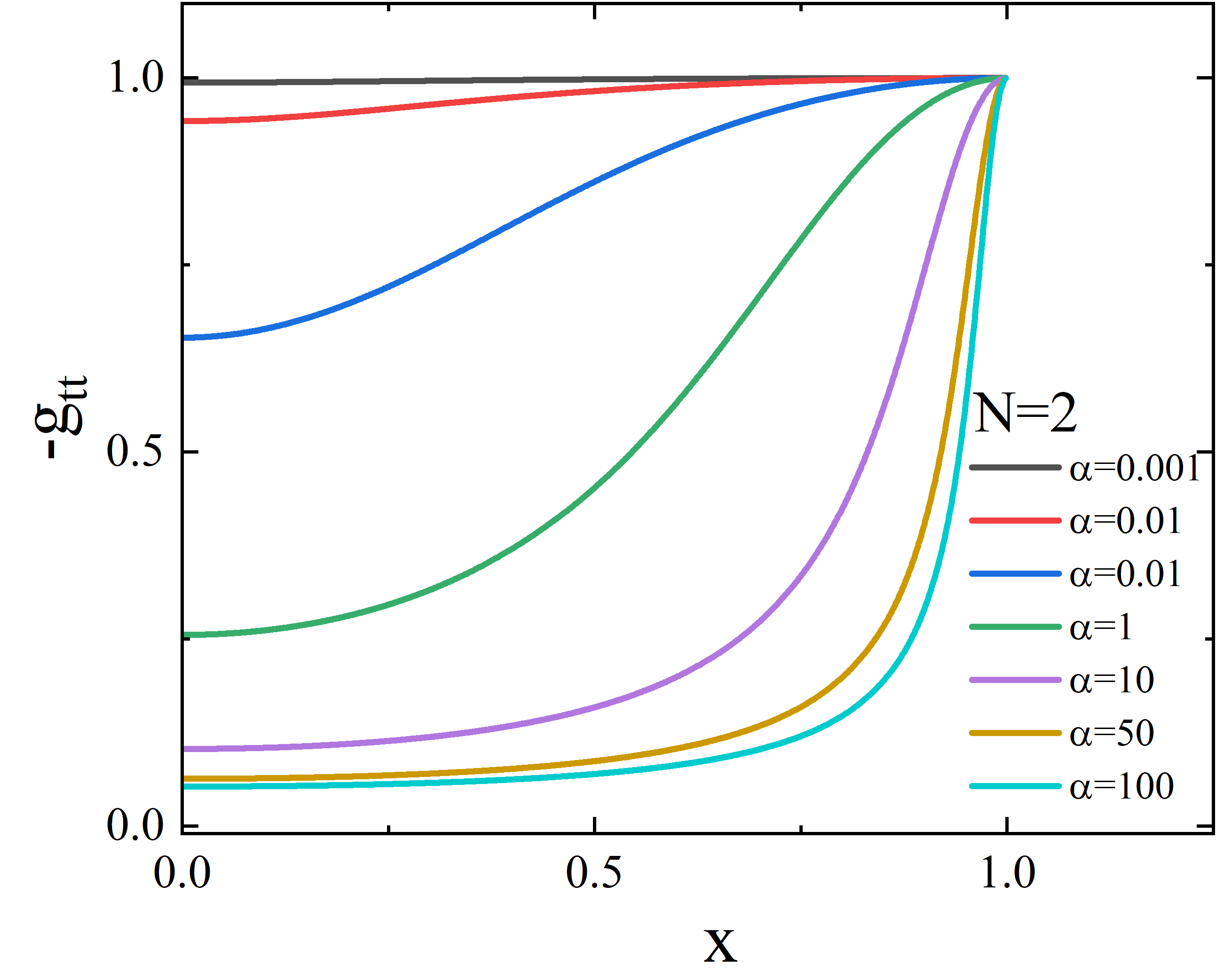} &
\includegraphics[width=0.45\textwidth]{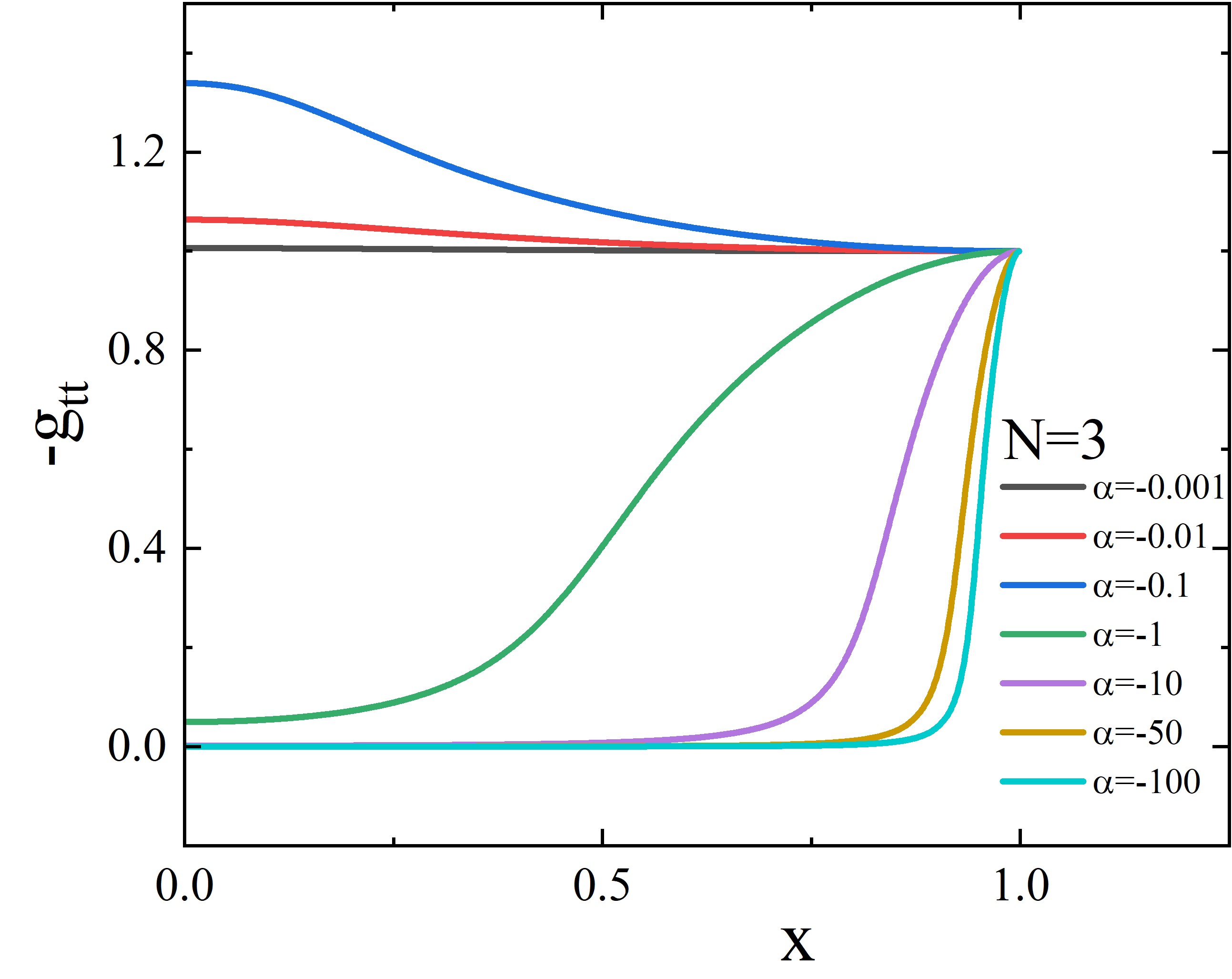} \\
\textit{(a)} & \textit{(b)} \\[0.7ex]
\includegraphics[width=0.44\textwidth]{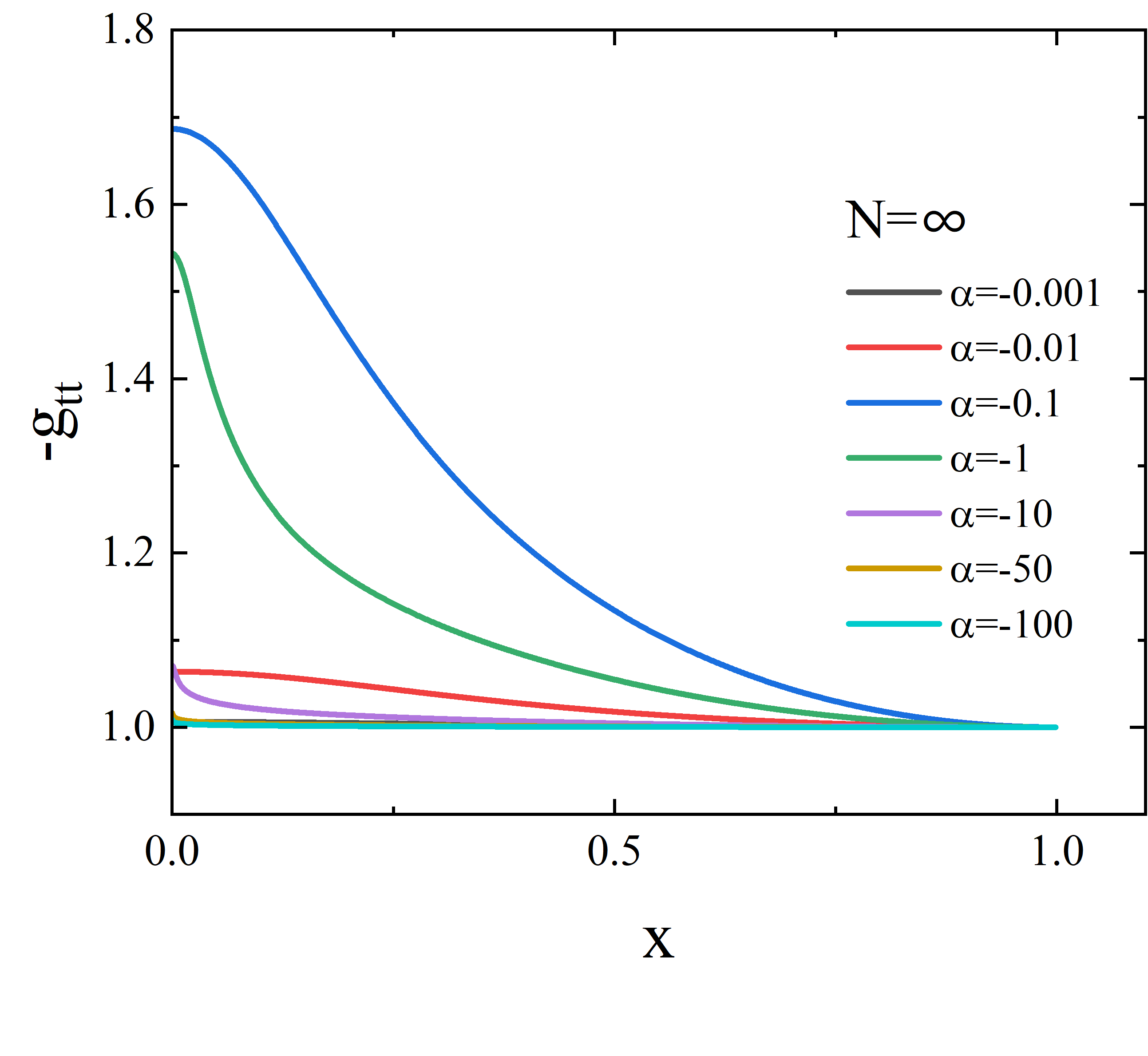} &
\includegraphics[width=0.44\textwidth]{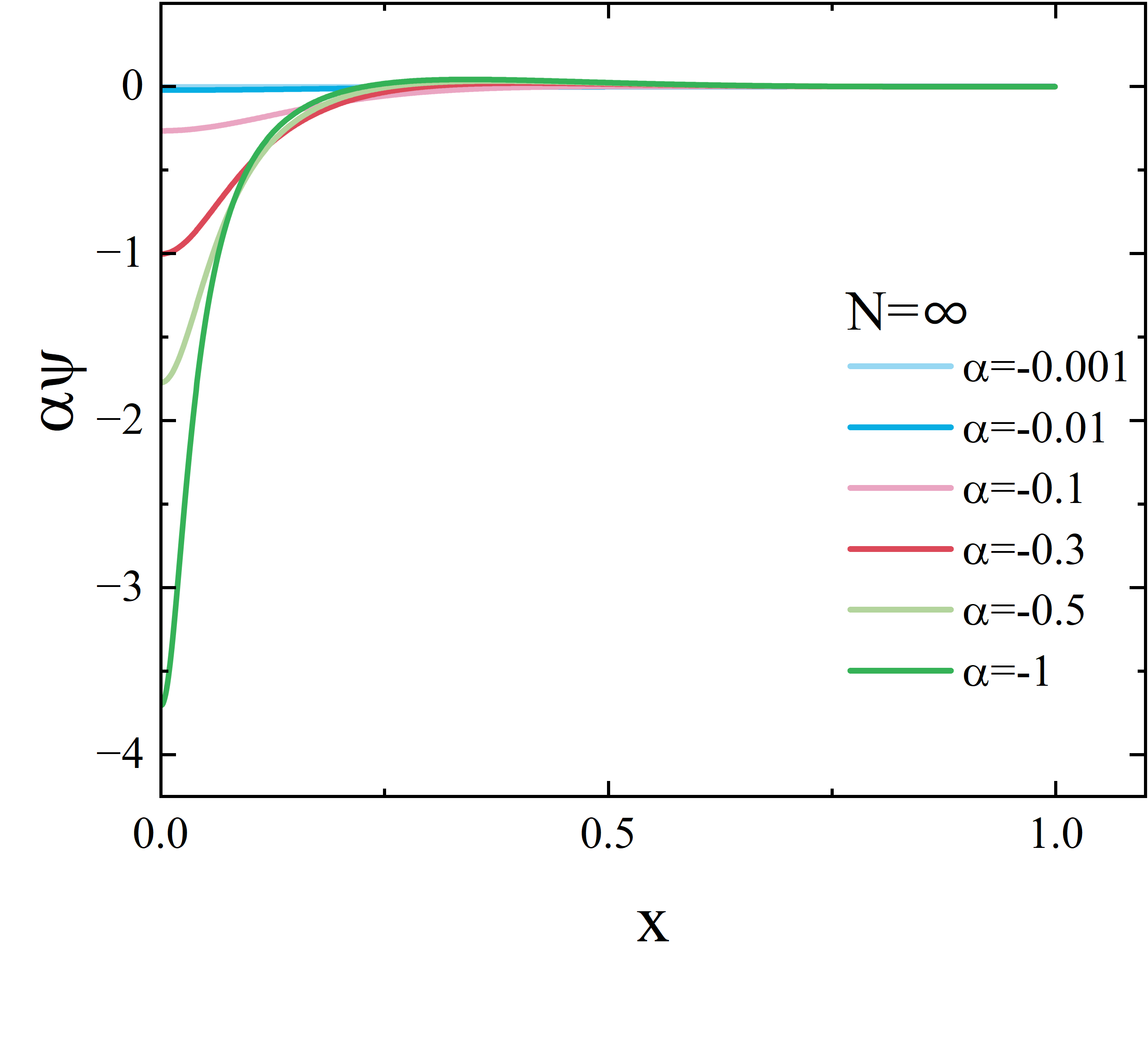} \\
\textit{(c)} & \textit{(d)}
\end{tabular}
\caption{The function $-g_{tt}\equiv e^{(D-3)A(r)}$ (panels (a)--(c)) and the quantity $\alpha\psi$ (panel (d)) as functions of the radial coordinate $x$. Different curves correspond to different truncation orders $N$ and representative values of $\alpha$, with the coefficients chosen as $\alpha_n=\alpha^{\,n-1}$.}
\label{fig:gtt_phi}
\end{figure}

 Fig.~\ref{fig:gtt_phi} shows the metric function $-g_{tt}\equiv e^{(D-3)A(r)}$ and of $\alpha\psi$ for various parameter choices. 
We begin with the case $N=2$ for a fixed coefficient configuration $\{\alpha_n\}$, where only $\alpha$ is varied (Fig.~\ref{fig:gtt_phi}(a)). 
For small $\alpha$, $-g_{tt}$ approaches $1$, consistent with the five-dimensional static, spherically symmetric Ellis wormhole in the absence of higher-curvature corrections. 
As $\alpha$ increases, the minimum value of $-g_{tt}$ decreases and gradually approaches $0$. 
Meanwhile, the variation becomes much more pronounced near $x\to 1$, where the curves become increasingly steep. 
A similar trend is also observed for $N=3,4,\infty$ and for other choices of $\{\alpha_n\}$.

On the $\alpha<0$ branch, the behavior changes. 
For $N=3$ (Fig.~\ref{fig:gtt_phi}(b)), as $|\alpha|$ increases the value of $-g_{tt}$ at the throat first increases and then decreases, and a region with $-g_{tt}>1$ emerges. 
This behavior correlates with the appearance of negative-mass solutions in Fig.~\ref{fig:MorD_vs_a}. 
When $|\alpha|$ is increased further, the overall behavior gradually approaches that observed in the positive-$\alpha$ case.

For the all-order case $N=\infty$ on the same $\alpha<0$ branch (Fig.~\ref{fig:gtt_phi}(c)), as $|\alpha|$ varies, $-g_{tt}$ near the throat again exhibits a trend of first increasing and then decreasing, while remaining above $1$ throughout. 
At the same time, the metric function $-g_{tt}$ becomes sharper near the throat, whereas it flattens out in the asymptotic region.

Motivated by this, we further ask whether, for $\alpha<0$, the $N=\infty$ theory can still be understood as an infinite sum over the $Z_n$ corrections. 
To this end, we examine $\alpha\psi$, shown in Fig.~\ref{fig:gtt_phi}(d). 
For $\alpha\lesssim -0.3$, regions with $\alpha\psi<-1$ appear, indicating that in this parameter range the all-order correction cannot be represented by the infinite-series summation over the $Z_n$ terms. 
Consequently, the corresponding wormhole solutions may require a different interpretation in terms of an all-order completion of the higher-curvature sector.

\begin{figure}[]
\centering
\setlength{\tabcolsep}{6pt}
\renewcommand{\arraystretch}{1.1}
\begin{tabular}{cc}
\includegraphics[width=0.45\textwidth]{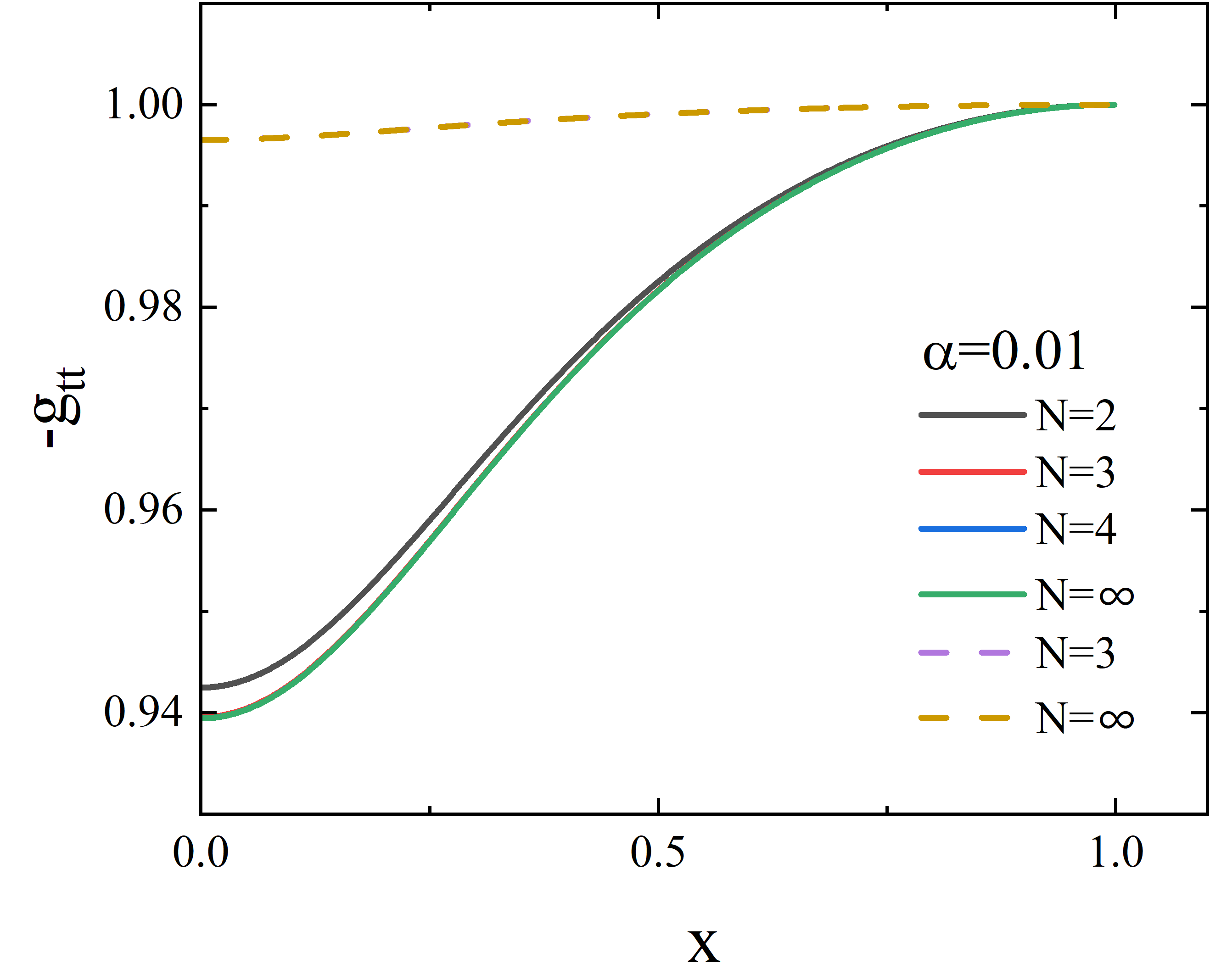} &
\includegraphics[width=0.45\textwidth]{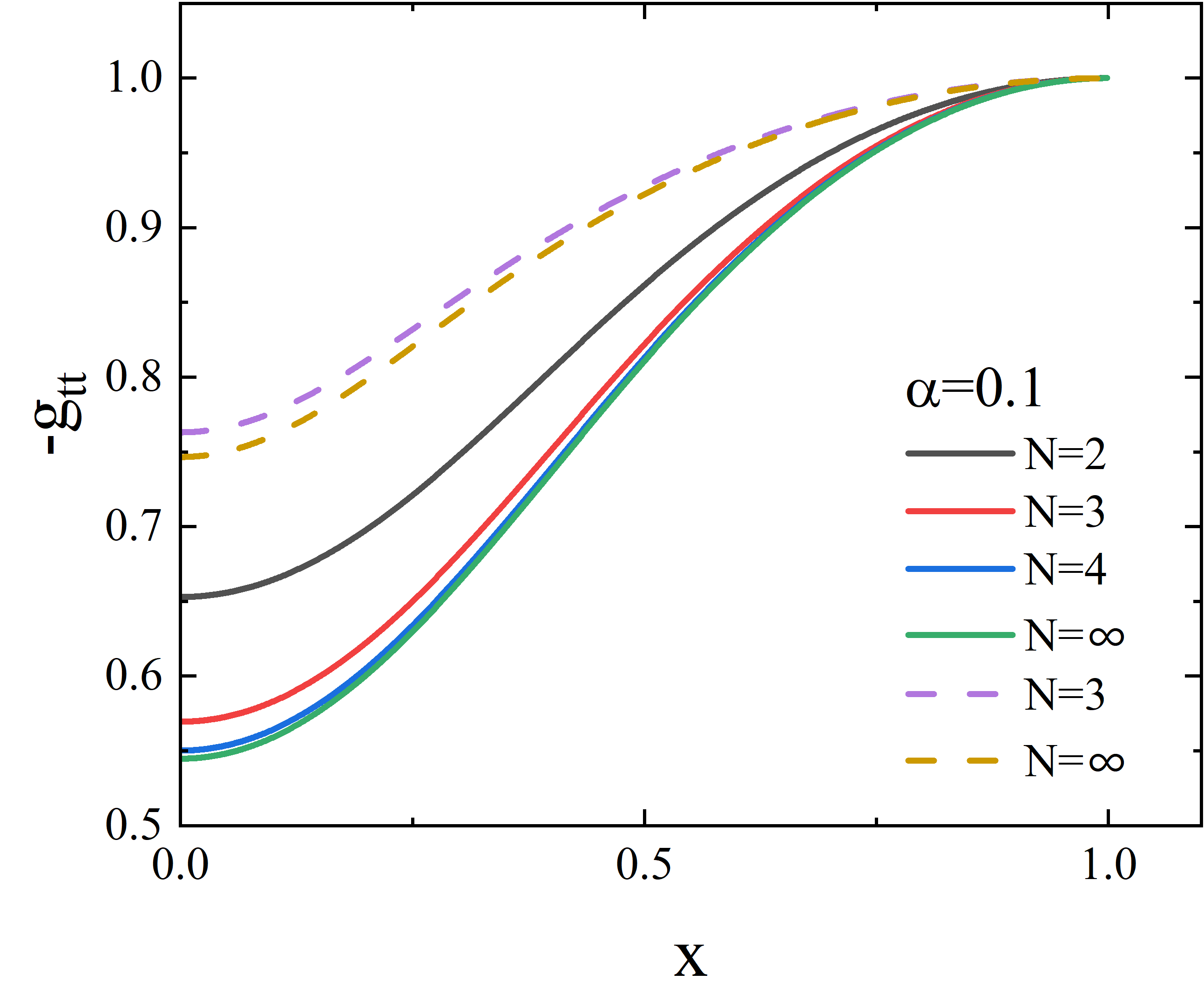} \\
\textit{(a)}& \textit{(b)}\\[0.7ex]
\includegraphics[width=0.45\textwidth]{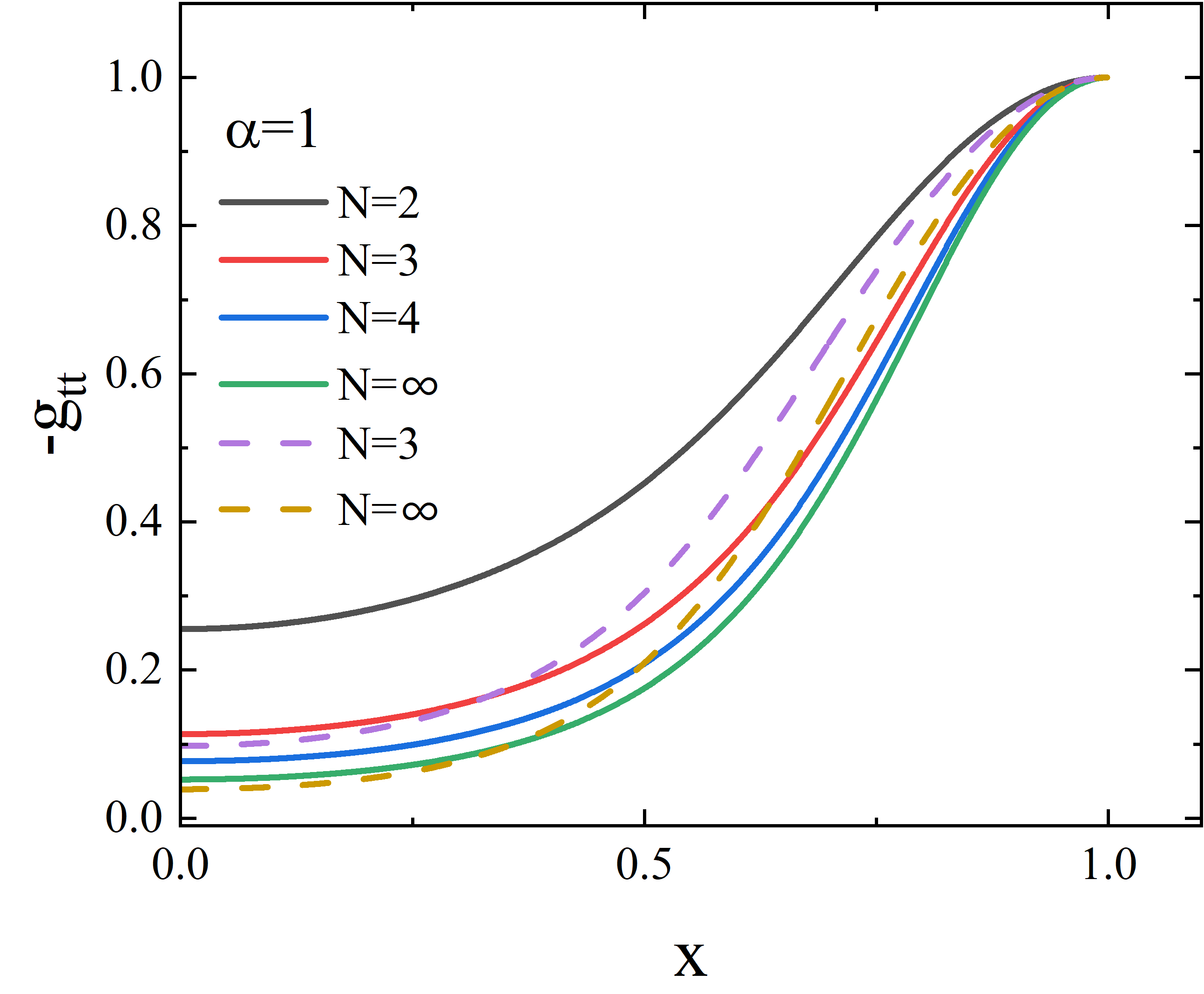} &
\includegraphics[width=0.45\textwidth]{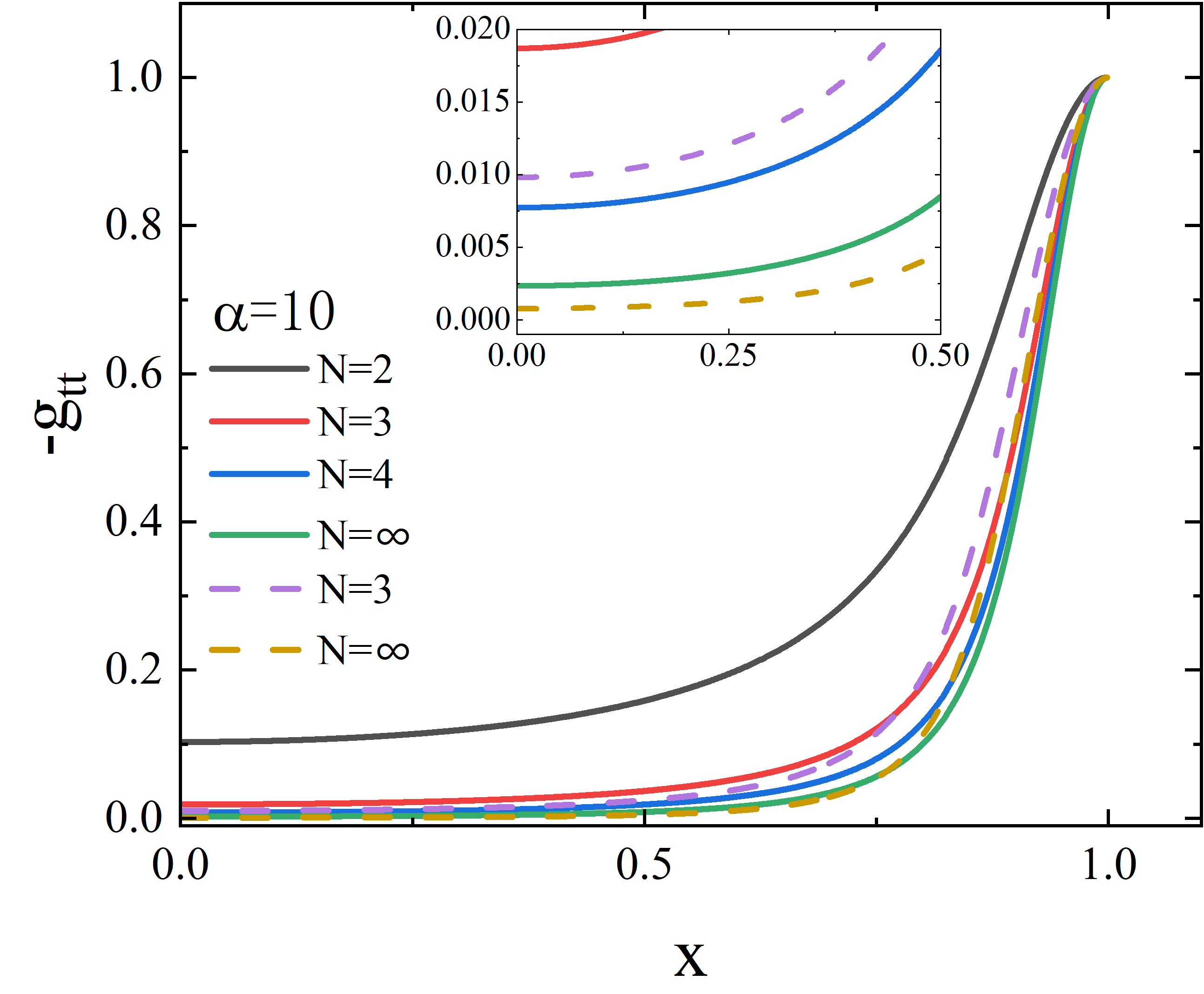} \\
\textit{(c)}& \textit{(d)}
\end{tabular}
\caption{The function $-g_{tt}$ for different values of $\alpha$. Solid curves correspond to $\alpha_n=\alpha^{\,n-1}$, while dashed curves correspond to $\alpha_n=\frac{1-(-1)^n}{2}\alpha^{\,n-1}$.}
\label{fig:gtt_N_varying_a}
\end{figure}

Fig.~\ref{fig:gtt_N_varying_a} compares how the radial profile of $-g_{tt}$ depends on the truncation order $N$ ($N=2,3,4,\infty$) and on the choice of coefficient configuration, for several representative values of $\alpha$.

For small $\alpha$ (Figs.~\ref{fig:gtt_N_varying_a}(a) and \ref{fig:gtt_N_varying_a}(b)), the curves corresponding to different truncation orders are still close to each other and do not separate significantly. 
In particular, for the dashed-sequence configuration, the $N=3$ and $N=\infty$ curves are nearly indistinguishable over most of the domain. 
By contrast, the two coefficient configurations already exhibit an overall shift: the solid-curve configuration typically yields smaller values of $-g_{tt}$ throughout the radial range.

For larger values of $\alpha$ (Figs.~\ref{fig:gtt_N_varying_a}(c) and \ref{fig:gtt_N_varying_a}(d)), the dependence on the coefficient configuration becomes more pronounced. 
In this regime, the dashed curves for $N=3$ and $N=\infty$ lie below their solid-curve counterparts. 
Moreover, as $\alpha$ grows, all the metric component $-g_{tt}$ develop a pronounced dip near the throat, with the minimum approaching zero; the smallest values can reach the level of $10^{-3}$. 
This behavior suggests the emergence of an approximate ``horizon'' structure, reminiscent of the ``black bounce'' geometries discussed in Ref.~\cite{Simpson:2018tsi}. 
For a more explicit comparison, Table~\ref{tab:gtt_N_alpha100} lists the minimum values of $-g_{tt}$ at the throat.

\begin{table}[H]
\centering
\begin{tabular}{|c|c|c|c|c|}
\hline\hline
$N$ & $2$ & $3$ & $4$ & $\infty$ \\
\hline
$-g_{tt}^{\min}$ 
& $5.2\times10^{-2}$ 
& $3.0\times10^{-3}$ 
& $7.5\times10^{-4}$ 
& $9.5\times10^{-5}$ \\
\hline\hline
\end{tabular}
\caption{Minimum values of $-g_{tt}$ at the throat for different truncation orders $N$ at fixed coupling $\alpha=100$.}
\label{tab:gtt_N_alpha100}
\end{table}

\begin{figure}[]
\centering
\subfigure[]{\includegraphics[width=0.45\textwidth]{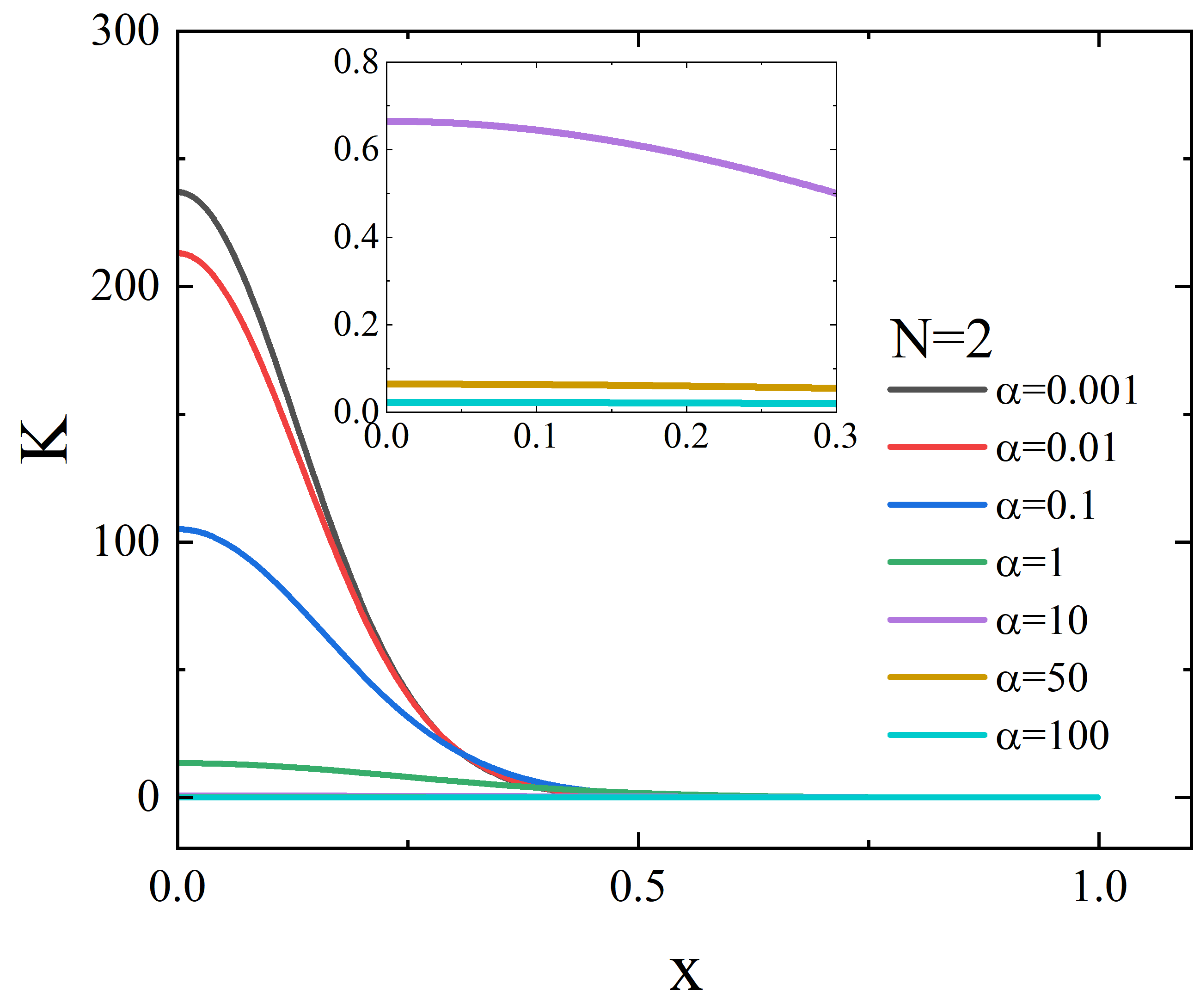}\label{fig:kk_a}}
\subfigure[]{\includegraphics[width=0.45\textwidth]{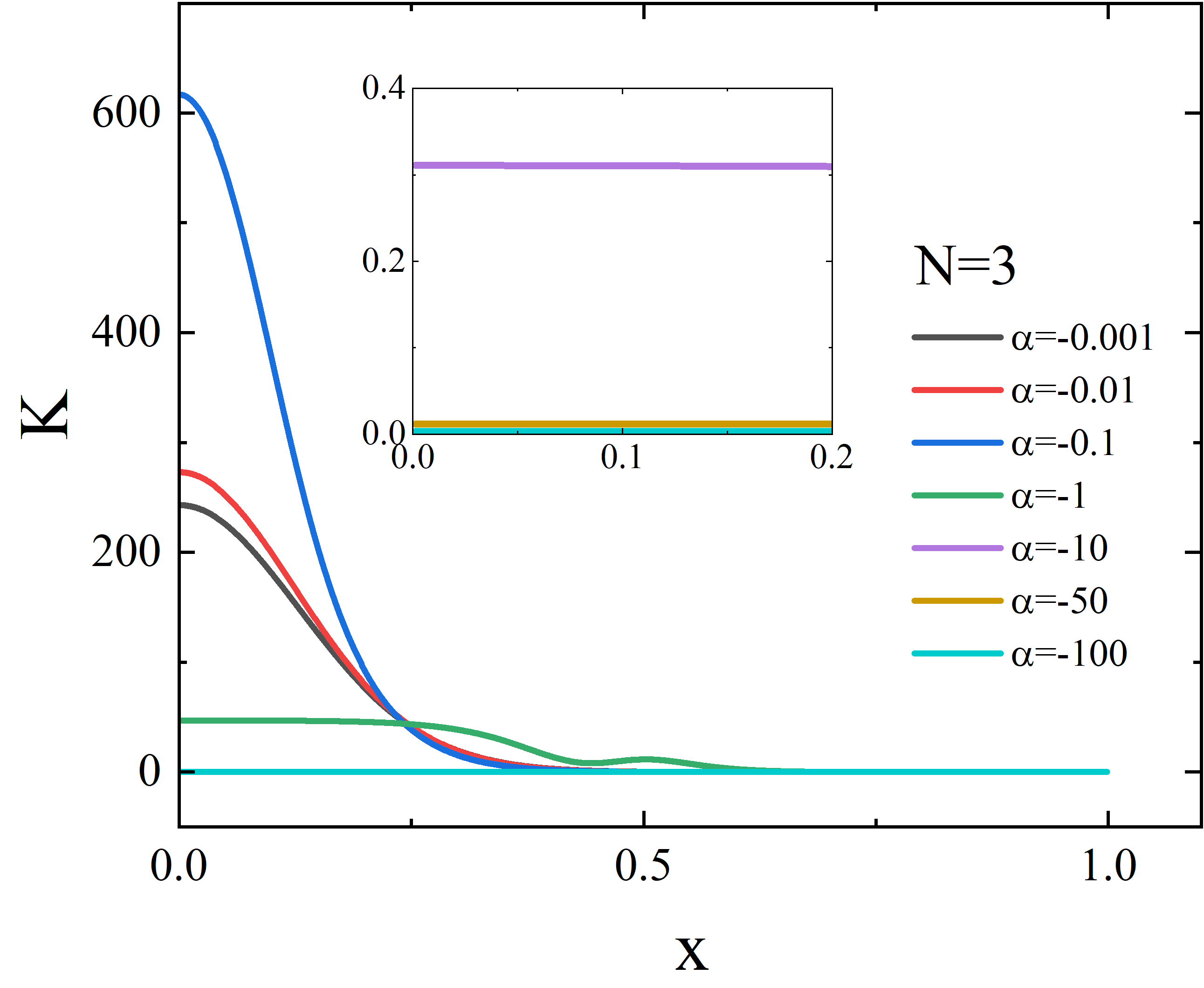}\label{fig:kk_b}}\\
\subfigure[]{\includegraphics[width=0.45\textwidth]{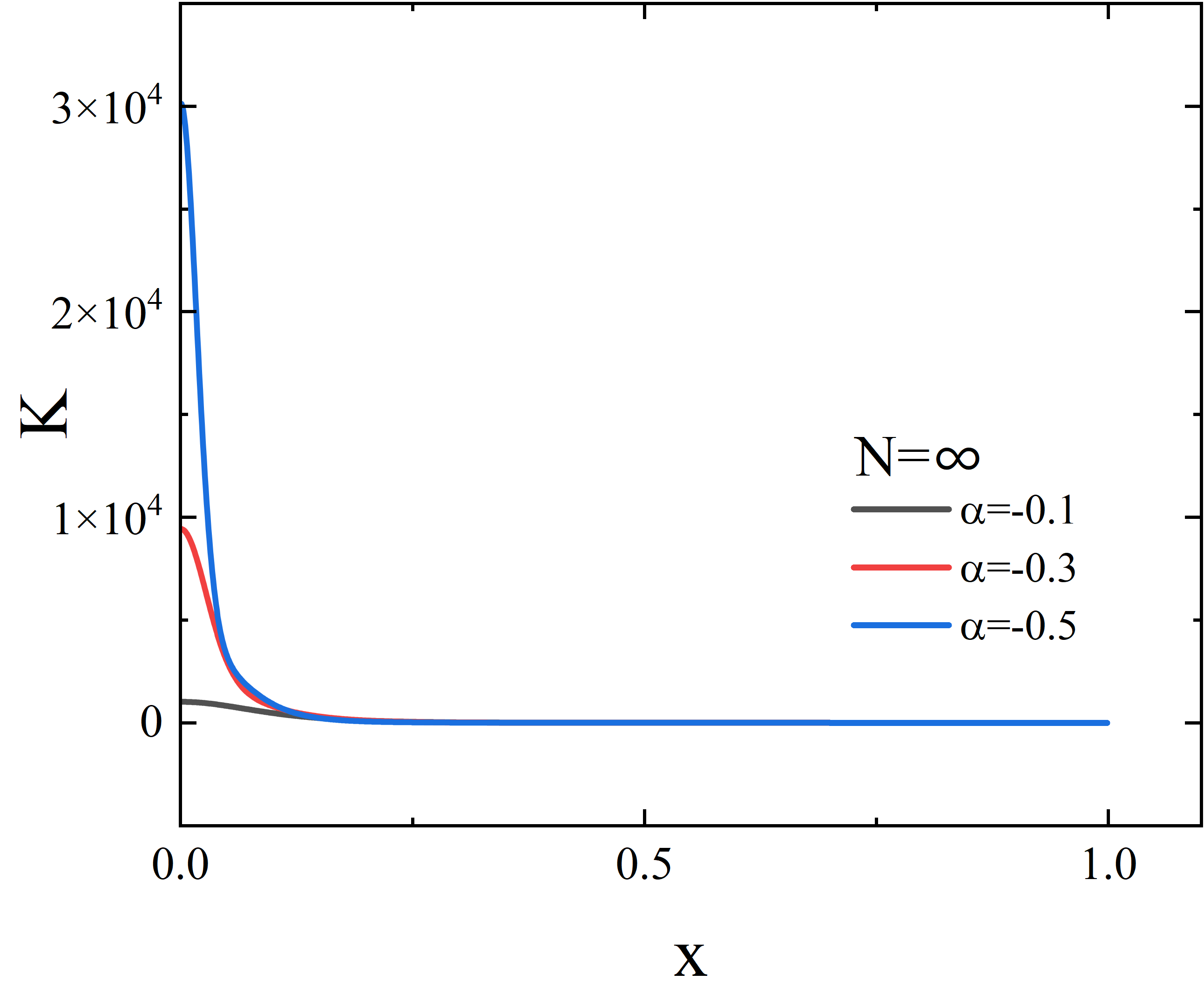}\label{fig:kk_c}}
\subfigure[]{\includegraphics[width=0.45\textwidth]{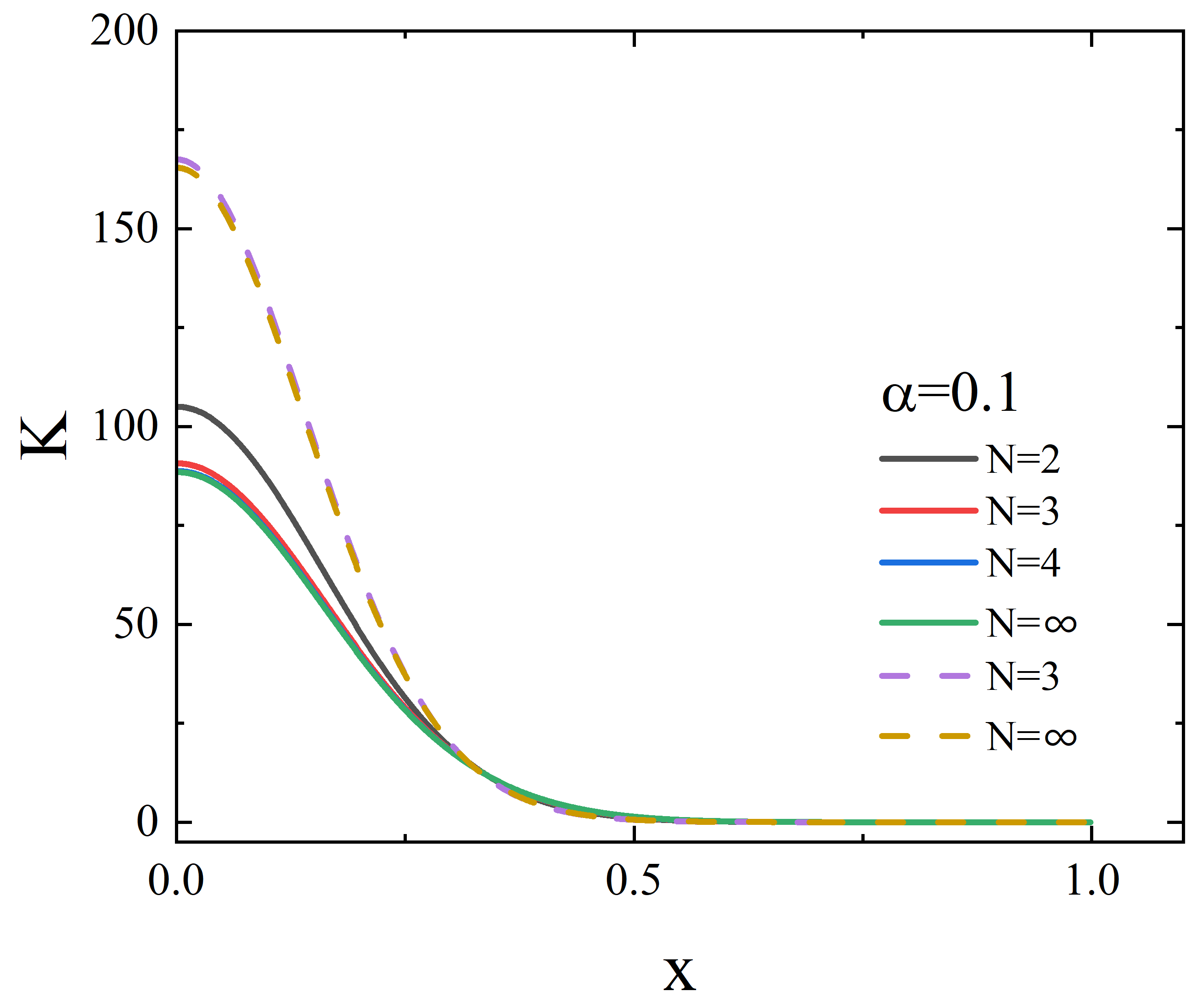}\label{fig:kk_d}}\\
\subfigure[]{\includegraphics[width=0.45\textwidth]{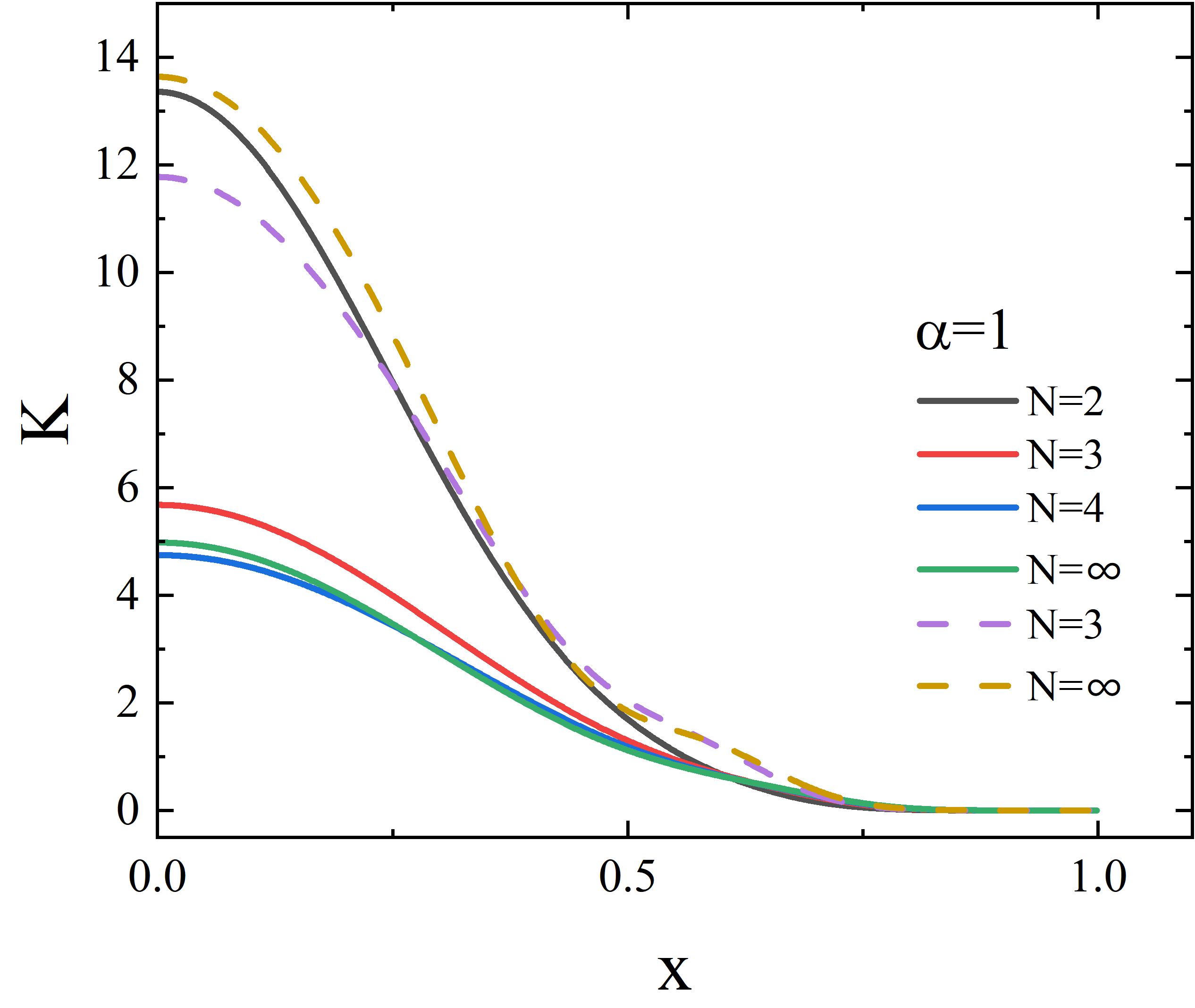}\label{fig:kk_e}}
\subfigure[]{\includegraphics[width=0.45\textwidth]{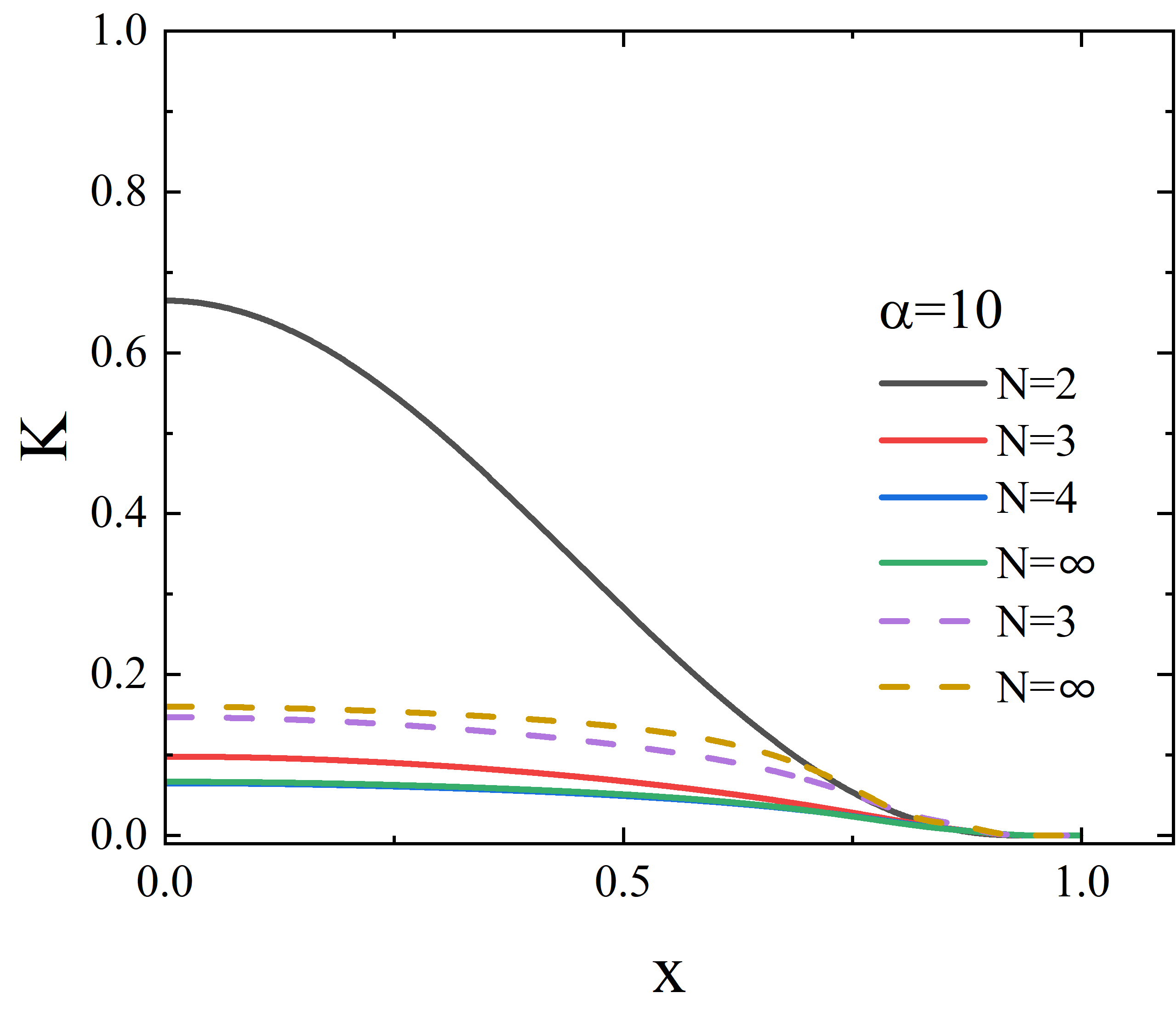}\label{fig:kk_f}}
\caption{The distribution of the Kretschmann scalar $K$ as a function of the radial coordinate $x$. Solid curves correspond to $\alpha_n=\alpha^{\,n-1}$, while dashed curves correspond to $\alpha_n=\frac{1-(-1)^n}{2}\alpha^{\,n-1}$.}
\label{fig:kk}
\end{figure}

Since the metric component $-g_{tt}$ becomes very small near the throat, we further examine the curvature properties of the spacetime by computing the Kretschmann scalar,
$K \equiv R_{\mu\nu\alpha\beta}R^{\mu\nu\alpha\beta}$.
As shown in Fig.~\ref{fig:kk} we investigate how $K$ depends on the coupling parameter $\alpha$ and the truncation order $N$.

We first consider Fig.~\ref{fig:kk}(a), corresponding to the case $N=2$. 
One can clearly see that, as $\alpha$ increases, the magnitude of the Kretschmann scalar decreases noticeably. 
For $\alpha<0$, however, the behavior can be qualitatively different. 
In Fig.~\ref{fig:kk}(b), where the truncation order is fixed to $N=3$, the dependence of $K$ on decreasing $\alpha$ becomes more intricate: the overall amplitude first increases up to a peak and then decreases. 
During this process, the Kretschmann scalar may develop two local maxima; for instance, at $\alpha=-1$ a clear double-peak structure is visible. 
When $\alpha$ is decreased further to more negative values, this multi-peak feature disappears and the global maximum of $K$ is again located near the throat.

By contrast, Fig.~\ref{fig:kk}(c) exhibits a different trend in the $\alpha<0$ regime: in this case, the value of $K$ at the throat increases monotonically as $\alpha$ decreases, and its maximum can reach values as large as $10^{4}$. Such large curvature invariants typically indicate stronger spacetime curvature, which may in turn impose more stringent constraints on the traversability of these solutions.

To further assess the impact of the coefficient choice, we compare the two $\{\alpha_n\}$ configurations at fixed values of $\alpha$. For relatively small coupling (Fig.~\ref{fig:kk}(d)), the curves for different $N$ within a given configuration have not yet fully separated; nevertheless, the dashed configuration yields systematically larger values of $K$ than the solid one. As $\alpha$ increases (Fig.~\ref{fig:kk}(e)), the curves within each configuration begin to separate, and larger truncation order $N$ generally corresponds to smaller values of $K$. Finally, at even larger coupling (Fig.~\ref{fig:kk}(f)), the dashed curves for $N=3$ and $N=\infty$ lie between the solid curves for $N=2$ and $N=3$.

To complement this analysis and to assess the matter requirements underlying these geometries, we now examine the associated energy conditions. 
In particular, the null energy condition (NEC) provides a natural criterion for quantifying the exotic matter content necessary to sustain a traversable wormhole.

\begin{figure}[]
\centering
\subfigure[]{\includegraphics[width=0.45\textwidth]{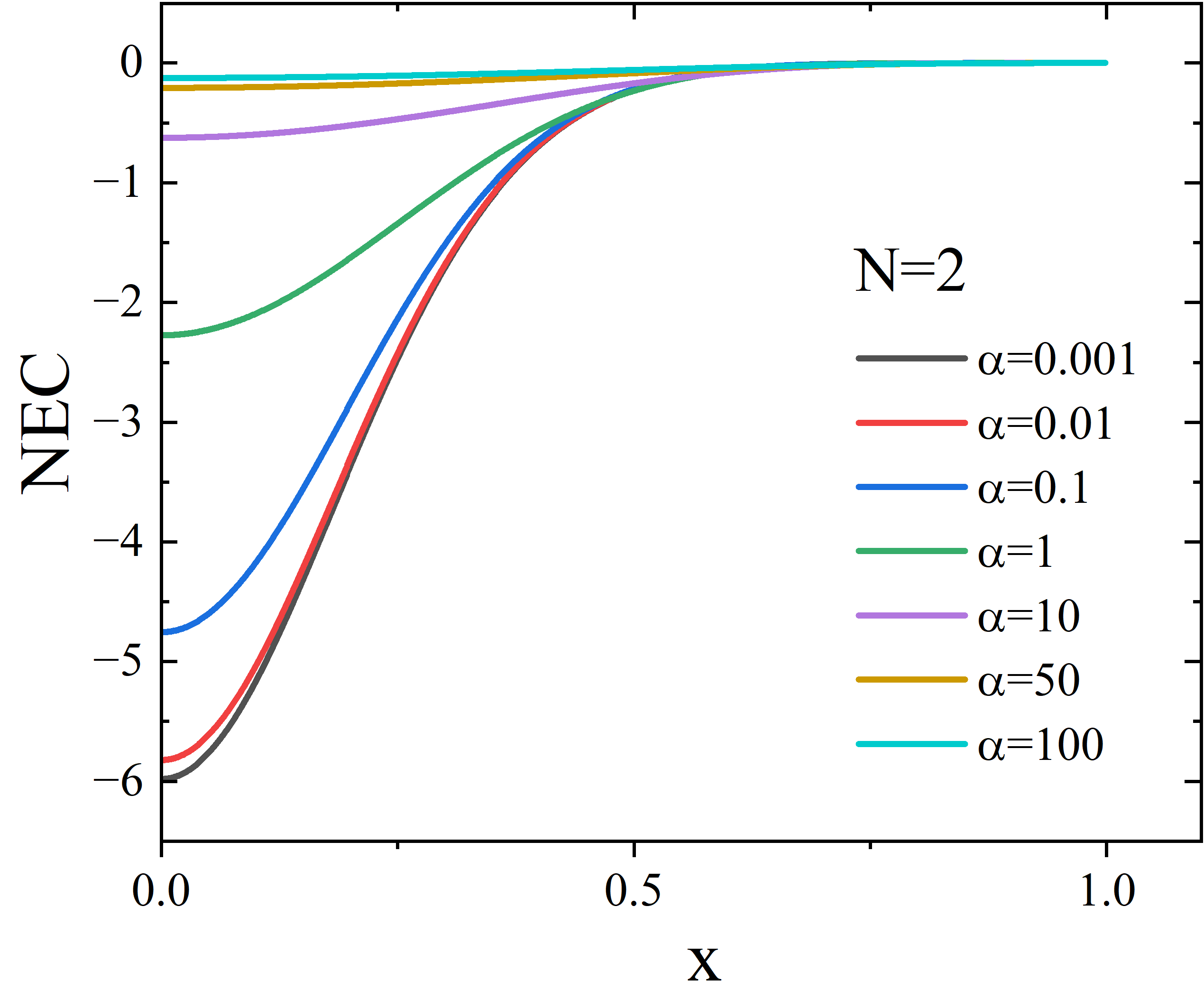}\label{fig:nec_a}}
\subfigure[]{\includegraphics[width=0.45\textwidth]{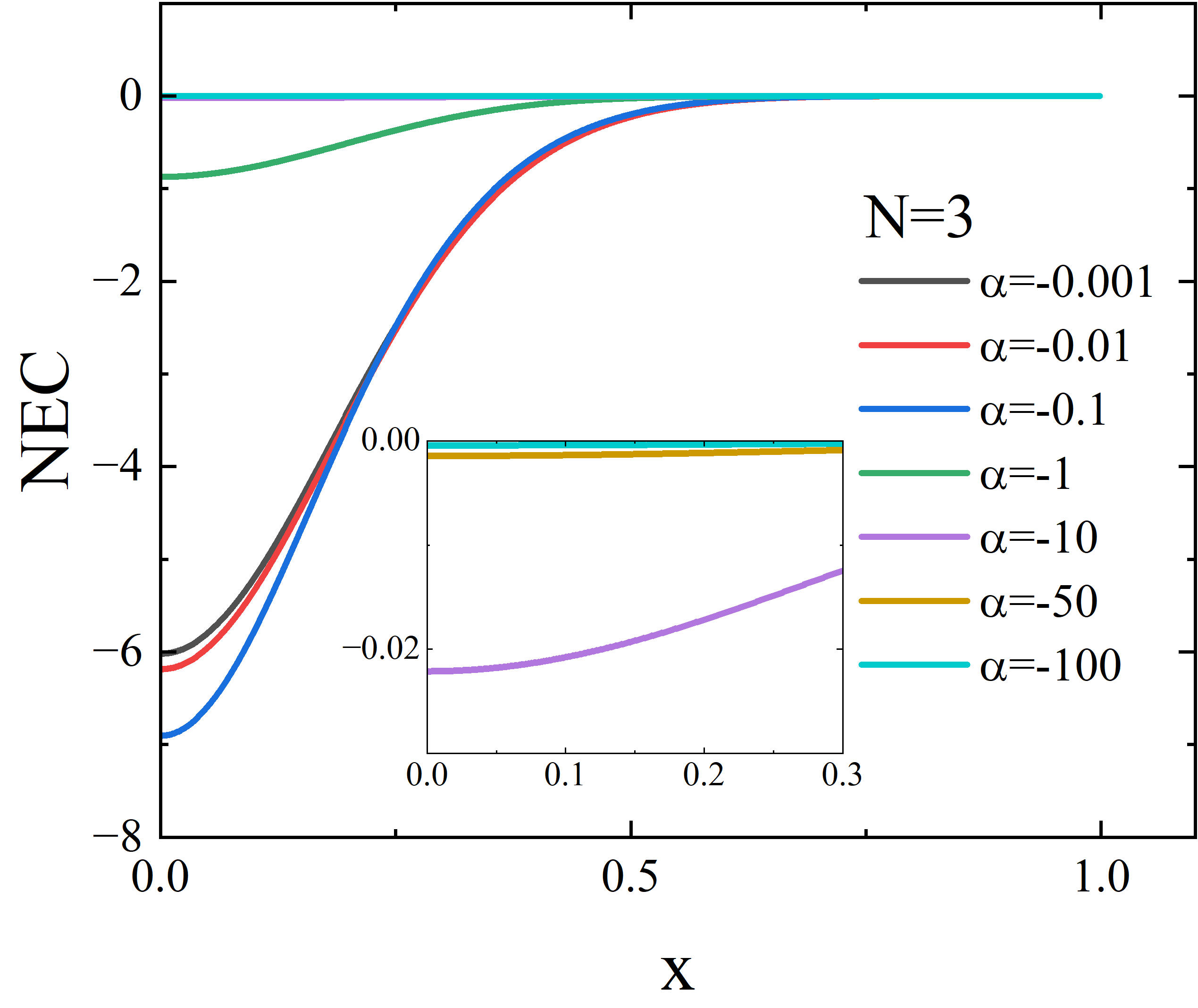}\label{fig:nec_b}}
\subfigure[]{\includegraphics[width=0.45\textwidth]{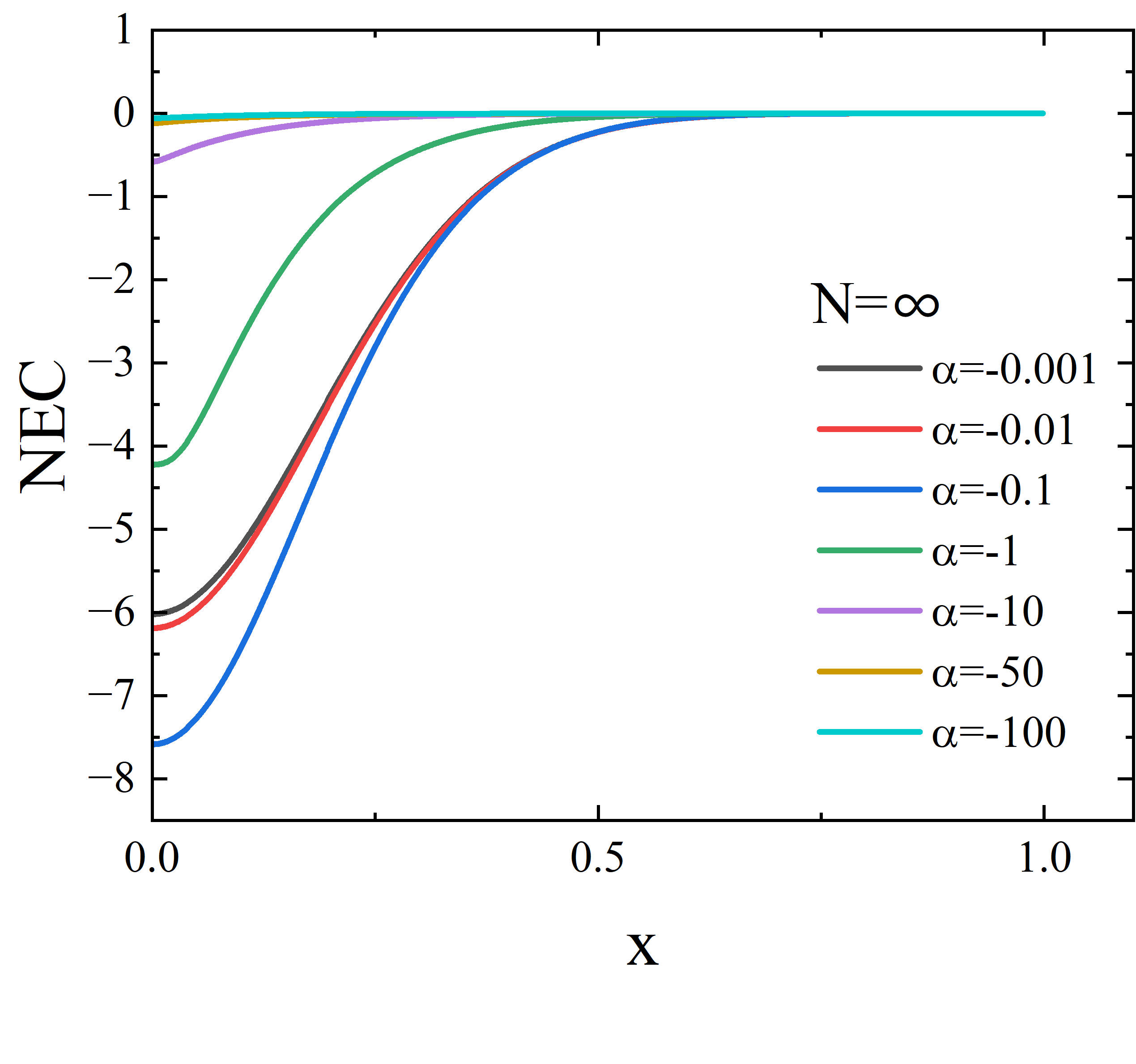}\label{fig:nec_c}}
\caption{The variation of NEC ($-T^t_t+T^r_r$) violation with the radial coordinate $x$. Panels (a)--(c) correspond to different truncation orders $N$, with the coefficients chosen as $\alpha_n=\alpha^{\,n-1}$.}
\label{fig:nec}
\end{figure}

In $D$-dimensional spherical coordinates $(t,r,\theta_1,\theta_2,\dots,\theta_{D-3},\varphi)$, we take two null vector
\begin{equation}
k^\mu=\left(\frac{1}{\sqrt{-g_{tt}}},\,\frac{\sqrt{-g_{tt}}}{g_{rr}},\,0,\,\dots,\,0\right),
\qquad
k^\mu=\left(\frac{1}{\sqrt{-g_{tt}}},\,0,\,\frac{\sqrt{-g_{tt}}}{g_{\theta_1\theta_1}},\,0,\,\dots,\,0\right).
\end{equation}

Substituting these into $T_{\mu\nu}k^\mu k^\nu$ and using the stress-energy components associated with the metric \eqref{eq:wormhole_metric}, we obtain
\begin{equation}
-T^t_t+T^r_r=-\frac{e^{A(r)}}{p(r)}\,\phi'{(r)}^2,
\qquad
-T^t_t+T^{\theta_1}_{\theta_1}=0.
\end{equation}
The first relation shows explicitly that the null energy condition is violated in the radial direction, which is a typical signature of the exotic matter required to keep the throat open. 
The second relation indicates that the null energy condition is satisfied in the transverse direction. 
Although the wormhole still exhibits NEC violation, it is of interest to determine whether quasi-topological corrections can reduce the degree of the matter-sector violation.

Building on the results above, we now investigate the degree of violation of the null energy condition, characterized here by the quantity $-T^t_t+T^r_r$. 
As shown in Fig.~\ref{fig:nec}, we focus on the five-dimensional case. Fig~\ref{fig:nec}(a) shows the dependence of the NEC violation measure on $\alpha$ for $N=2$. 
One observes that, as $\alpha$ increases, the violation of NEC decreases steadily and gradually approaches zero.

We next discuss a qualitatively different behavior. 
In Fig.~\ref{fig:nec}(b), we fix $N=3$ and consider the branch with $\alpha<0$, varying $\alpha$ toward more negative values. 
In this case, the NEC violation exhibits a non-monotonic trend: it first decreases and then increases as $\alpha$ decreases. 
This pattern is further corroborated by Fig.~\ref{fig:nec}(c). 
Taken together, Figs.~\ref{fig:nec}(b) and \ref{fig:nec}(c) indicate that for sufficiently large $|\alpha|$ (either positive or negative), the violation of the NEC becomes very small and tends to zero.

\begin{figure}[H]
\centering
\subfigure[]{\includegraphics[width=0.45\textwidth]{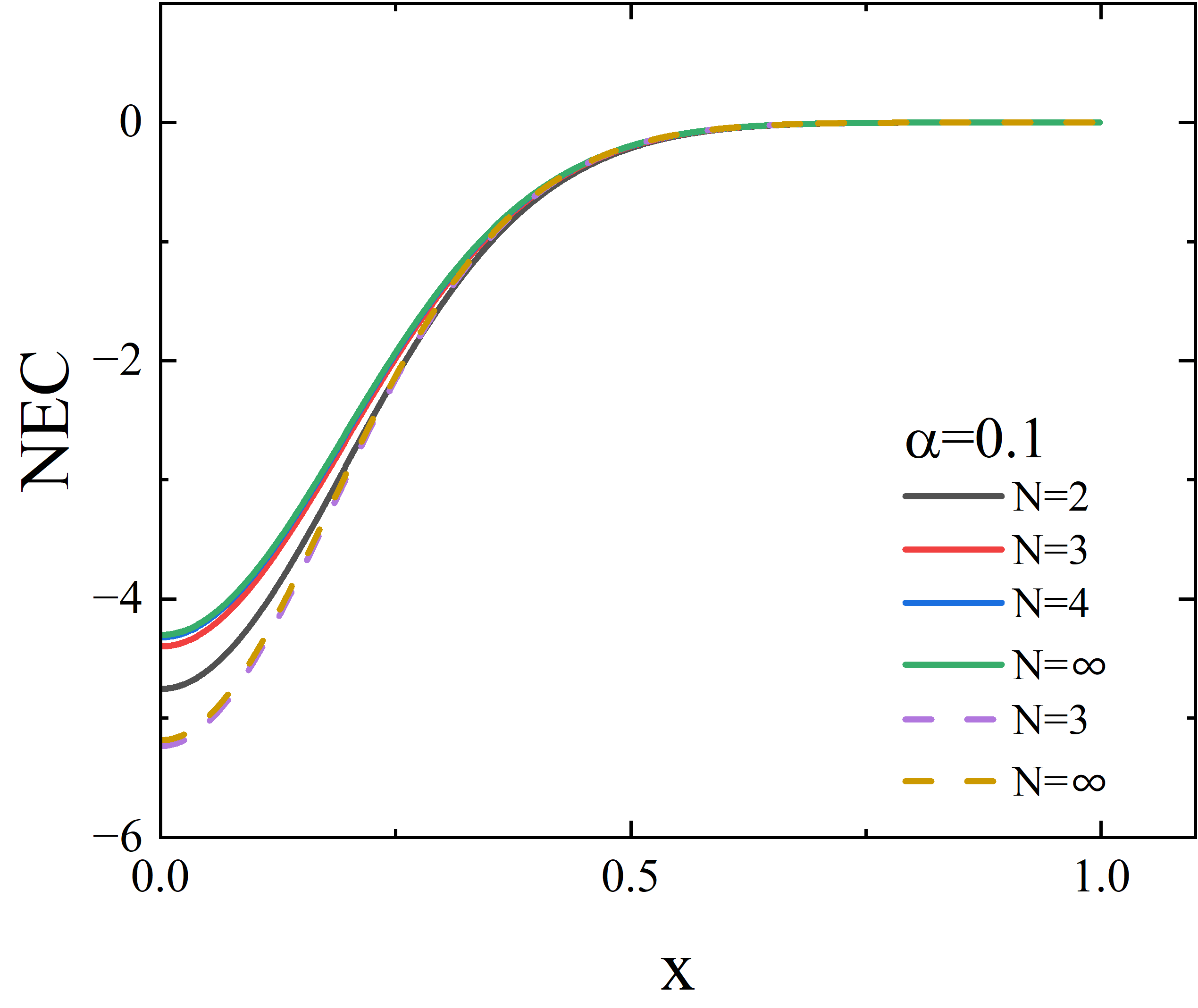}\label{fig:nec2_a}}
\subfigure[]{\includegraphics[width=0.45\textwidth]{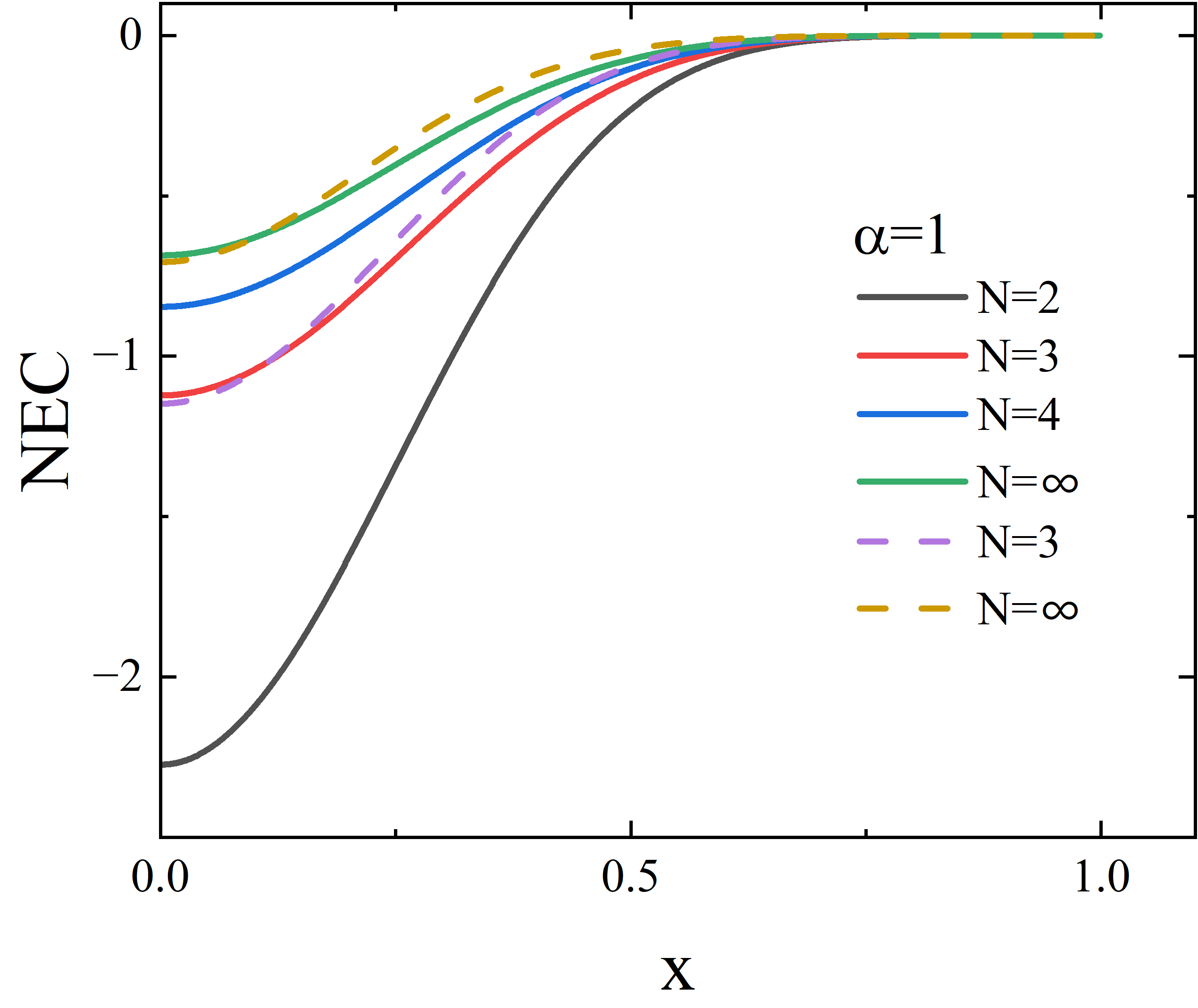}\label{fig:nec2_b}}\\
\subfigure[]{\includegraphics[width=0.45\textwidth]{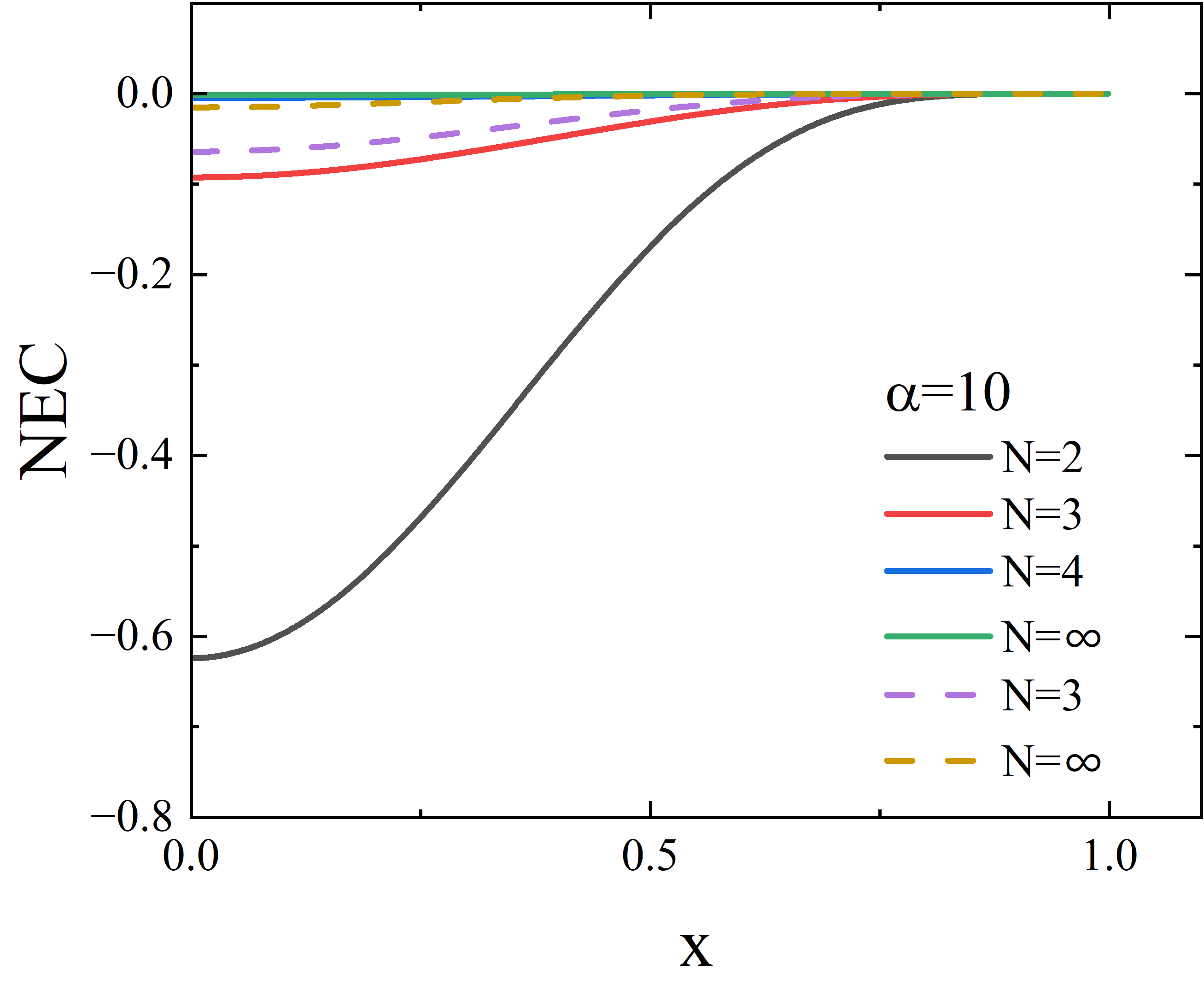}\label{fig:nec2_c}}
\subfigure[]{\includegraphics[width=0.45\textwidth]{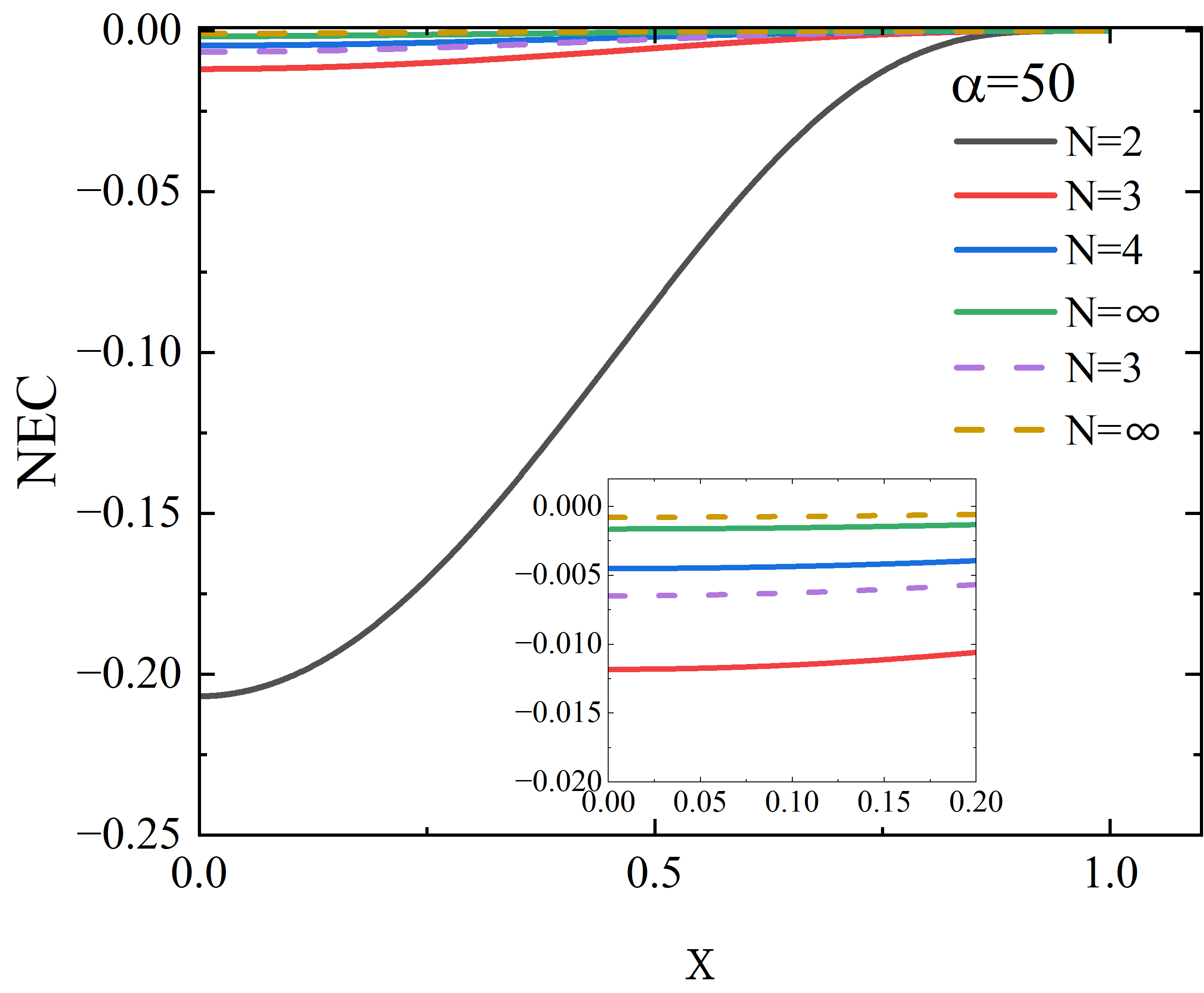}\label{fig:nec2_d}}
\caption{The variation of NEC violation with the radial coordinate $x$. Solid curves correspond to $\alpha_n=\alpha^{\,n-1}$, while dashed curves correspond to $\alpha_n=\frac{1-(-1)^n}{2}\alpha^{\,n-1}$.}
\label{fig:nec2}
\end{figure}

Fig~\ref{fig:nec} mainly captures the overall trend of how the NEC violation varies with the coupling parameter $\alpha$, whereas Fig.~\ref{fig:nec2} emphasizes the differences between the two coefficient configurations at fixed values of $\alpha$. 
Specifically, in Fig.~\ref{fig:nec2}, for relatively small $\alpha$ the dashed configuration exhibits a systematically larger NEC violation than the solid one. 
As $\alpha$ increases, the NEC violation in the dashed configuration drops rapidly and approaches zero. 
Moreover, at the same value of $\alpha$, the violation levels for $N=3$ and $N=\infty$ in the dashed configuration are smaller than the corresponding results with the same $N$ in the solid configuration.

\subsection{Comparison Between Different Spacetime Dimensions: $D=5$ and $D=6$}

After analyzing how different coefficient configurations affect the wormhole solutions, we next consider the extension of quasi-topological gravity to arbitrary spacetime dimensions $D>4$. 
To this end, we investigate how changing the dimension $D$ influences the wormhole solutions while keeping the coefficient configuration fixed.

\begin{figure}[]
\centering
\setlength{\tabcolsep}{6pt}
\renewcommand{\arraystretch}{1.1}
\begin{tabular}{cc}
\includegraphics[width=0.45\textwidth]{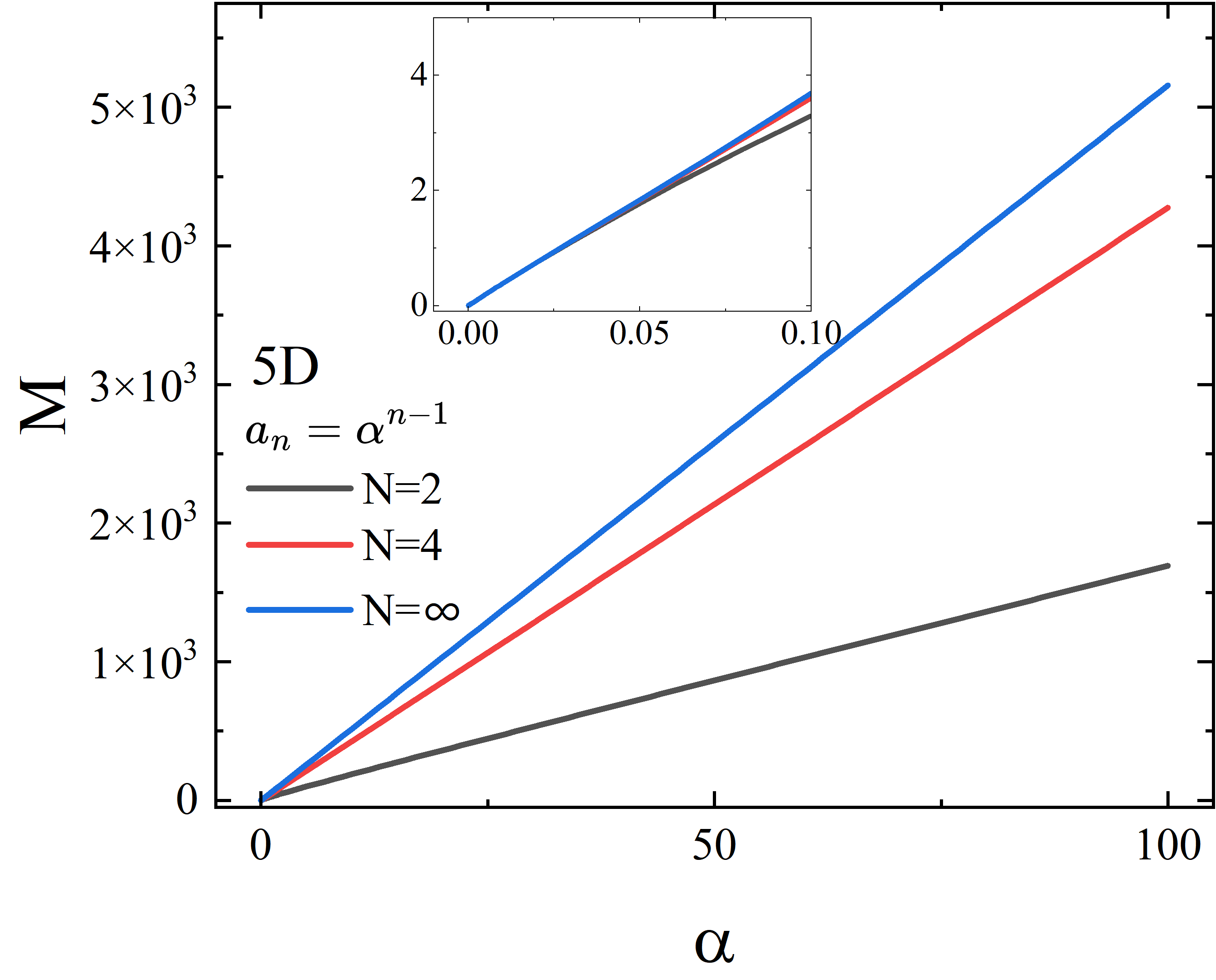} &
\includegraphics[width=0.45\textwidth]{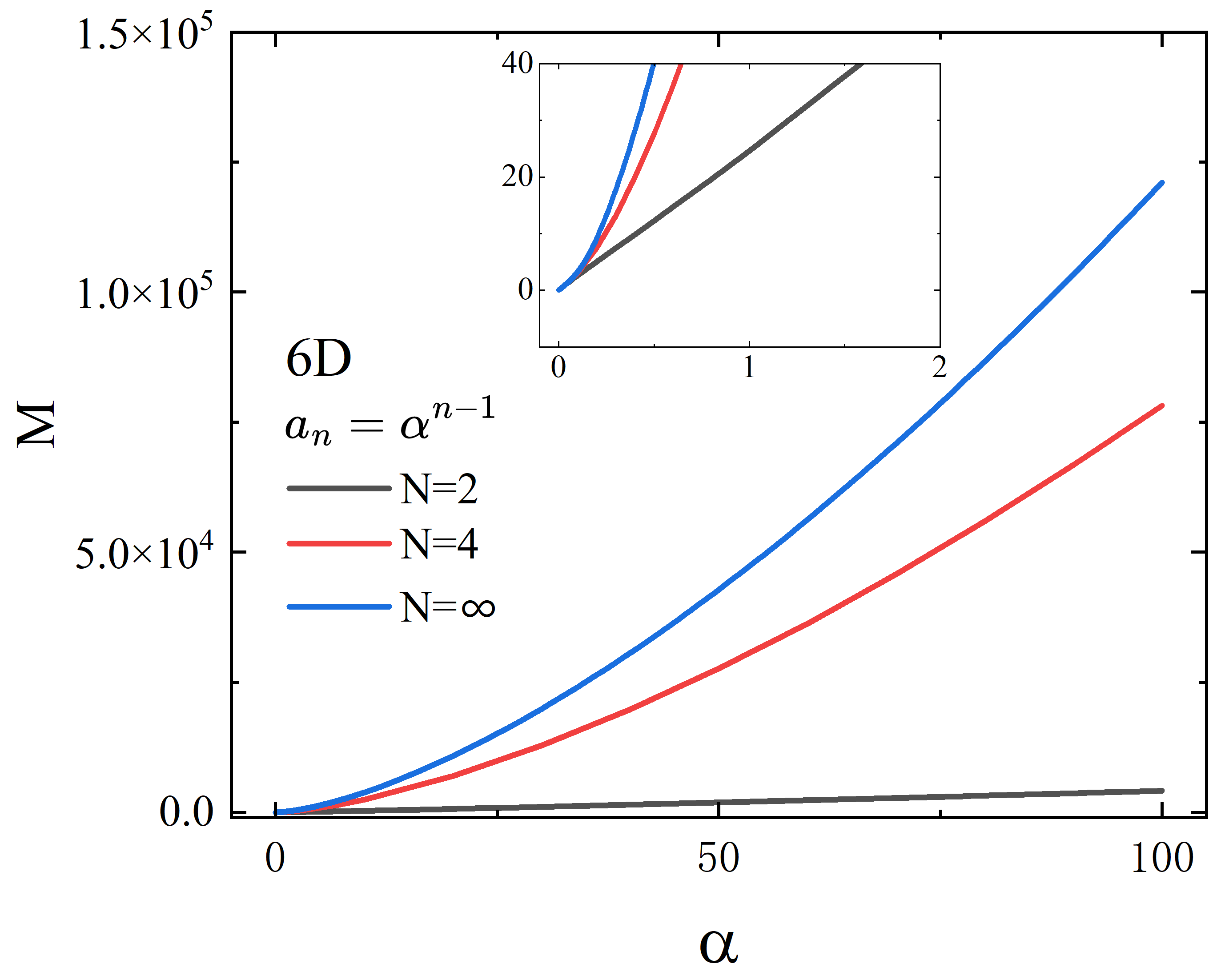} \\
\textit{(a)} & \textit{(b)} \\[0.7ex]
\includegraphics[width=0.45\textwidth]{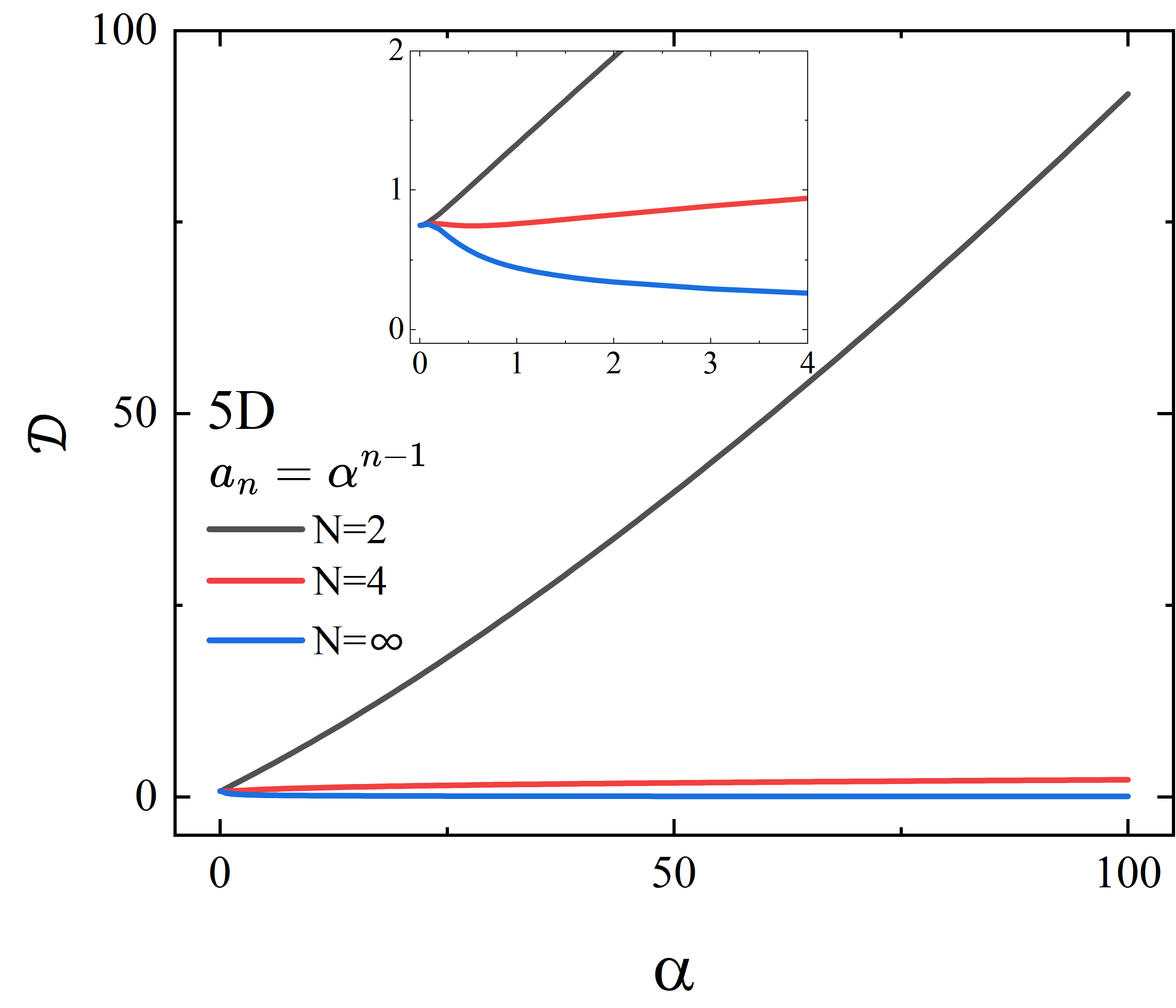} &
\includegraphics[width=0.45\textwidth]{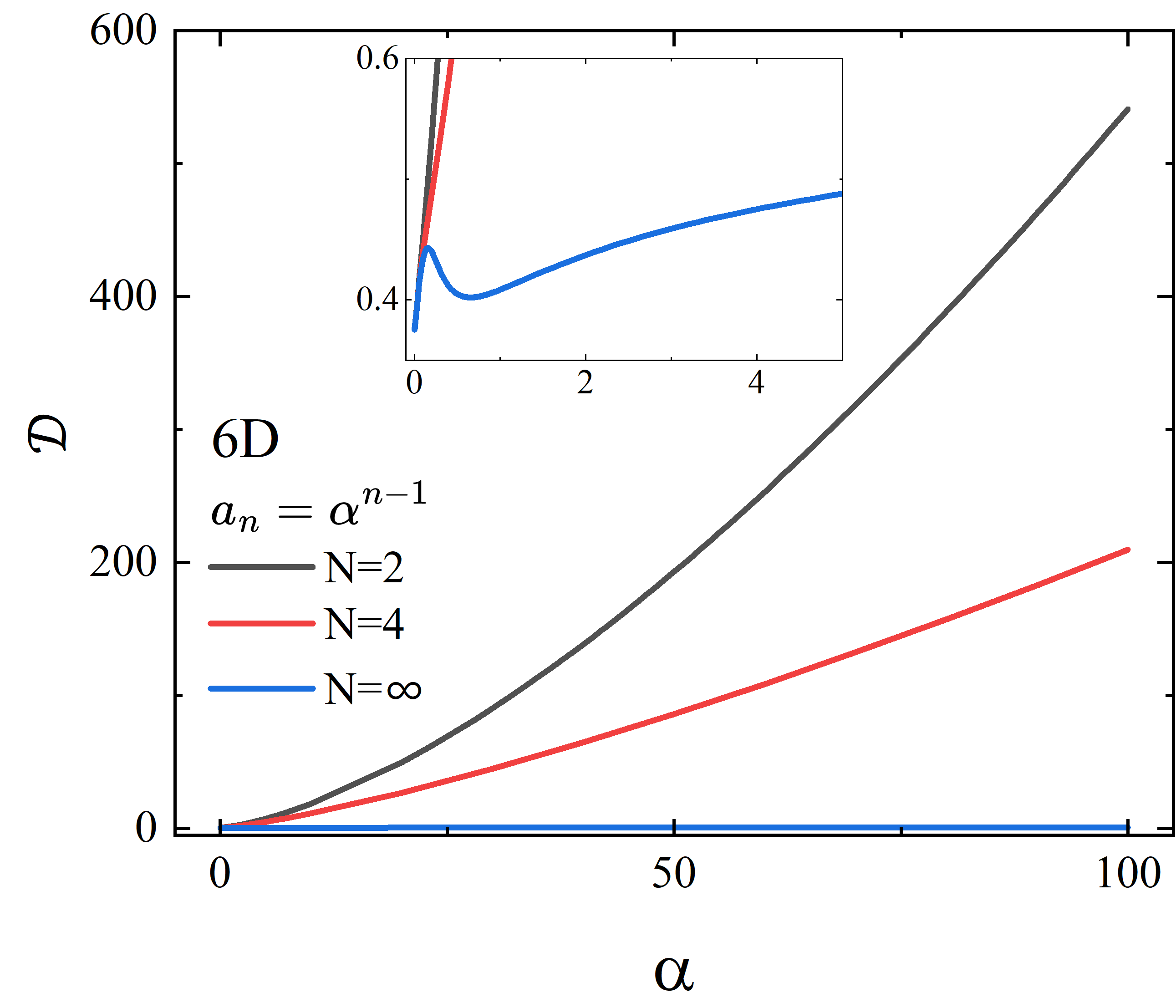} \\
\textit{(c)} & \textit{(d)}
\end{tabular}
\caption{Total mass $M$ (top row) and scalar charge $\mathcal{D}$ (bottom row) as functions of the coupling parameter $\alpha$. The left (right) column corresponds to $D=5$ ($D=6$). The higher-curvature couplings are chosen as $\alpha_n=\alpha^{\,n-1}$.}
\label{fig:MD_5D6D}
\end{figure}

Before presenting the results, we note the following technical point. 
In six dimensions ($D=6$), the cubic quasi-topological term does not contribute to the gravitational dynamics for our static, spherically symmetric background. 
In order to perform a meaningful comparison within the same coefficient configuration, we therefore also remove the $\mathcal{Z}_3$ contribution from the field equations in the five-dimensional ($D=5$) case. 
Accordingly, the curves labeled by $N=\infty$ in the following figures represent the all-order corrections with the $\mathcal{Z}_3$ contribution excluded. 
For simplicity, we continue to denote this case by $N=\infty$, and the same convention is adopted for the finite-$N$ truncations.

As can be seen from Fig.~\ref{fig:MD_5D6D}, in contrast to the situation in Fig.~\ref{fig:MorD_vs_a}, the admissible range of $\alpha$ is more strongly restricted on the $\alpha<0$ side in both $D=5$ and $D=6$. 
For this reason, our discussion below focuses mainly on the $\alpha>0$ regime.

Specifically, Fig.~\ref{fig:MD_5D6D}(a) shows a trend consistent with our previous observations: the wormhole mass $M$ increases as $\alpha$ grows. 
At the same time, for fixed $\alpha$ the mass tends to decrease as the truncation order $N$ increases. 
A similar behavior is also seen in Fig.~\ref{fig:MD_5D6D}(b). 
We emphasize, however, that in $D=6$ the mass can reach values as large as $10^{5}$, and the dependence of $M$ on $\alpha$ is no longer approximately linear as in Fig.~\ref{fig:MD_5D6D}(a), but instead exhibits a more pronounced nonlinearity.

In five dimensions, $\mathcal{D}$ is shown in Fig.~\ref{fig:MD_5D6D}(c). 
Compared with Fig.~\ref{fig:MorD_vs_a}(c), the values of $\mathcal{D}$ for $N=4$ are significantly smaller here and, rather than exhibiting a decrease followed by an increase, they now grow monotonically with $\alpha$. 
For the $N=\infty$ case, $\mathcal{D}$ first increases with $\alpha$ and then decreases. 
This difference originates from the unified prescription adopted in the cross-dimensional comparison, namely the removal of the $\mathcal{Z}_3$ contribution. 
In Fig.~\ref{fig:MD_5D6D}(d), the overall trend is similar, but the $N=\infty$ branch becomes more intricate: as $\alpha$ increases, $\mathcal{D}$ first grows, then decreases, and subsequently increases again. 
This indicates that in $D=6$, changing $\alpha$ alone may not be sufficient to reduce $\mathcal{D}$ to arbitrarily small values.

\begin{figure}[]
\centering
\setlength{\tabcolsep}{6pt}
\renewcommand{\arraystretch}{1.1}
\begin{tabular}{cc}
\includegraphics[width=0.45\textwidth]{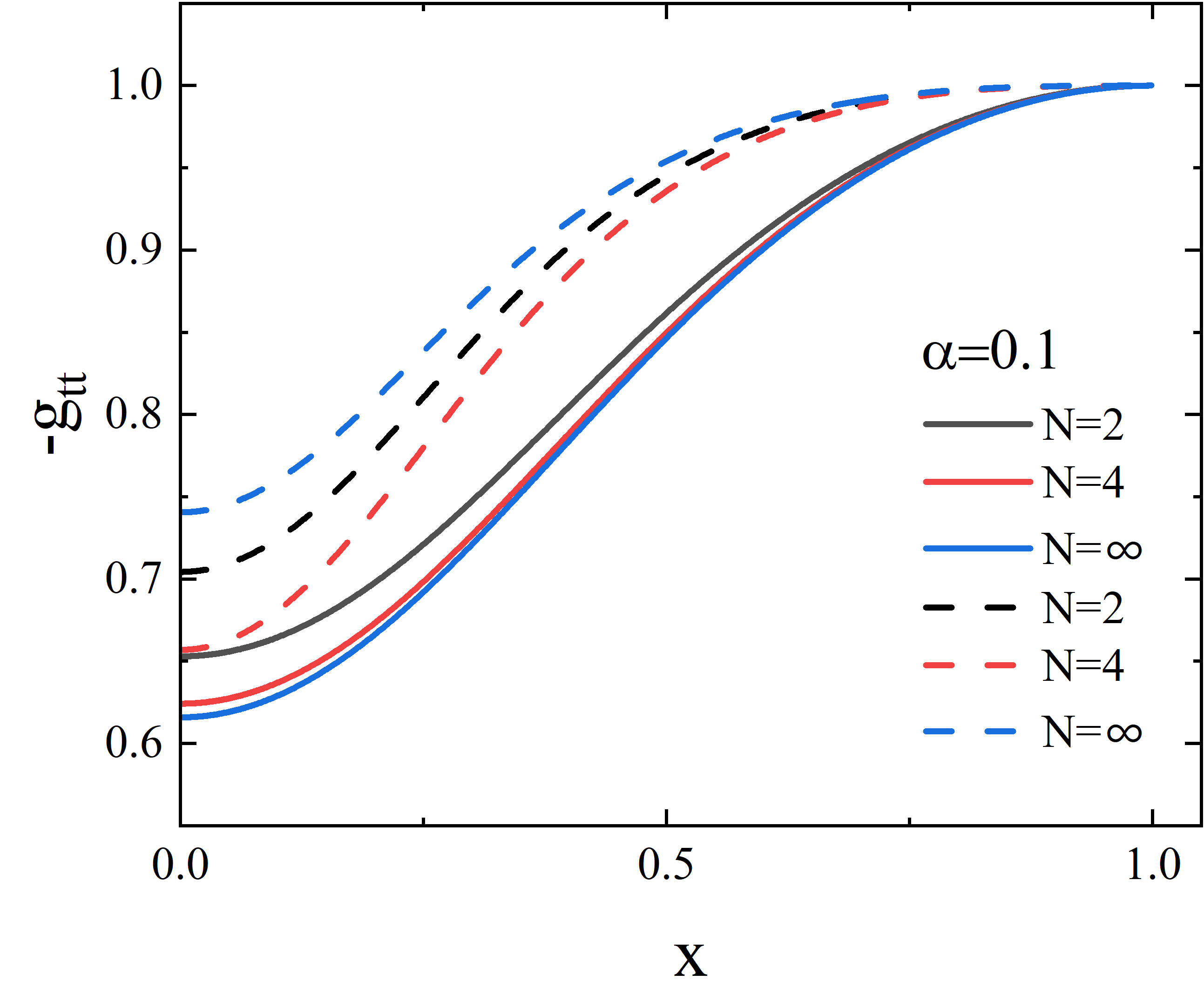} &
\includegraphics[width=0.45\textwidth]{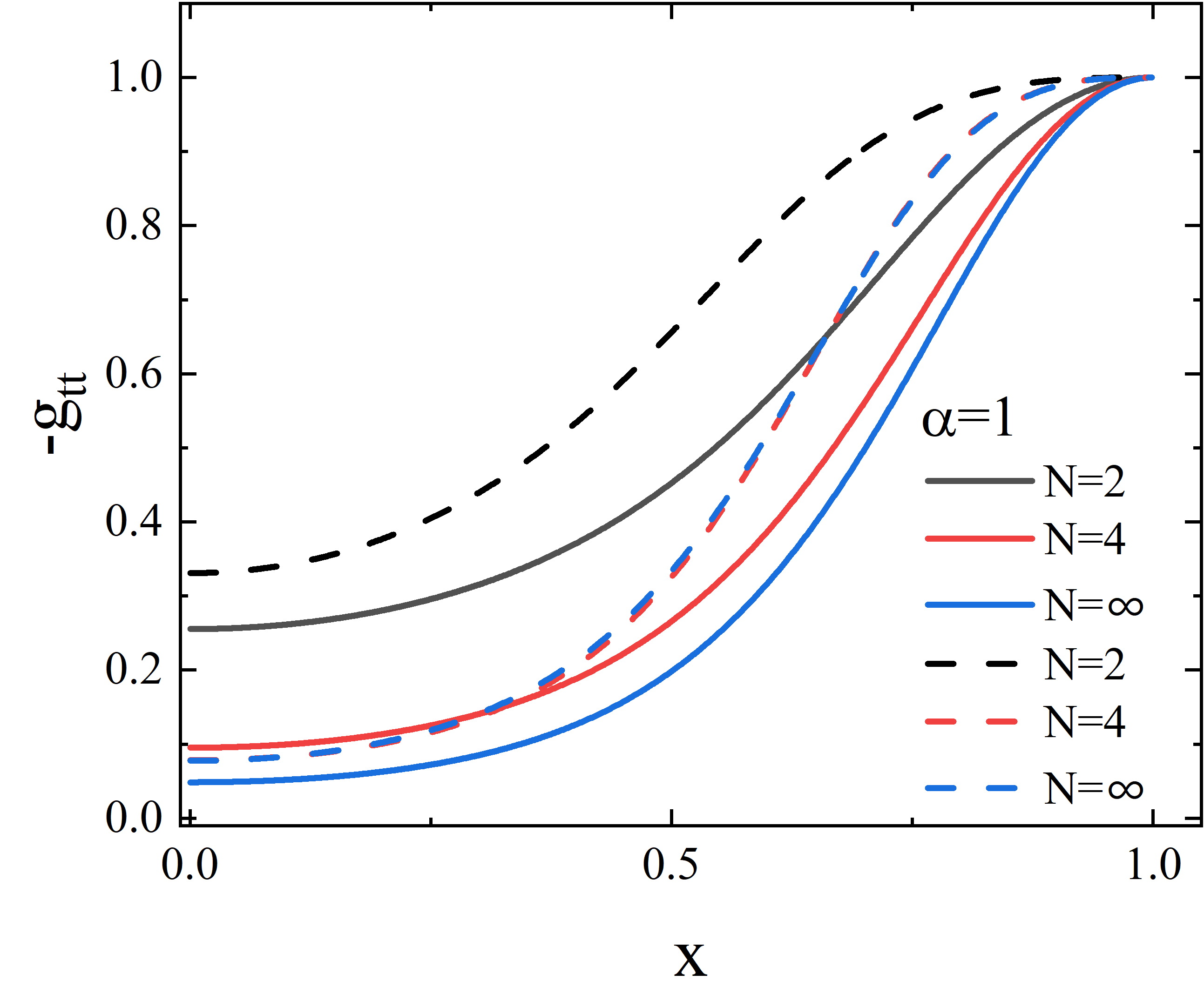} \\
\textit{(a)}  & \textit{(b)}  \\[0.7ex]
\includegraphics[width=0.45\textwidth]{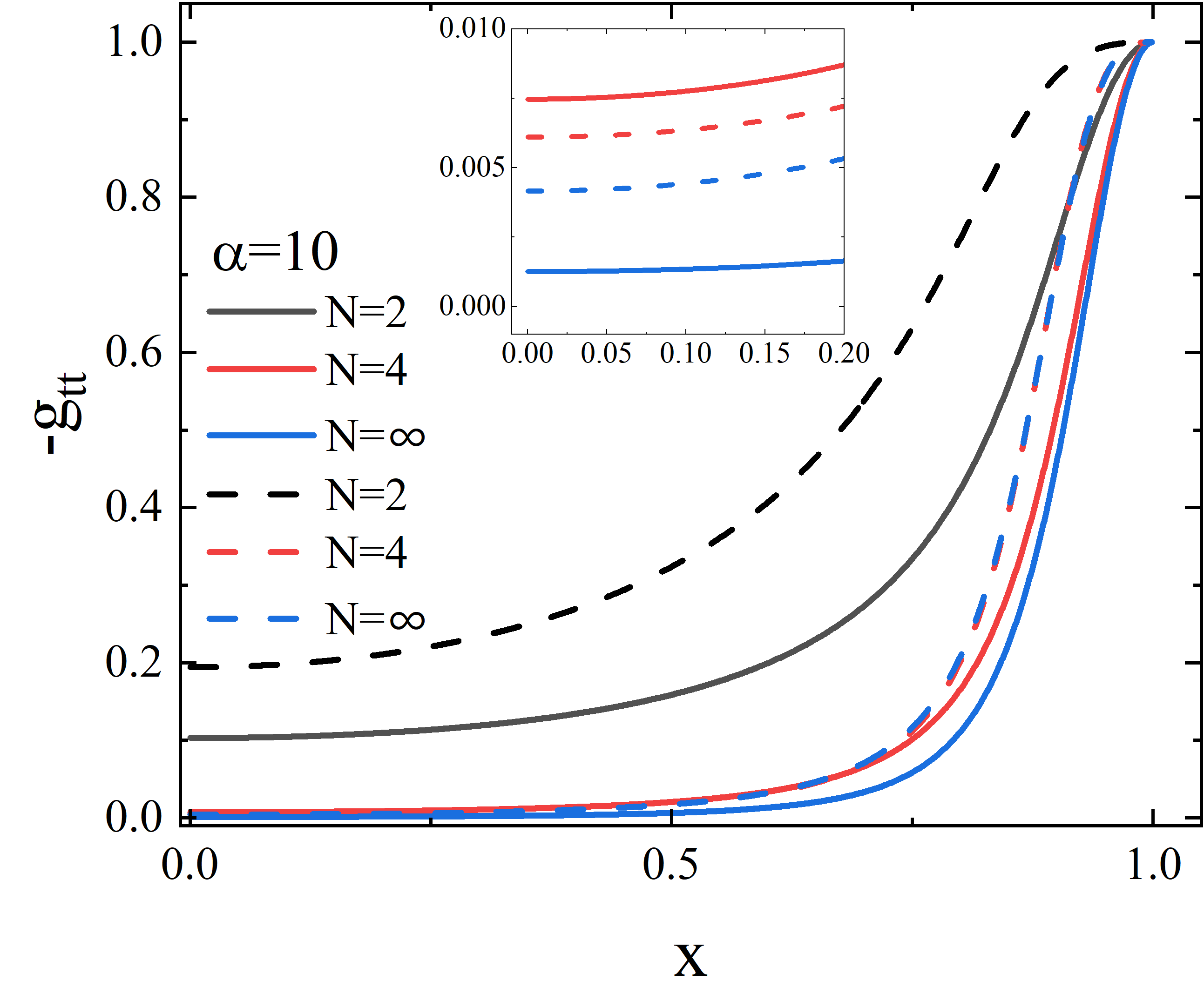} &
\includegraphics[width=0.45\textwidth]{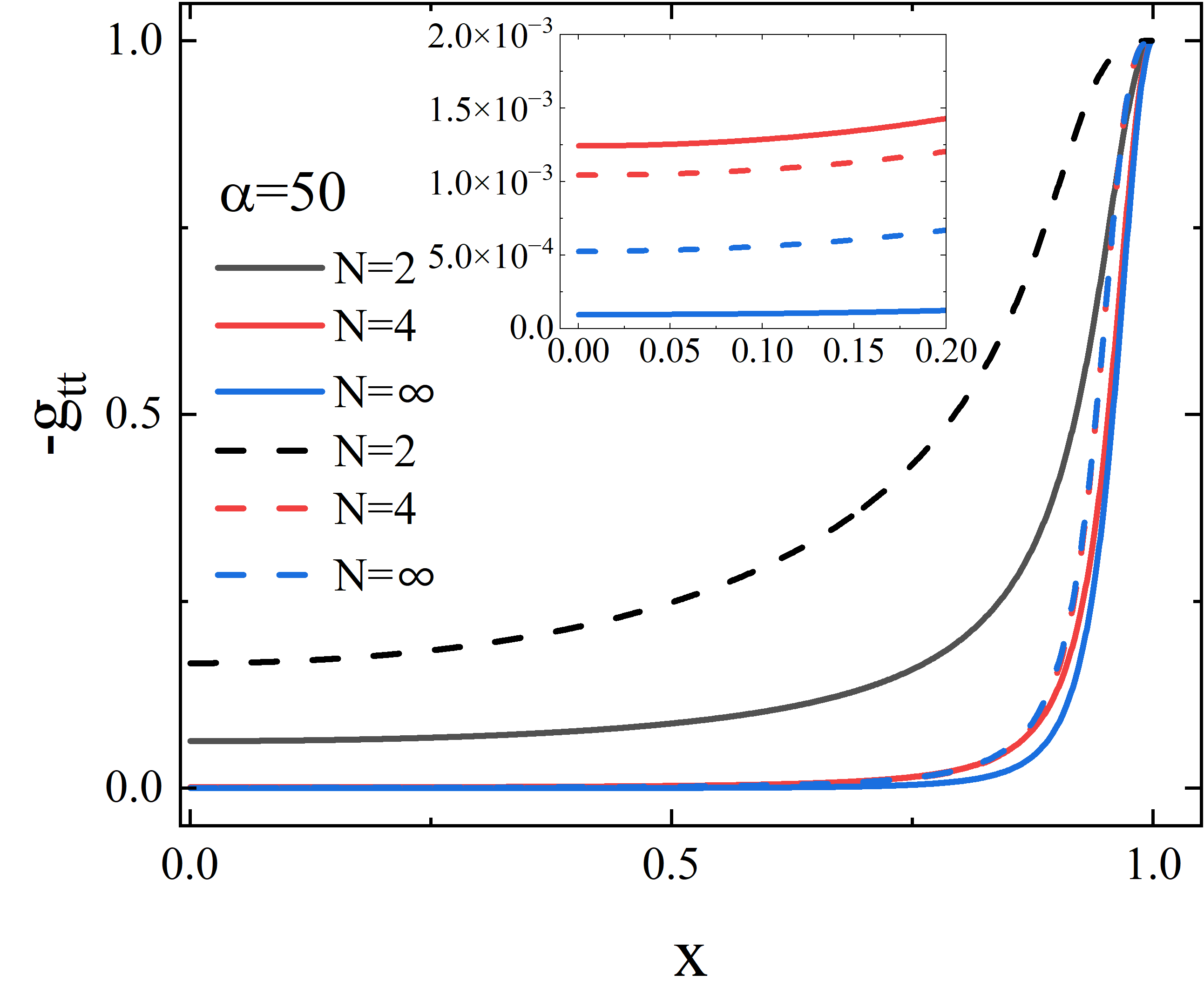} \\
\textit{(c)}  & \textit{(d)} 
\end{tabular}
\caption{The metric functions $-g_{tt}$ as functions of the radial coordinate $x$. Solid curves correspond to $D=5$, while dashed curves correspond to $D=6$. The higher-curvature couplings are chosen as $\alpha_n=\alpha^{\,n-1}$.}
\label{fig:gtt_highD_alpha}
\end{figure}

To compare the dimensional dependence of $-g_{tt}$, we plot in Fig.~\ref{fig:gtt_highD_alpha} the distribution characteristics of the metric function $-g_{tt}$, obtained in $D=5$ (solid curves) and $D=6$ (dashed curves) while varying the coupling parameter $\alpha$. 
For relatively small coupling (Fig.~\ref{fig:gtt_highD_alpha}(a)), the $D=6$ curves lie above the $D=5$ ones. 
As $\alpha$ is increased further (Figs.~\ref{fig:gtt_highD_alpha}(b)--\ref{fig:gtt_highD_alpha}(d)), the $N=2$ curve in $D=6$ remains above its $D=5$ counterpart throughout. 
For higher truncation orders (e.g., $N=4$ and $N=\infty$), we find that $-g_{tt}$ near the throat tends to approach zero, signaling the emergence of a ``horizon''-like structure, irrespective of the spacetime dimension.

\begin{figure}[]
\centering
\setlength{\tabcolsep}{6pt}
\renewcommand{\arraystretch}{1.1}
\begin{tabular}{cc}
\includegraphics[width=0.45\textwidth]{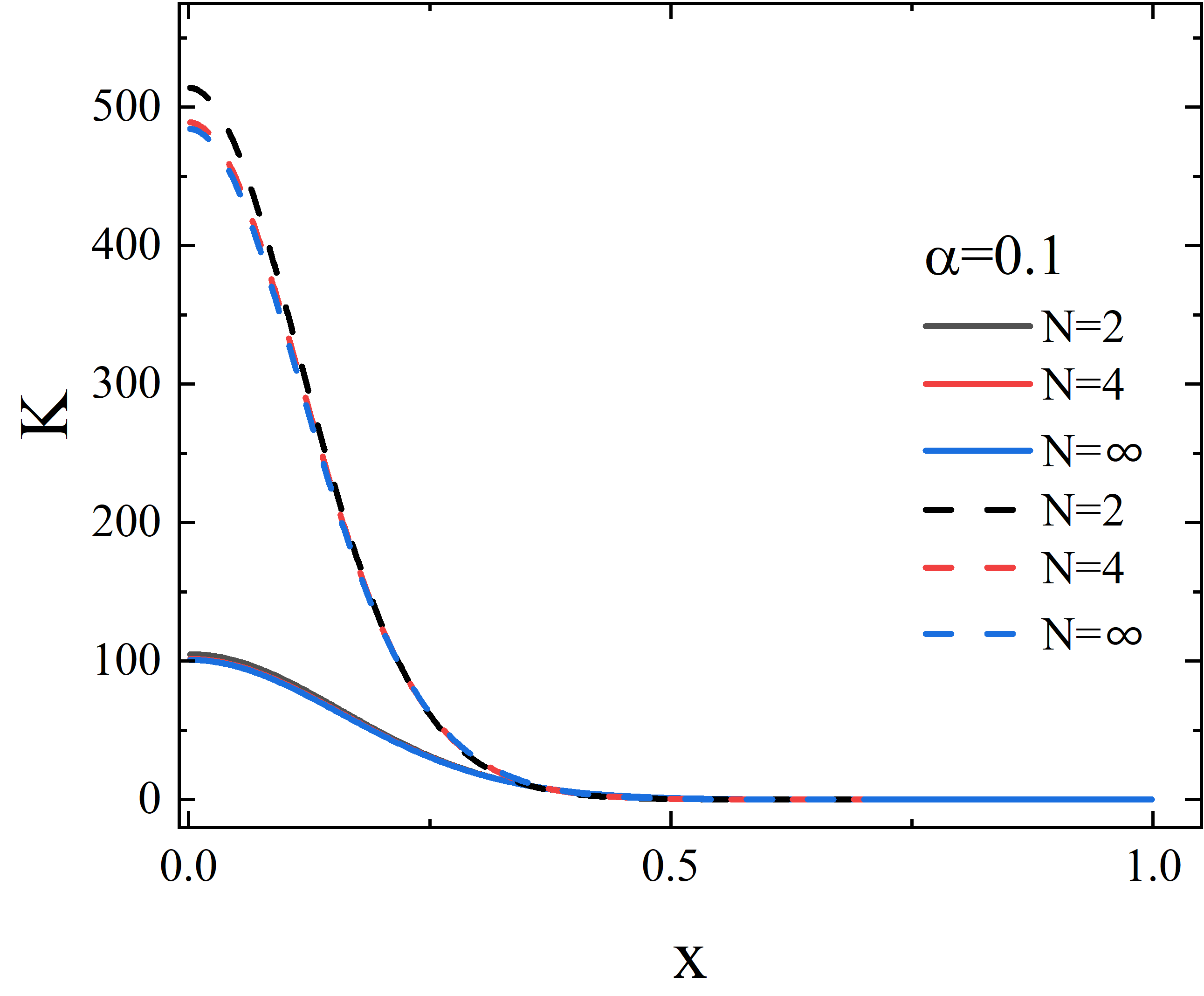} &
\includegraphics[width=0.45\textwidth]{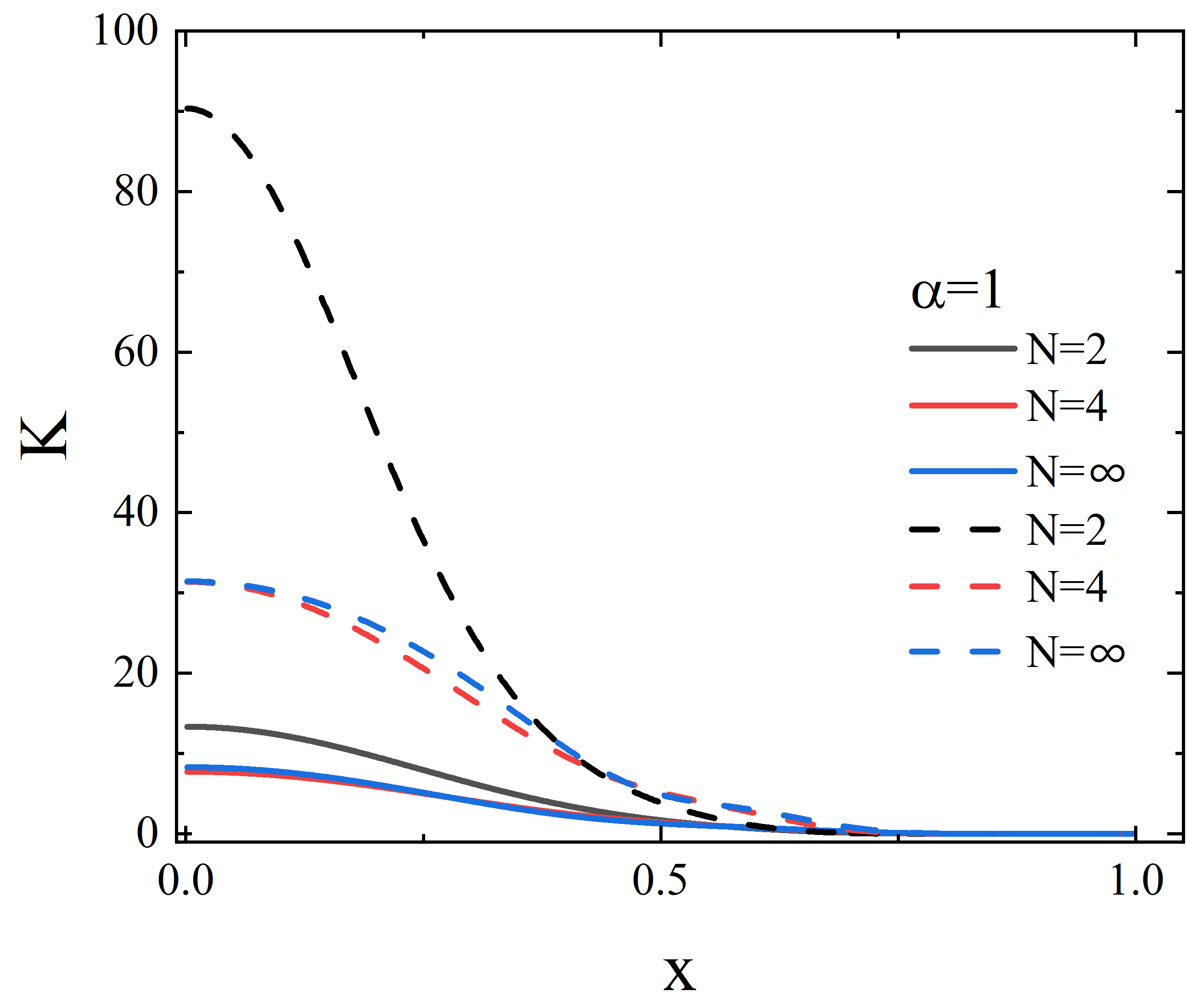} \\
\textit{(a)}  & \textit{(b)}  \\[0.7ex]
\includegraphics[width=0.45\textwidth]{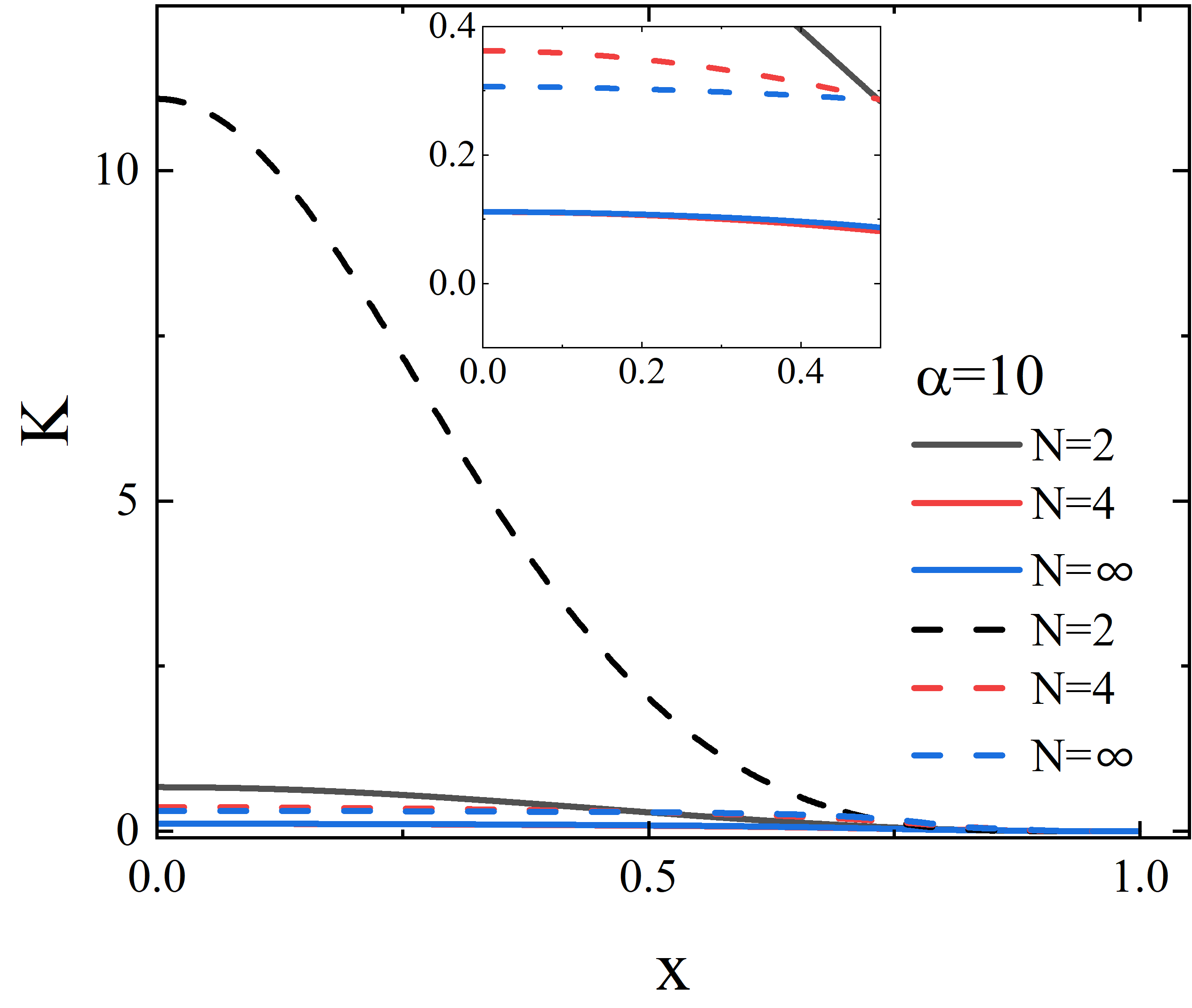} &
\includegraphics[width=0.45\textwidth]{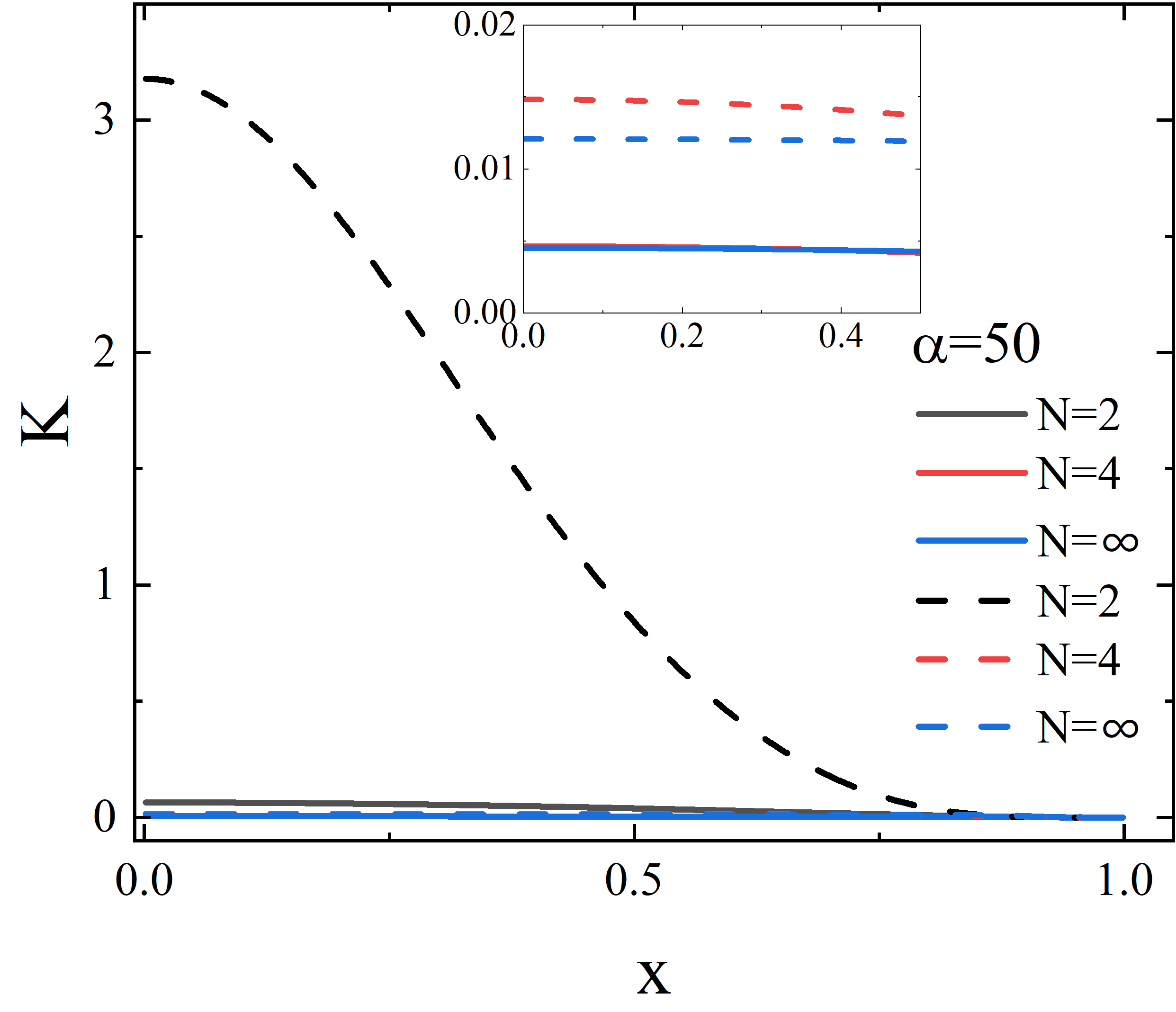} \\
\textit{(c)}  & \textit{(d)} 
\end{tabular}
\caption{The distribution of the Kretschmann scalar $K$ as a function of the radial coordinate $x$. Solid and dashed curves correspond to $D=5$ and $D=6$, respectively.}
\label{fig:K_highD_alpha}
\end{figure}

We compare the Kretschmann scalar $K$ as $\alpha$ is varied for $D=5$ and $D=6$ in Fig.~\ref{fig:K_highD_alpha}. 
The overall trend is consistent with our previous observations: as $\alpha$ increases, $K$ decreases, and this behavior is clearly present in both $D=5$ and $D=6$. 
Moreover, increasing the truncation order $N$ further lowers the value of $K$.
A direct comparison between five and six dimensions shows that the main difference is quantitative: for the same truncation order $N$, the values of $K$ in six dimensions are generally larger than the corresponding values in five dimensions. 
A plausible explanation is that even in the absence of higher-curvature corrections ($\alpha=0$), the corresponding solutions in $D=5$ and $D=6$ already differ in their Kretschmann scalar. 
This initial discrepancy is then carried over as $\alpha$ is increased, resulting in systematically larger values of $K$ in the six-dimensional case.

\begin{figure}[H]
\centering
\setlength{\tabcolsep}{6pt}
\renewcommand{\arraystretch}{1.1}
\begin{tabular}{cc}
\includegraphics[width=0.45\textwidth]{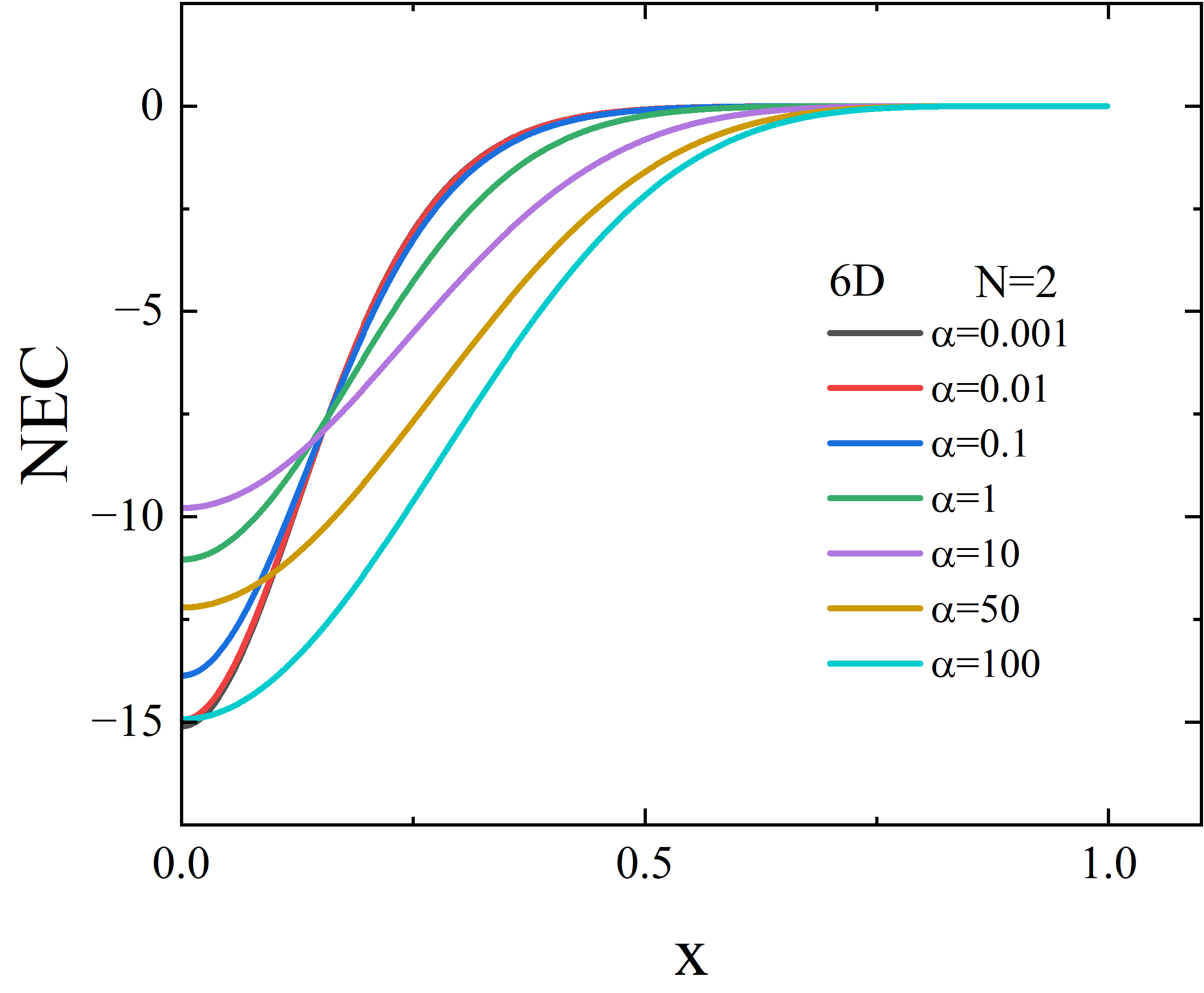} &
\includegraphics[width=0.45\textwidth]{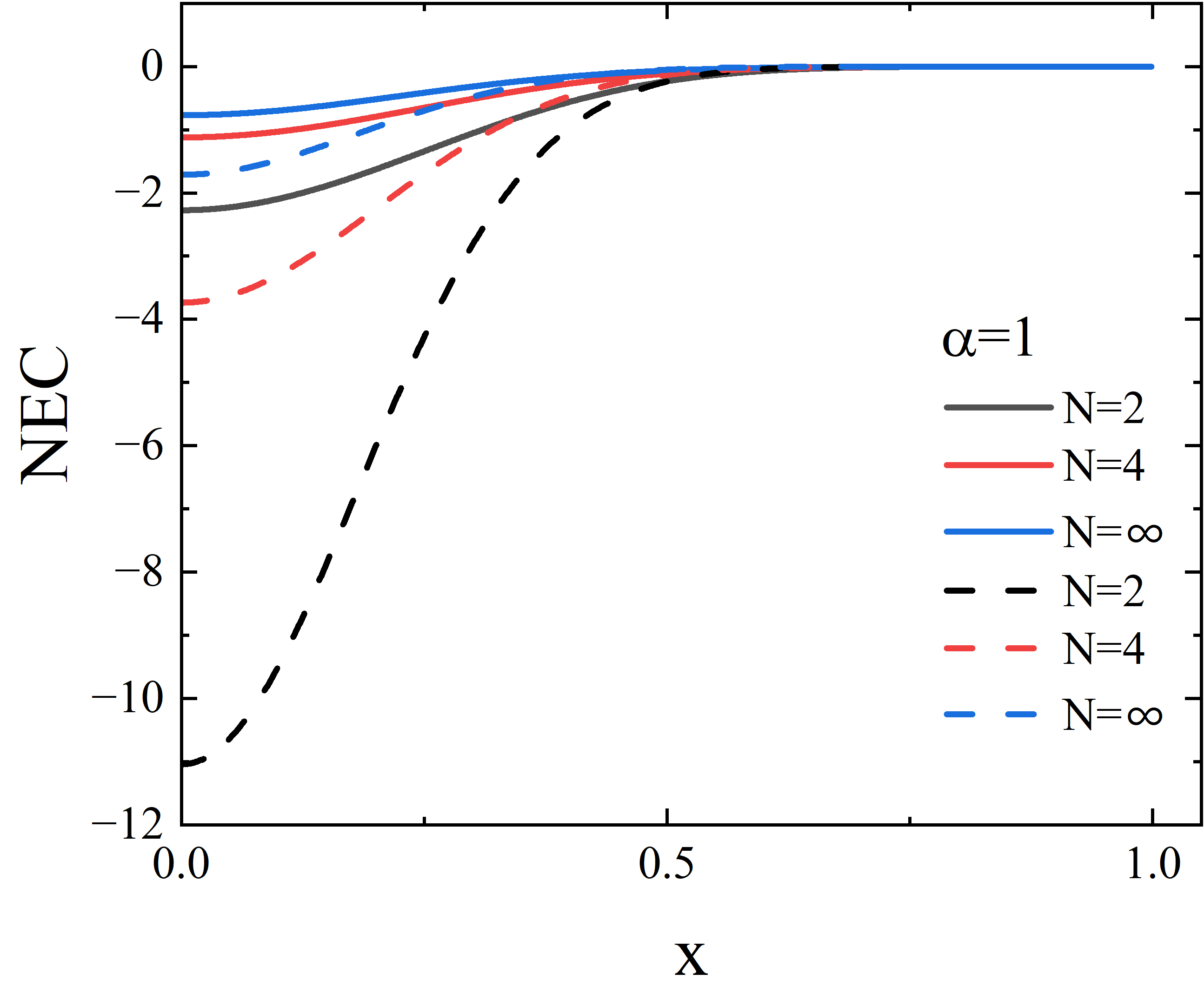} \\
\textit{(a)} & \textit{(b)} \\[0.7ex]
\includegraphics[width=0.45\textwidth]{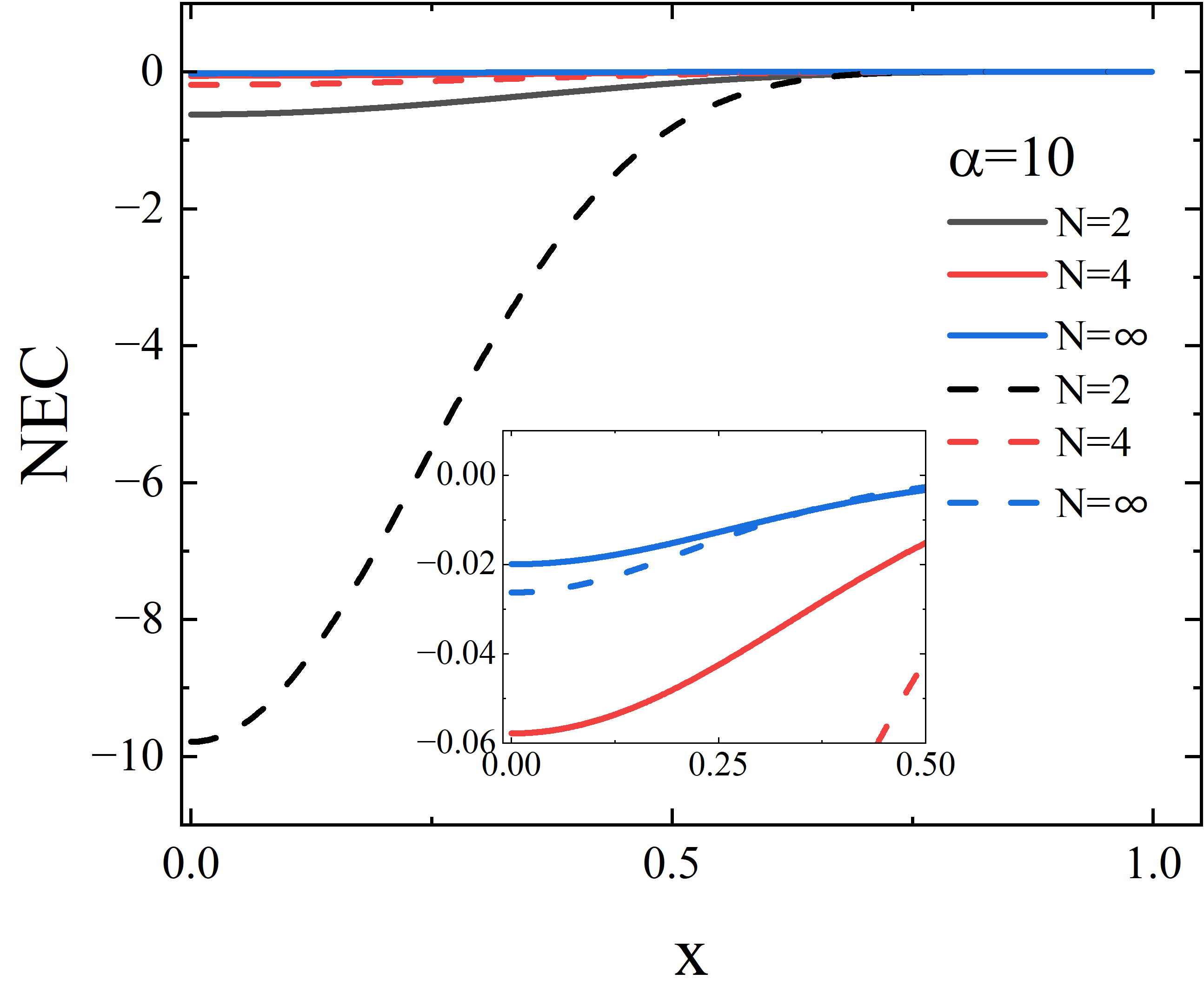} &
\includegraphics[width=0.45\textwidth]{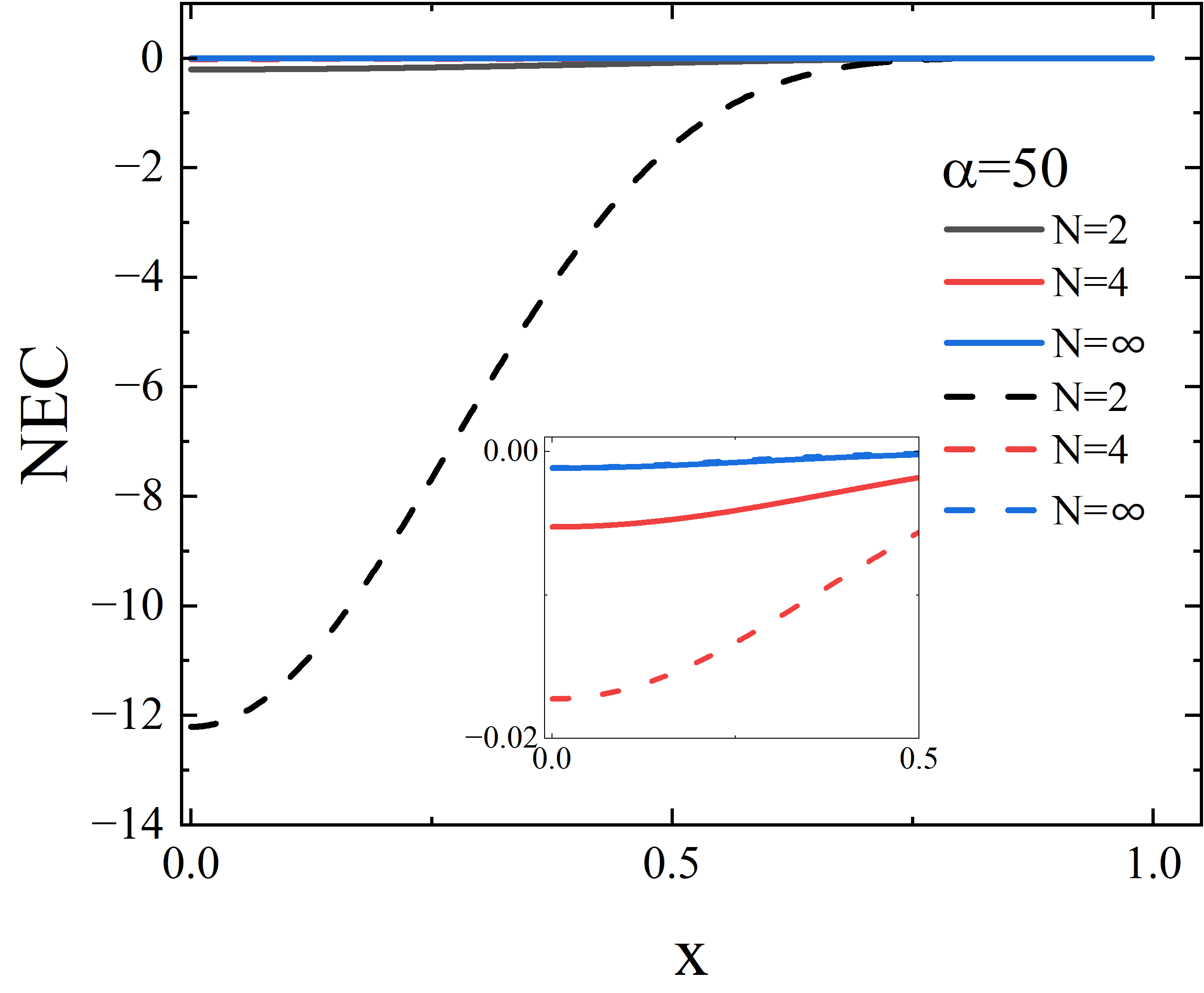} \\
\textit{(c)} & \textit{(d)}
\end{tabular}
\caption{The variation of NEC violation with the radial coordinate $x$. Solid and dashed curves correspond to $D=5$ and $D=6$, respectively. The higher-curvature couplings are chosen as $\alpha_n=\alpha^{\,n-1}$.}
\label{fig:NEC_6D_alpha}
\end{figure}

To illustrate the $\alpha$-dependence of the NEC violation in six dimensions, we plot the corresponding quantity in Fig.~\ref{fig:NEC_6D_alpha}. 
For $N=2$ (Fig.~\ref{fig:NEC_6D_alpha}(a)), the behavior in $D=6$ differs from the five-dimensional case: as $\alpha$ increases, the NEC violation first becomes stronger and then weakens.
By contrast, the other six-dimensional cases (e.g., $N=4$ and $N=\infty$) follow the trend observed previously: increasing $\alpha$ leads to a monotonic suppression of the NEC violation, which gradually approaches zero.

\subsection{Geometric Properties}

To visualize the geometry of the wormhole throat and its variation across the parameter space, we employ an isometric embedding of a two-dimensional spatial section.
We first provide an intuitive picture of the wormhole geometry using an isometric embedding.
Starting from the line element \eqref{eq:wormhole_metric}, we fix the time coordinate $t$ and restrict to the ``equatorial'' section by setting
\begin{equation}
    \theta_1=\theta_2=\cdots=\theta_{D-3}=\frac{\pi}{2}. 
\end{equation}
The resulting two-dimensional surface can be embedded into three-dimensional Euclidean space with cylindrical coordinates $(\rho,\varphi,z)$. 
On this slice, the induced line element can be written in two equivalent forms,
\begin{equation}
\mathrm{d}s^2
=\frac{p(r)}{e^{A(r)}}\,\mathrm{d}r^2
+\frac{p(r)}{e^{A(r)}}(r^2+r_0^2)\,\mathrm{d}\varphi^2
=\mathrm{d}\rho^2+\mathrm{d}z^2+\rho^2\,\mathrm{d}\varphi^2 .
\end{equation}
By comparing the metric coefficients, we obtain the embedding relations and hence the functions $\rho(r)$ and $z(r)$ describing the wormhole profile:
\begin{align}
\rho(r) &= \sqrt{\frac{p(r)}{e^{A(r)}}\,(r^2+r_0^2)}, \\
z(r) &= \pm \int \sqrt{\frac{p(r)}{e^{A(r)}}-\left(\frac{\mathrm{d}\rho}{\mathrm{d}r}\right)^2}\,\mathrm{d}r .
\end{align}
The quantity $\rho$ is the circumferential radius, i.e., the radius of the circle at fixed $r$ on the equatorial plane. 
The behavior of $\rho(r)$ characterizes the key geometric features of the wormhole. 
In particular, at an extremum of $\rho(r)$ (where $\mathrm{d}\rho/\mathrm{d}r=0$), a positive second derivative corresponds to a minimal surface, which we identify as the \textbf{throat}. 
Conversely, a negative second derivative corresponds to a maximal surface, which we refer to as the \textbf{equator}.

\begin{figure}[]
\centering
\setlength{\tabcolsep}{6pt}
\renewcommand{\arraystretch}{1.1}
\begin{tabular}{cc}
\includegraphics[width=0.45\textwidth]{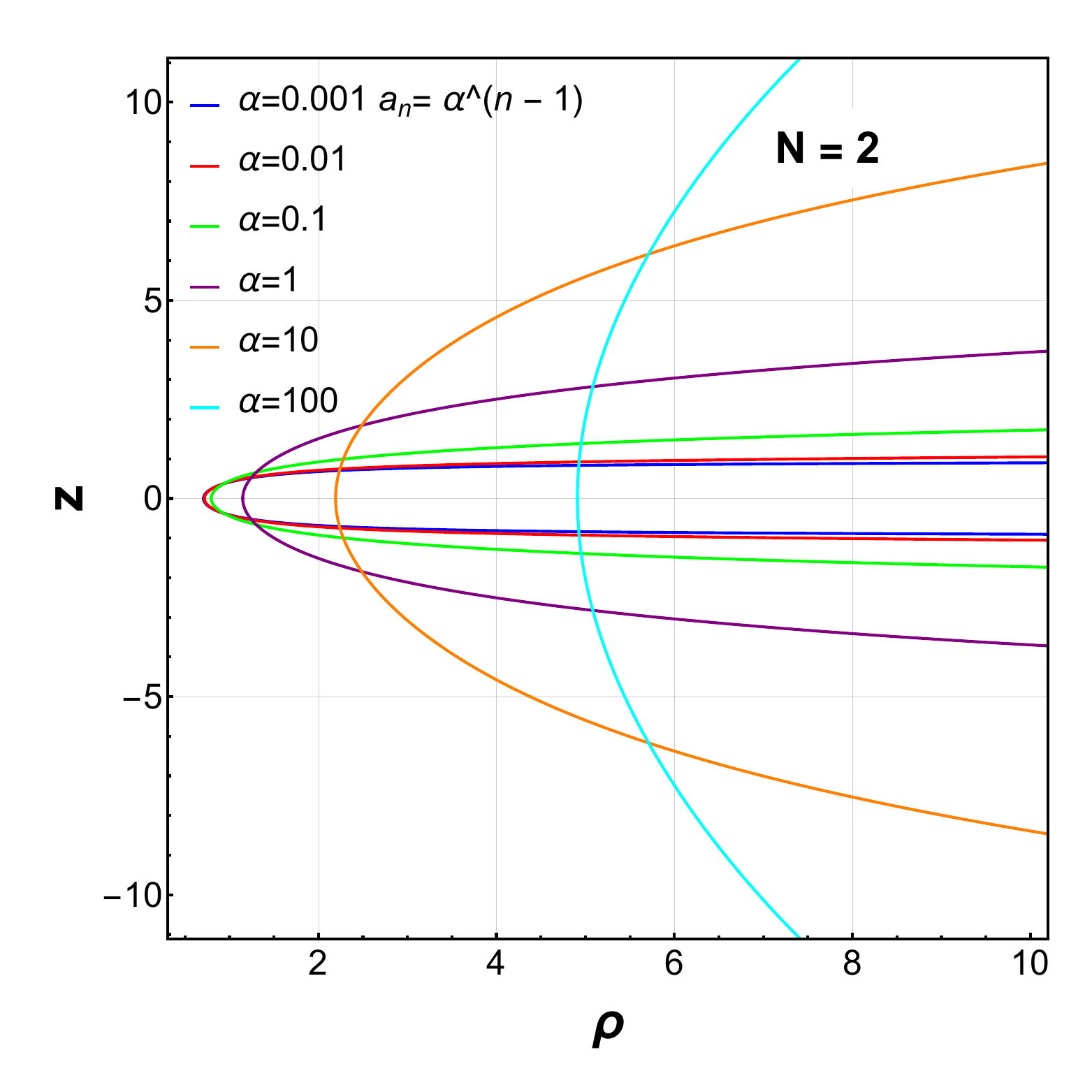} &
\includegraphics[width=0.45\textwidth]{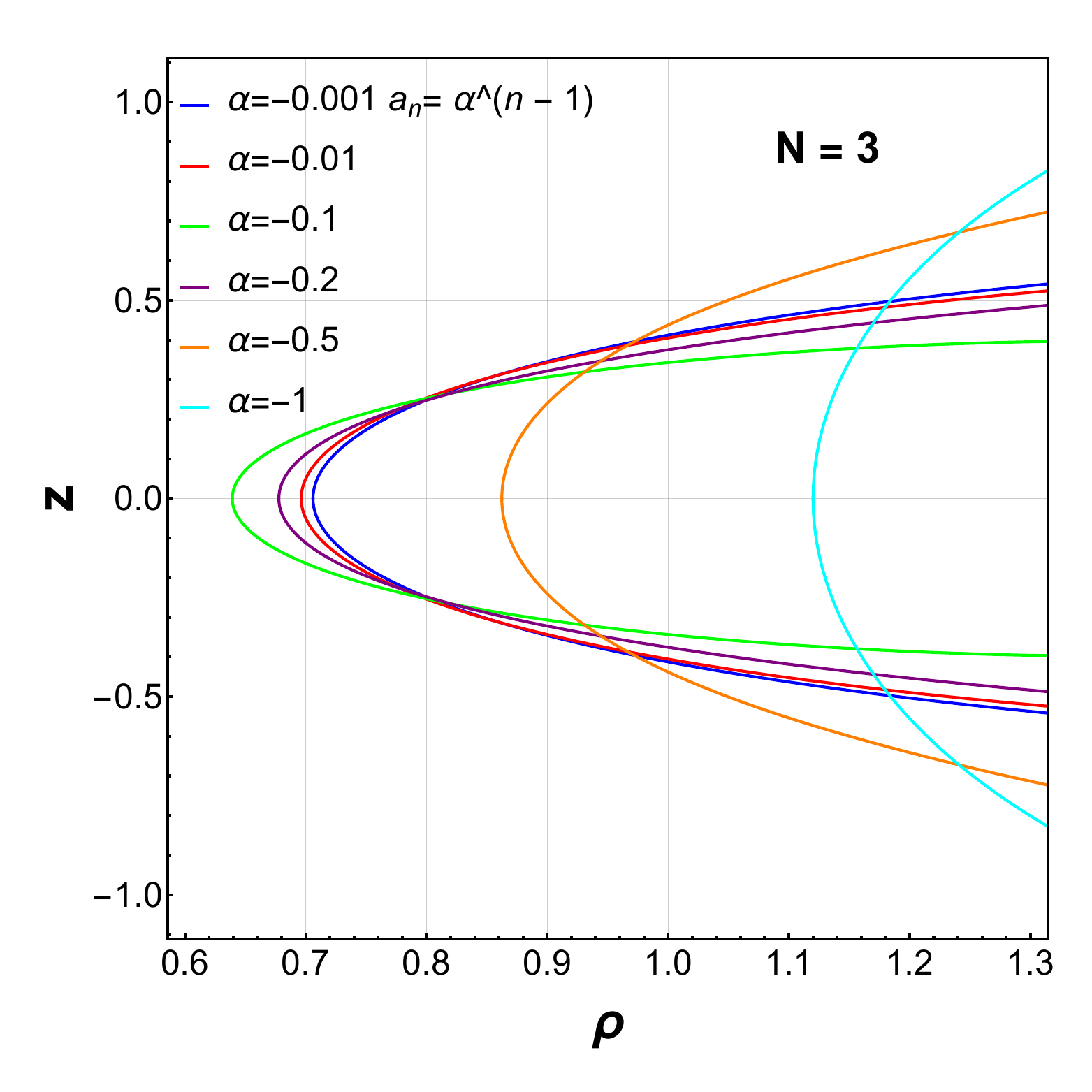} \\
\textit{(a)} & \textit{(b)} \\[0.7ex]
\multicolumn{2}{c}{\includegraphics[width=0.45\textwidth]{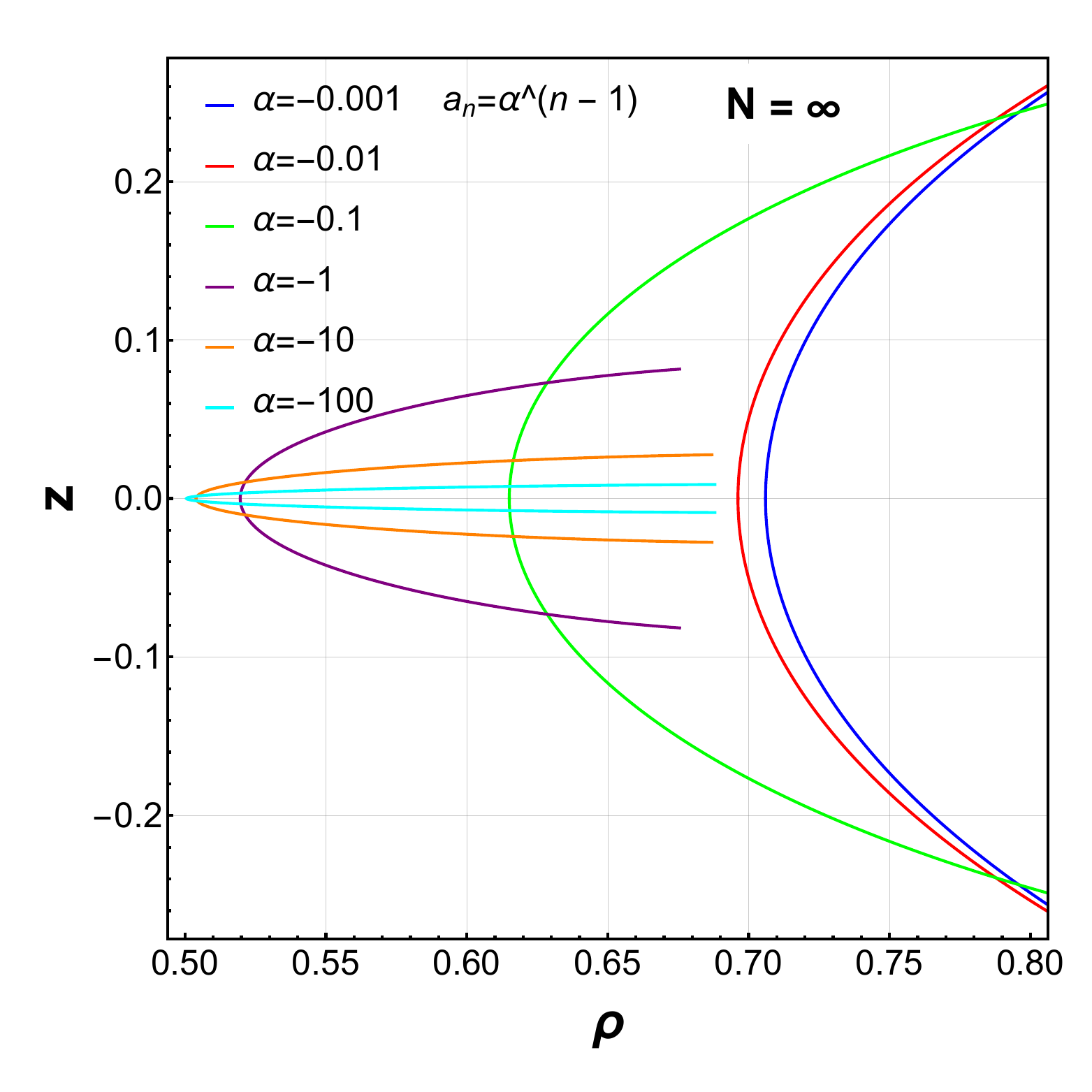}} \\
\multicolumn{2}{c}{\textit{(c)}}
\end{tabular}
\caption{Two-dimensional wormhole embedding diagrams ($D=5$). Panels (a)--(c) correspond to different truncation orders $N$. The coefficients are chosen as $\alpha_n=\alpha^{\,n-1}$.}
\label{fig:geo}
\end{figure}

\begin{figure}[]
\centering
\setlength{\tabcolsep}{6pt}
\renewcommand{\arraystretch}{1.1}
\begin{tabular}{cc}
\includegraphics[width=0.45\textwidth]{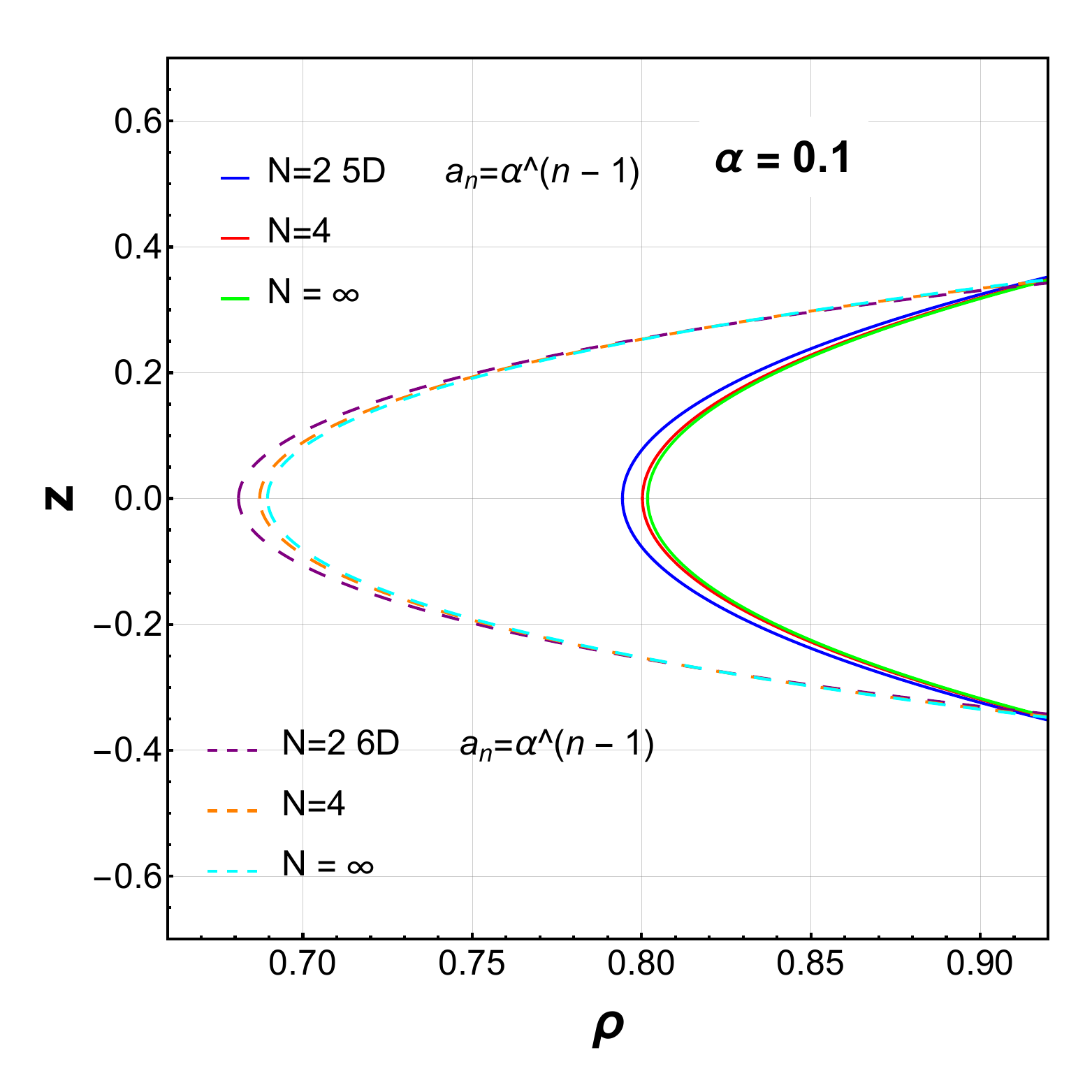} &
\includegraphics[width=0.45\textwidth]{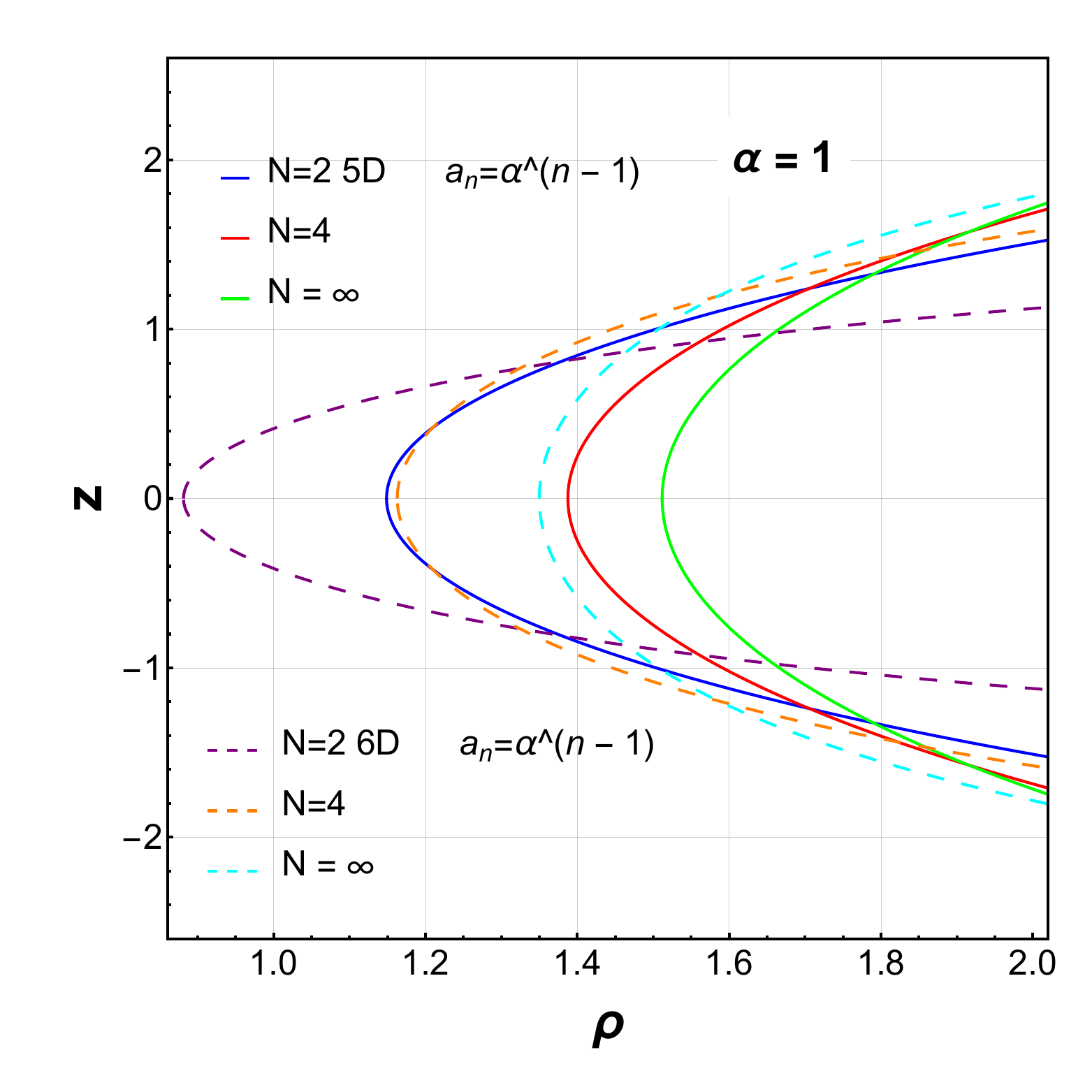} \\
\textit{(a)}  & \textit{(b)}  \\[0.7ex]
\multicolumn{2}{c}{\includegraphics[width=0.45\textwidth]{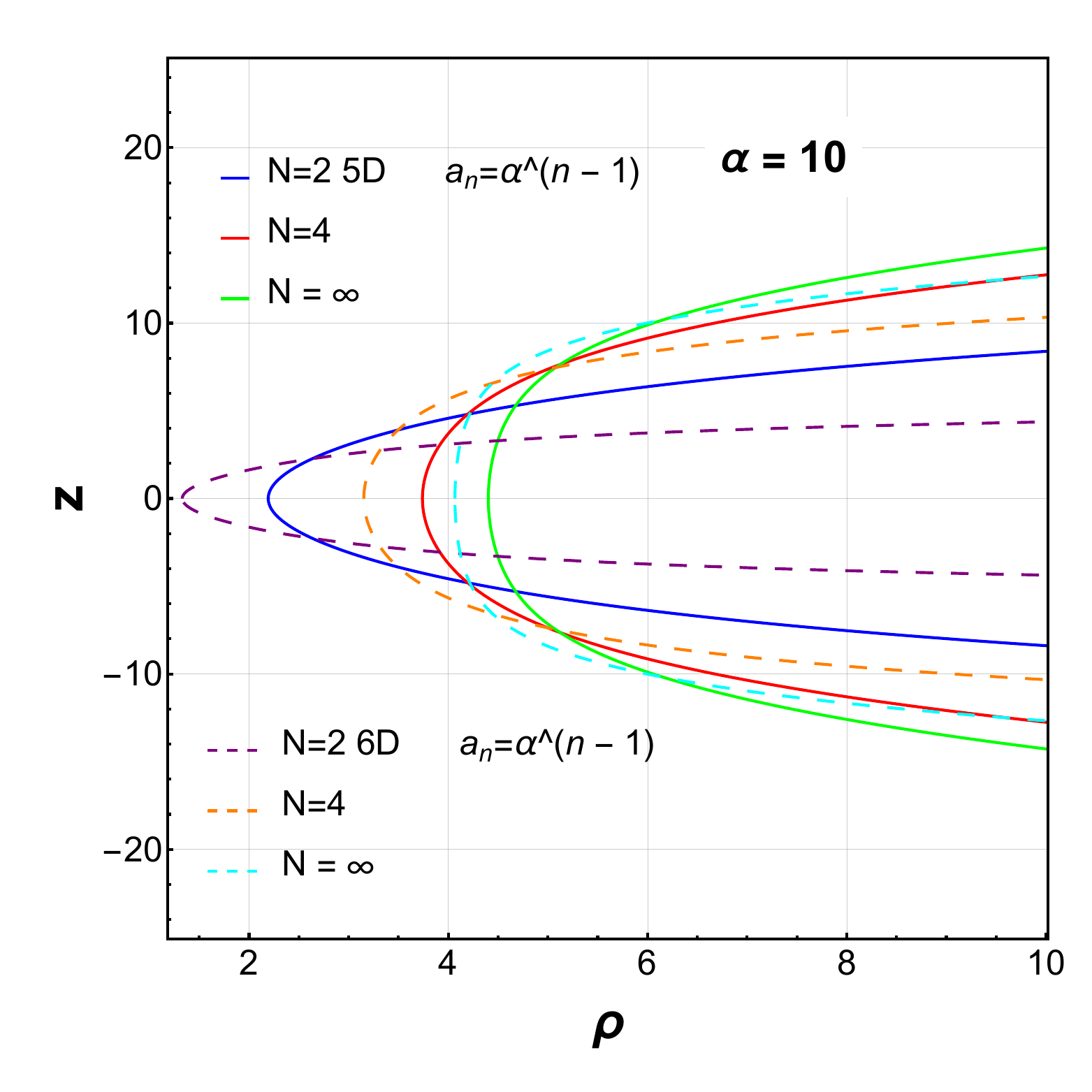}} \\
\multicolumn{2}{c}{\textit{(c)}}
\end{tabular}
\caption{Two-dimensional wormhole embedding diagrams ($D=5$) for two different coefficient choices: $\alpha_n=\alpha^{\,n-1}$ (solid curves) and $\alpha_n=\frac{1-(-1)^n}{2}\,\alpha^{\,n-1}$ (dashed curves).}
\label{fig:geo2}
\end{figure}

\begin{figure}[]
\centering
\setlength{\tabcolsep}{6pt}
\renewcommand{\arraystretch}{1.1}
\begin{tabular}{cc}
\includegraphics[width=0.45\textwidth]{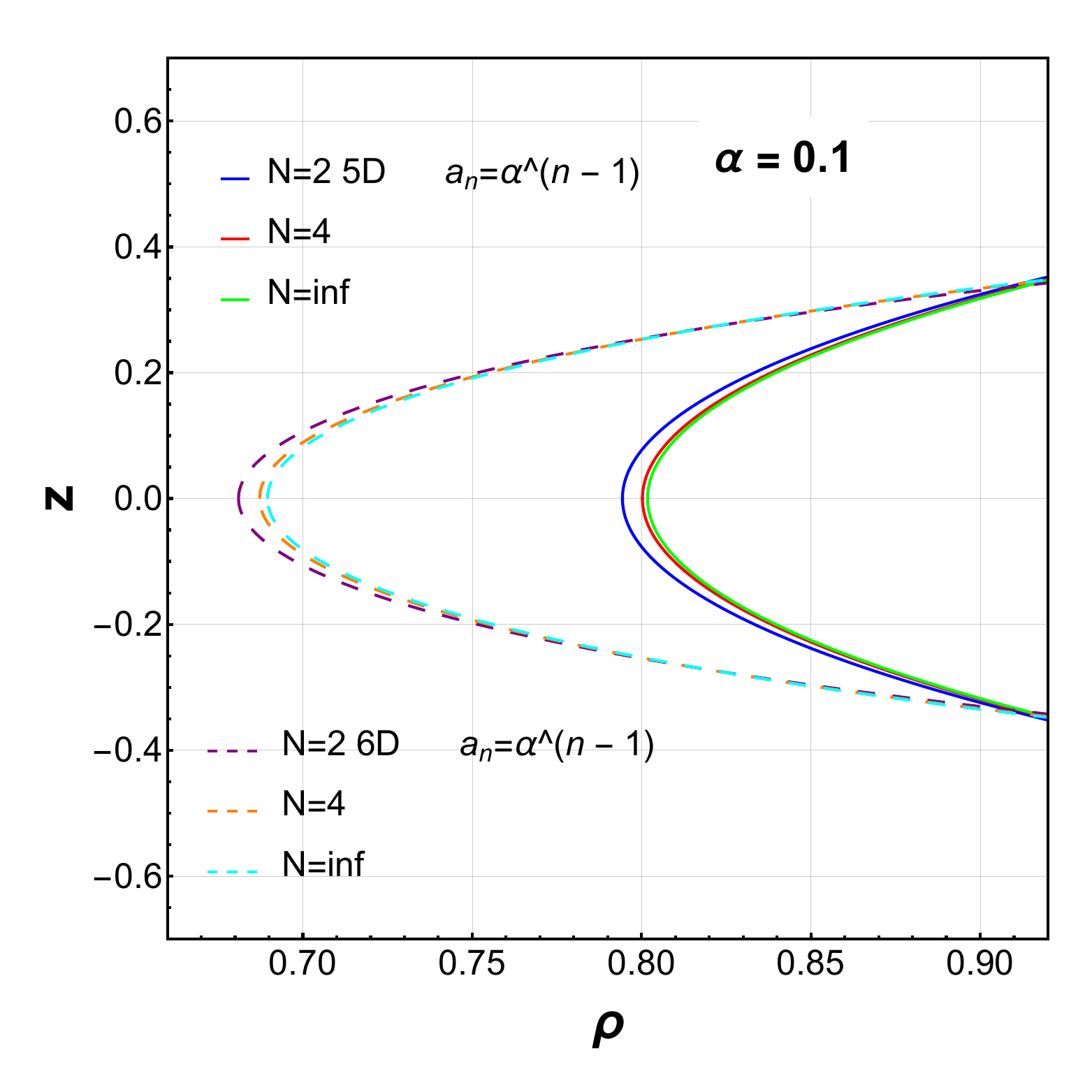} &
\includegraphics[width=0.45\textwidth]{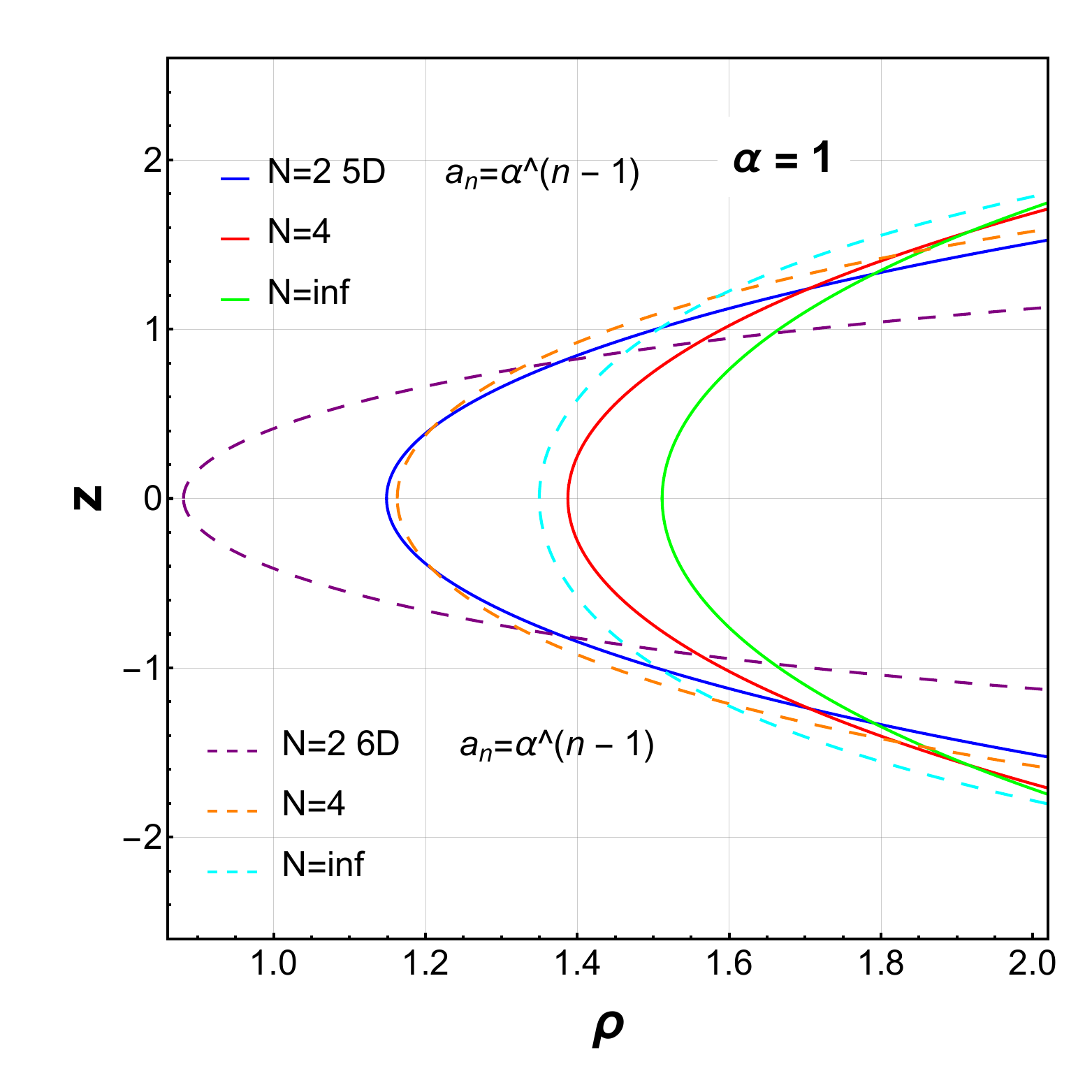} \\
\textit{(a)}  & \textit{(b)} \\[0.7ex]
\includegraphics[width=0.45\textwidth]{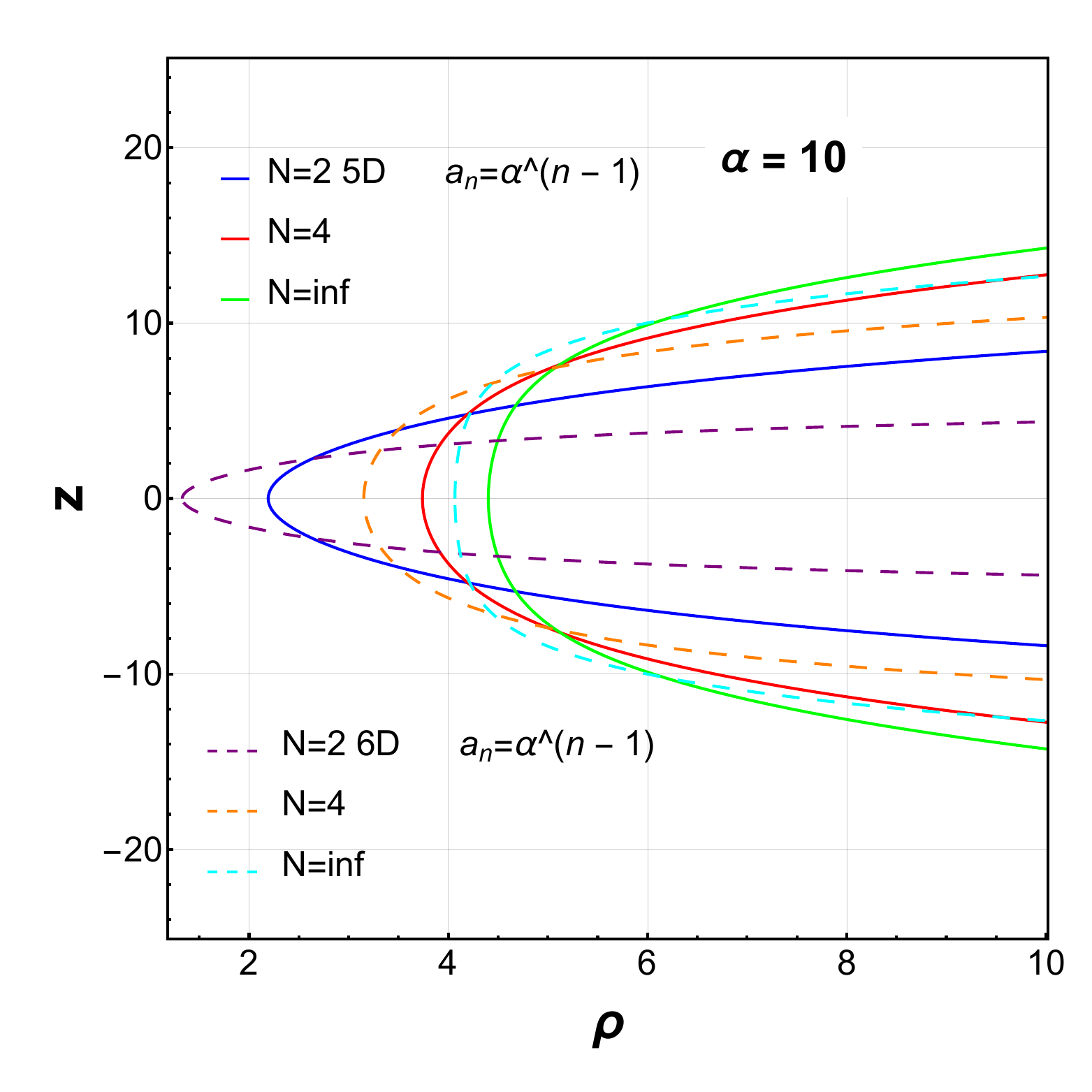} &
\includegraphics[width=0.45\textwidth]{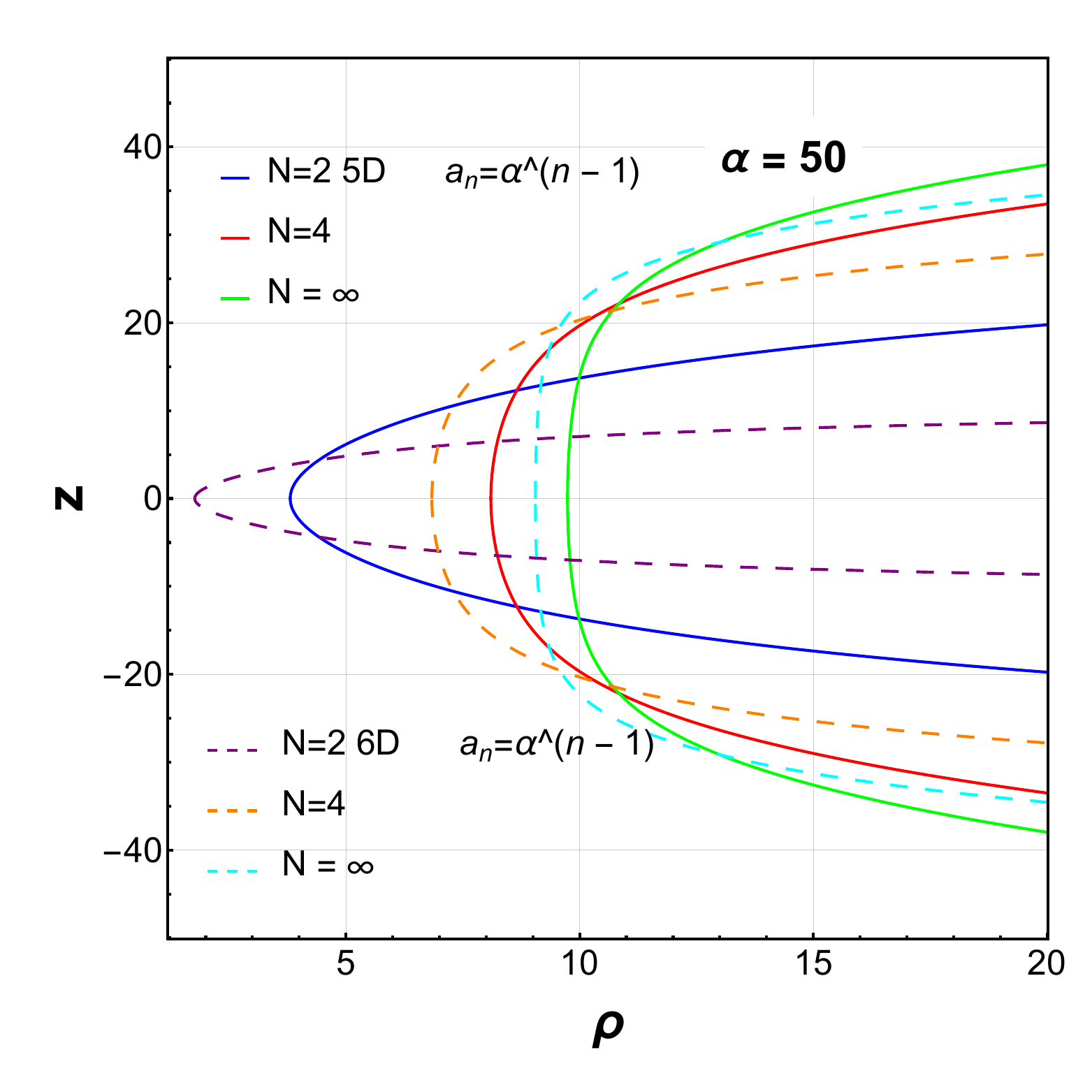} \\
\textit{(c)}  & \textit{(d)} 
\end{tabular}
\caption{Wormhole geometry for different values of the coupling parameter $\alpha$. Solid and dashed curves correspond to $D=5$ and $D=6$, respectively.}
\label{fig:geo_5D6D_alpha}
\end{figure}

We now study the geometric properties of the wormhole throat and examine how they respond to the coupling parameter $\alpha$, the truncation order $N$, and the spacetime dimension. 
An overview of the throat variation is provided by the isometric embedding profiles shown in Figs.~\ref{fig:geo}, \ref{fig:geo2}, and \ref{fig:geo_5D6D_alpha}.

For finite truncation orders, the response of the throat geometry depends sensitively on whether $\alpha$ is positive or negative, as well as on the specific value of $N$.
In the $N=2$ case, increasing $\alpha$ leads to a clear enlargement of the circumferential radius and a progressive smoothing of the embedding profile in the vicinity of the throat (Fig.~\ref{fig:geo}(a)). 
By contrast, along the $\alpha<0$ branch for $N=3$, the geometric response of the throat becomes non-monotonic. 
As $\alpha$ decreases from $0$ toward negative values, the circumferential radius of the throat first decreases and the embedding profile sharpens in the vicinity of the throat. 
When $\alpha$ is decreased further to larger negative values, this trend reverses, and the geometry gradually transitions toward a smoother behavior similar to that observed in the $N=2$ case (see Fig.~\ref{fig:geo}(b)). 
In the all-order limit $N=\infty$, this effect is even more pronounced: for sufficiently large negative $\alpha$, the throat radius decreases rapidly and the throat region develops a markedly pointed structure, as shown in Fig.~\ref{fig:geo}(c).

The influence of the coefficient configuration $\{\alpha_n\}$ is illustrated by comparing the solid and dashed sequences in Fig.~\ref{fig:geo2}. 
For relatively small coupling, the two configurations yield very similar embedding diagrams, with the dashed sequence remaining closer to the $\alpha=0$ geometry. 
As $\alpha$ increases, differences between the two configurations become more apparent: for larger truncation orders such as $N=3$ and $N=\infty$, the solid configuration consistently produces a larger throat radius and a flatter overall embedding shape at fixed $\alpha$, a feature that is particularly evident at $\alpha=10$.

\begin{figure}[H] \centering \setlength{\tabcolsep}{10pt} \renewcommand{\arraystretch}{1.1} \begin{tabular}{cc} \includegraphics[width=0.45\textwidth]{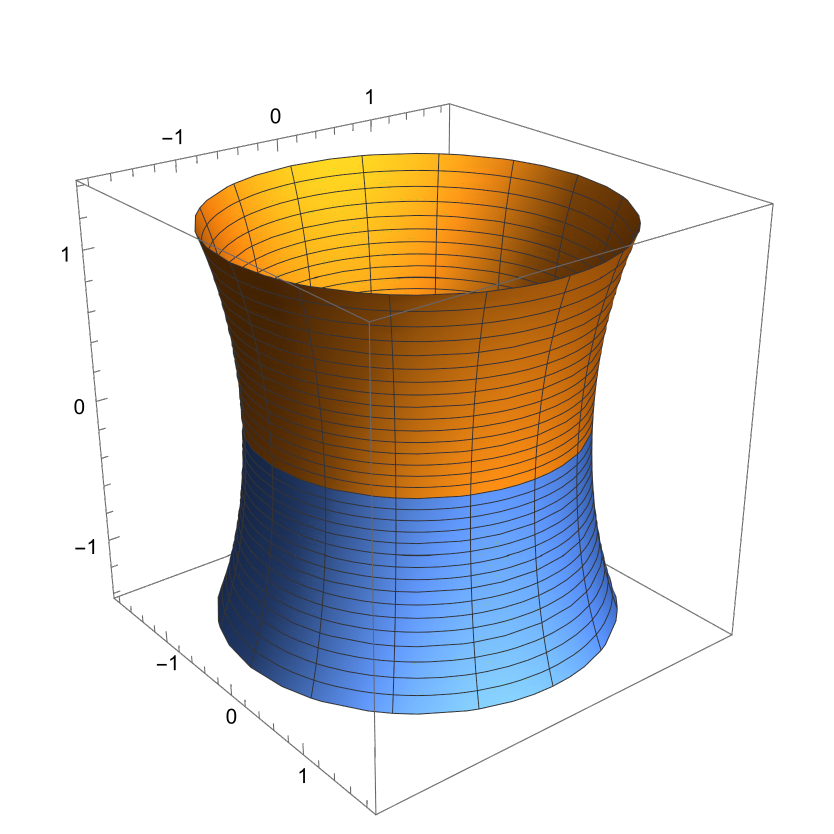} & \includegraphics[width=0.45\textwidth]{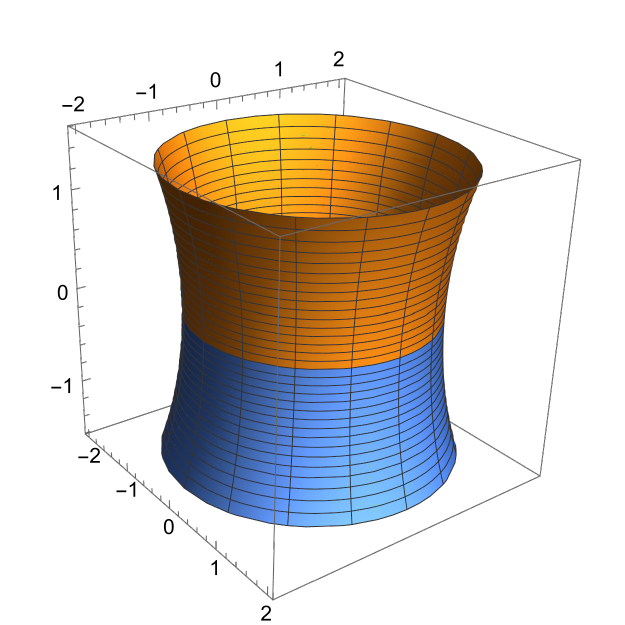} \\ \textit{(a)} & \textit{(b)} \end{tabular} \caption{Three-dimensional embedding diagrams of the wormhole geometry in five dimensions ($D=5$) for the all-order case $N=\infty$ with throat radius $r_0=1$. Panels (a) and (b) correspond to two different coefficient configurations $\{\alpha_n\}$, namely $\alpha_n=\alpha^{\,n-1}$ and $\alpha_n=\frac{1-(-1)^n}{2}\,\alpha^{\,n-1}$, respectively, at the same coupling $\alpha=1$.}
 \label{fig:embed3d} \end{figure}

Finally, the dependence on the spacetime dimension is shown in Fig.~\ref{fig:geo_5D6D_alpha}. 
In the absence of higher-curvature corrections, the six-dimensional wormhole exhibits a smaller throat radius and a more pointed embedding shape than its five-dimensional counterpart. 
As $\alpha$ increases, however, the throat radius grows in both dimensions and the throat region becomes progressively flatter, an effect that is further enhanced at higher truncation orders. 
At sufficiently large coupling, the geometric differences between $D=5$ and $D=6$ are substantially reduced, with the corresponding embedding diagrams approaching each other. Figure~\ref{fig:embed3d} shows the corresponding three-dimensional embedding diagram.

\section{Conclusion and Outlook}\label{sec5}

In this paper, we have constructed and numerically investigated a class of higher-dimensional Ellis--Bronnikov traversable wormhole solutions within quasi-topological gravity with higher-curvature corrections. 
For spacetime dimensions $D>4$ and prescribed coefficient configurations $\{\alpha_n\}$, the overall strength of the higher-curvature sector is characterized by a single coupling parameter $\alpha$, while the truncation order $N$ is varied to explore the impact of higher-order corrections. 
The resulting solutions exhibit a clear reflection symmetry under $r\to -r$ and possess a single throat located at r = 0.

First, our analysis of the total mass $M$ and the scalar charge $\mathcal{D}$ shows that a larger $\alpha$ can significantly increase the mass of the wormhole spacetime, whereas increasing $N$ tends to weaken this mass-growth trend and can keep the scalar charge $\mathcal{D}$ at a very small value, thereby reducing the dependence of the wormhole on the phantom field. However, when considering $\alpha<0$, negative-mass solutions may appear; in this case, $-g_{tt}$ can become larger than $1$. In general, as the higher-curvature effects become stronger, $-g_{tt}$ near the throat becomes extremely small and exhibits an approximate ``horizon''-like structure. Notably, this approximate ``horizon''-like structure is usually not accompanied by an enhancement of the spacetime curvature Hao:2023kvf,Su:2024gxp. On the contrary, the overall magnitude of the Kretschmann scalar $K$ typically decreases as the higher-curvature corrections become stronger, indicating that the geometry near the wormhole throat is effectively smoothed. Concerning the energy conditions, although the violation of the null energy condition is still unavoidable, it becomes numerically very small when $|\alpha|$ is large, indicating that higher-curvature corrections can effectively reduce the level to which the wormhole spacetime violates the null energy conditions. The embedding diagrams also provide a more intuitive illustration of the impact of quasi-topological gravity on the wormhole geometry. We find that higher-curvature corrections typically soften the throat region and increase its circumferential radius, although non-monotonic behavior may occur in specific parameter regions.

In summary, our results demonstrate that quasi-topological gravity provides a viable framework for traversable wormholes. Based on the present study, several directions merit further investigation.
First, it would be interesting to explore viable quasi-topological (or effectively equivalent) frameworks in four dimensions \cite{Glavan:2019inb,Bueno:2025zaj} and construct Ellis-type wormhole models with a more direct connection to realistic physics. 
Second, one may search for wormhole solutions in other quasi-topological settings involving compact stars \cite{Ma:2024olw,Hao:2025utc,Tan:2025hht,Chen:2025iuy}, and study how the presence of compact matter configurations affects the wormhole geometry. 
Third, a systematic stability analysis---including linear perturbations and fully nonlinear numerical evolutions---is essential to assess the dynamical viability and physical realizability of these wormhole solutions.

\section*{Acknowledgements}

This work is supported by the National Key Research and Development Program of China (Grant No. 2022YFC2204101 and 2020YFC2201503) and the National Natural Science Foundation of China (Grant No. 12275110 and No. 12247101).

\end{CJK}
\end{document}